\documentclass{aastex62}

\usepackage{threeparttable}
\usepackage{makecell}
\usepackage{hyperref}
\usepackage{natbib}

\graphicspath{{./}{figures/}}

\begin{document}
\title{The Hubble PanCET Program: A Metal-rich Atmosphere for the Inflated Hot Jupiter HAT-P-41b}
\author[0000-0003-4552-9541]{Kyle B. Sheppard}
\affil{Department of Astronomy, University of Maryland at College Park, College Park, MD, 20742, USA}
\affil{Solar System Exploration Division, NASA's Goddard Space Flight Center, Greenbelt, MD 20771, USA}
\author[0000-0003-0156-4564]{Luis Welbanks}
\affil{Institute of Astronomy, University of Cambridge, Madingley Road Cambridge CB3 0HA, UK}
\author[0000-0002-8119-3355]{Avi M. Mandell}
\affil{Solar System Exploration Division, NASA's Goddard Space Flight Center, Greenbelt, MD 20771, USA}
\author[0000-0002-4869-000X]{Nikku Madhusudhan}
\affil{Institute of Astronomy, University of Cambridge, Madingley Road Cambridge CB3 0HA, UK}
\author[0000-0002-6500-3574]{Nikolay Nikolov}
\affil{Space Telescope Science Institute, 3700 San Martin Drive, Baltimore, MD 21218, USA}
\author[0000-0001-5727-4094]{Drake Deming}
\affil{Department of Astronomy, University of Maryland at College Park, College Park, MD, 20742, USA}
\author[0000-0003-4155-8513]{Gregory W. Henry}
\affil{Center of Excellence in Information Systems, Tennessee State University, Nashville, TN 37209, USA}
\author{Michael H. Williamson}
\affil{Center of Excellence in Information Systems, Tennessee State University, Nashville, TN 37209, USA}
\author[0000-0001-6050-7645]{David K. Sing}
\affil{Department of Physics \& Astronomy, Johns Hopkins University, Baltimore, MD 21218, USA}
\affil{Department of Earth \& Planetary Sciences, Johns Hopkins University, Baltimore, MD 21218, USA}
\author[0000-0003-3204-8183]{Mercedes L\'opez-Morales}
\affil{Center for Astrophysics ${\rm \mid}$ Harvard {\rm \&} Smithsonian, 60 Garden Street, Cambridge, MA 01238, USA}
\author[0000-0003-2775-653X]{Jegug Ih}
\affil{Department of Astronomy, University of Maryland at College Park, College Park, MD, 20742, USA}
\author[0000-0002-1600-7835]{Jorge Sanz-Forcada}
\affil{Centro de Astrobiolog\'{i}a (CSIC-INTA), Spain}
\author[0000-0002-5360-3660]{Panayotis Lavvas}
\affil{Groupe de Spectrom\'etrie Moleculaire et Atmosph\'erique, Universit\'e de Reims Champagne Ardenne, Reims, France}
\author{Gilda E. Ballester}
\affil{Lunar \& Planetary Laboratory, Department of Planetary Sciences, University of Arizona, Tucson, AZ, USA}
\author[0000-0001-5442-1300]{Thomas M.\ Evans}
\affil{Kavli Institute for Astrophysics and Space Research, Massachusetts Institute of Technology, 77 Massachusetts Avenue, 37-241, Cambridge, MA 02139, USA}
\author[0000-0003-1756-4825]{Antonio Garc\'ia Mu\~noz}
\affil{Zentrum f\"ur Astronomie und Astrophysik, Technische
Universit\"at Berlin, Berlin, Germany}
\author[0000-0002-2248-3838]{Leonardo A. dos Santos}
\affil{Observatoire Astronomique de l\'Universit\'e de Geneve, 51 chemin des Maillettes, 1290 Versoix, Switzerland}

\correspondingauthor{Kyle Sheppard}
\email{kshep23@umd.edu}

\begin{abstract}

We present a comprehensive analysis of the 0.3--5\,$\mu$m transit spectrum for the inflated hot Jupiter HAT-P-41b. The planet was observed in transit with Hubble STIS and WFC3 as part of the Hubble Panchromatic Comparative Exoplanet Treasury (PanCET) program, and we combine those data with warm \textit{Spitzer} transit observations. We extract transit depths from each of the data sets, presenting the STIS transit spectrum (0.29--0.93\,$\mu$m) for the first time. We retrieve the transit spectrum both with a free-chemistry retrieval suite (AURA) and a complementary chemical equilibrium retrieval suite (PLATON) to constrain the atmospheric properties at the day-night terminator. Both methods provide an excellent fit to the observed spectrum. Both AURA and PLATON retrieve a metal-rich atmosphere for almost all model assumptions (most likely O/H ratio of $\log_{10}{Z/Z_{\odot}} = 1.46^{+0.53}_{-0.68}$ and $\log_{10}{Z/Z_{\odot}} = 2.33^{+0.23}_{-0.25}$, respectively); this is driven by a 4.9-$\sigma$ detection of H$_2$O as well as evidence of gas absorption in the optical ($>$2.7-$\sigma$ detection) due to Na, AlO and/or VO/TiO, though no individual species is strongly detected. Both retrievals determine the transit spectrum to be consistent with a clear atmosphere, with no evidence of haze or high-altitude clouds. Interior modeling constraints on the maximum atmospheric metallicity ($\log_{10}{Z/Z_{\odot}} < 1.7$) favor the AURA results. The inferred elemental oxygen abundance suggests that HAT-P-41b has one of the most metal-rich atmospheres of any hot Jupiters known to date. Overall, the inferred high metallicity and high inflation make HAT-P-41b an interesting test case for planet formation theories.

\end{abstract}

\section{Introduction}\label{sec:intro}
Transit spectroscopy has been fundamental in understanding the physics and chemistry of hot exoplanet atmospheres. Transit observations with the Hubble Space Telescope (HST) and the \textit{Spitzer} Space Telescope have been especially fruitful in illuminating the composition and atmospheric structure of close-in planets, starting with the first measurements of sodium absorption \citep{charbonneau2002} and the first detection of thermal emission \citep{deming2005} for the atmosphere of HD209458b.

The installation of the Wide Field Camera 3 (WFC3) instrument and the refurbishment of the Space Telescope Imaging Spectrograph (STIS) on HST opened up a new era of transit spectroscopy measurements for hot Jupiters.  WFC3 has provided the first repeatable and well-validated detections of the presence of water vapor \citep{deming2013,huitson2013,wakeford2013,mandell2013}, and has opened the field to population studies looking at H$_2$O abundance and metallicity as a function of stellar and planetary properties \citep{sing2016,tsiaras2018,pinhas2019, Welbanks2019b}. The upgraded STIS instrument has been a key contributor in illuminating the critical role that aerosols play in driving the continuum opacity for transit measurements of hot planets \citep{pont2013,nikolov2014,sing2016,chachan2019}.

One of the most intriguing topics from these studies is the question of atmospheric metallicity. Studies of individual planets suggested a wide diversity of atmospheric metallicity as a function of planetary mass \citep[e.g.,][]{Madhusudhan2014a,Kreidberg2014,wakeford2017,wakeford2018}. However, recent homogenous statistical analyses of many planets reveal that a paucity of water vapor in hot planet atmospheres is the norm \citep{barstow2017,pinhas2018,Welbanks2019b}. One strategy to investigate the relationships between mass and atmospheric metallicity is to study the best targets within the Saturn and Jupiter mass range, in order to achieve high S/N and leverage the expectation of a high primordial gas fraction and large transit signals.  

First discovered in 2012 \citep{hartman2012}, the inflated hot Jupiter HAT-P-41b (T$_{eq}$=1940K, P=2.7 days) is a strong candidate to inform these trends. It is among the most inflated hot Jupiters (R=$1.69 R_{Jup}$, M=$0.8M_{Jup}$), and it orbits a relatively inactive mid-F dwarf (R=$1.68 R_{\odot}$, T$_{\textrm{eff}}=6390$~K). HAT-P-41b's extended atmosphere and its host star's lack of significant variability make it highly amenable to characterization through transit spectroscopy. \citet{johnson2017} determined the spin-orbit misalignment of the system to be moderate (-22$^{\circ}$), while the host star appears to be part of a multi-stellar system, with a wide-orbit late-type companion discovered at $\sim1000$ AU \citep{hartman2012,wollert2015,evans2016}. 

\citet{tsiaras2018} retrieved the WFC3 G141 grism spectrum (1.1--1.7\,$\mu$m) with $\tau$-Rex \citep{waldmann2015b, waldmann2015}, confirming an $4.2\sigma$ atmospheric detection and finding no evidence of contributions from either high-altitude clouds or photochemical hazes \citep[e.g.,][]{zahnle2009}. \citet{tsiaras2018} also found evidence of and abundance constraints for water vapor ($\log_{10}$(X$_{\text{H}_2\text{O}})=-2.77 \pm{1.09}$), though due to narrow wavelength coverage abundance uncertainties are large. Still, they are able rule out upper atmospheric water depletion. \citet{fisher2018} built upon this result by retrieving on the same dataset with a focus on cloud opacity and other near-infrared opacity sources (NH$_3$, HCN). They find a water abundance of $-0.9 ^{+ 0.28 }_{- 1.20 }$ which they note is generally consistent with that of \citet{tsiaras2018}. However, this reported abundance is for a cloud-free model with only H$_2$O and NH$_3$ as opacity sources, and this simplified treatment is not necessarily directly comparable with more comprehensive atmospheric models. They also find weak evidence of NH$_3$, though they are unable to favor the NH$_3$ and H$_2$O model over a model with grey clouds and H$_2$O.

Wide spectral baselines provide the potential for a more complete and constrained understanding of atmospheric properties \citep{benneke2012,griffith2014, Welbanks2019}. For example, \citet{line2016} demonstrated how individual WFC3 spectra are unable to constrain mean molecular weight due to a degeneracy with partial clouds. Furthermore, for a fully homogeneous cloud cover the cloud-top pressure is degenerate with the chemical abundance \citep[e.g.,][]{deming2013}. \citet{Welbanks2019} showed that optical data help alleviate such degeneracies and improve the precision with which planetary radius, cloud properties, and molecular/atomic abundances are inferred. As a practical example, \citet{sing2016} utilized optical-to-infrared spectra to jointly constrain cloud, haze, and chemistry parameters for a sample of ten hot Jupiters. STIS data have been specifically useful in constraing the atmospheric metallicity of giant exoplanets \citep{wakeford2017, wakeford2018, chachaan2019}.

Bayesian spectral retrievals are the most reliable way to interpret exoplanet spectra, due to their flexibility in describing diverse exoplanet atmospheres and their ability to evaluate the full posterior distribution of a forward model's parameters \citep{madhusudhan2018}. This allows for understanding not only the properties of an exoplanet's atmosphere, but also the uncertainties on those properties. Consequently, such retrieval codes are common in atmospheric characterization \citep[and many others]{Madhusudhan2009,madhusudhan2010, lee2012,benneke2012,line2013,amundsen2014, waldmann2015,barstow2017}. Nested sampling \citep{skilling2004} is a particularly powerful Bayesian sampler, as it naturally determines the Bayesian evidence of the fitted model (with the posterior distribution being a byproduct), which is necessary for correct model comparison (e.g., justifying more complicated models, reporting correct detection significances). 

Despite their ubiquity, each retrieval code is necessarily unique given the assumptions and modeling choices that must be made. Though these retrievals generally agree, subtle discrepancies can lead to different conclusions for the same data \citep{kilpatrick2017, fisher2018, barstow2020}. Examples include different chemical parameterizations (i.e., enforcing chemical equilibrium), cloud parameterizations, opacity sources, and prior assumptions. Therefore, it is important to understand the effect of modeling assumptions on the retrieved atmospheric parameters \citep[e.g.,][]{Welbanks2019}. Testing a suite of models for a given retrieval code --- and, even better, different modeling paradigms altogether --- more accurately captures the uncertainty in the atmospheric parameters. It is important to be transparent about the assumptions made in a retrieval analysis in order to best contextualize and understand the results. 

In this paper we derive the 0.3--5\,$\mu$m transit spectrum of HAT-P-41b using transit observations from HST/STIS, HST/WFC3, and \textit{Spitzer} (Sec.~\ref{sec:obs}). Section~\ref{sec:properties} characterizes both the variability (incorporating both X-ray and visible photometric monitoring; Sec.~\ref{sec:activity}) and the parameters (Sec.~\ref{sec:gaia}) of the host star. Section~\ref{sec:data} describes the data analysis to derive the transit spectrum. Sec.~\ref{sec:retrieve} describes that we use two different retrieval methods. First, we use a chemical-equlibirum framework \citep[PLanetary Atmospheric Transmission for Observer Noobs; PLATON;][Sec.~\ref{sec:platon_retrieval}]{zhang2019}, and those results are described in Sec.~\ref{sec:results}. We also explore a more flexible free-chemistry retrieval using the AURA framework \citep[Sec.~\ref{sec:AURA_retrieval}]{pinhas2018}. The two retrieval analyses were independently done by different members of the team to allow for an unbiased comparison. We conclude that a high, supersolar atmospheric metallicity (on the order of $30$--$200\times$ solar O/H) best describes the observed spectrum, and --- though median retrieved values differ  --- this result is not sensitive to model assumptions. Sec.~\ref{sec:discussion} discusses the comparison between retrievals (Sec.~\ref{subsec:compare}, the comparison to interior modeling constraints (Sec~\ref{subsec:interior}, and the implications for planet formation (Sec~\ref{sec:form}).  Sec.~\ref{sec:summary} provides a summary of our conclusions.

\section{Observations} \label{sec:obs}

\subsection{HST}
We observed one transit of HAT-P-41b with the WFC3 instrument on HST and three transits with the STIS instrument as part of the PanCET Program (ID 14767, P.I. Sing).  The WFC3 observations were taken on October 16, 2016 using the G141 prism, which covers a wavelength range of approximately 1.1-1.7\,$\mu$m with a spectral resolving power of R$\sim$150. The STIS observations were taken with the G430L and G750L grisms, which cover a wavelength range of approximately 0.3-1.0\,$\mu$m with a spectral resolving power of R$\sim$500. The STIS data were acquired on September 4 2017 (G430L, visit 83), May 7 2018 (G430L, visit 84) and June 11 2018 (G750L, visit 85).  For each visit, the target was observed for 7 hours over five consecutive HST orbits. An HST gyro issue prevented the acquisition of the third orbit for visit 85. 

For the WFC3 observations, data were taken in spatial scan spectroscopic mode with a forward scanning rate of 0.065 arcsec s$^{-1}$ along the cross-dispersion axis, resulting in scans across approximately 46 pixel rows. We utilized the 256 $\times$ 256 pixel subarray and the SPARS-10 sampling sequence, with 12 non-destructive reads (NSAMP = 12) resulting in a total integration time of 81 seconds for each exposure. We obtained a total of 17 exposures in the first HST orbit following acquisition and 19 exposures in each subsequent HST orbit. Typical peak frame counts were $\sim$33,000 electrons per pixel, which is within the linear regime of the WFC3 detector.

For the STIS observations, each visit consisted of 5 orbits, with $\sim45$\,min gaps due to Earth occultations. We utilized the wide 52 $\times$ 2 arcsec slit to minimize slit light losses and an integration time of $\sim$253 sec for each exposure, for a total of 48 spectra for each visit. Data acquisition overheads were minimized by reading-out a subarray of the CCD with a size of 1024 $\times$ 128 pixels.

\subsection{Spitzer}
The Spitzer Infrared Array Camera (IRAC) observations were taken in January and February 2017 as part of Program 13044 (P.I. D. Deming).  A single transit of HAT-P-41b was observed in each of the IRAC1 (3.6\,$\mu$m) and IRAC2 (4.5\,$\mu$m) channels.  Each transit was preceeded by a 30-minute peakup sequence that also mitigates the steepest portion of a temporal ramp due to the detector. The transit was observed over $\sim$12 hours, with equivalent in-transit and out-of-transit coverage.  338 exposures were obtained for each transit, and each exposure consisted of 64 subarray frames of 32x32 pixels, using an exposure time of 2.0 seconds per frame.  

\subsection{Photometric Monitoring Observations}\label{sec:phot}

To better diagnose the likelihood of stellar variability impacting the transit spectrum, we complemented our transit observations with monitoring observations at both visible (AIT) and X-ray wavelengths (XMM-Newton). XMM-Newton observed HAT-P-41 on 2017-April-07 with an overall 17 ks exposure time (Proposal ID 80479, P.I. J. Sanz-Forcada). The target was
not detected in any of the EPIC detectors; we discuss the implications of this in Section~\ref{sec:activity}.

We obtained nightly ground-based photometric observations of HAT-P-41 during its 2018 and
2019 observing seasons with the Tennessee State University Celestron 14-inch 
(C14) automated imaging telescope (AIT) located at Fairborn Observatory in the 
Patagonia Mountains of southern Arizona \citep[see, e.g.,][]{h1999,ehf2003}.
The AIT is equipped with an SBIG STL-1001E CCD camera; observations were made
through a Cousins R filter.  Details of our observing, data reduction, and 
analysis procedures are described in \citet{sws+2015}.

We collected a total of 207 successful nightly observations (excluding a few 
isolated transit observations) over the two observing seasons.  Our observing
activities at Fairborn must come to a halt each year during the southern
Arizona rainy season, which typically lasts from approximately July 1 to 
September 10.  Since HAT-P-41 comes to opposition around July 18, each 
observing season is broken into two intervals, which we designate as 
intervals A and B. Information for a portion of the AIT observations are shown in Table~\ref{tab:ait_full}; the full table is available in the electronic edition of {\it ApJ}.

\begin{deluxetable}{ccccc}
\tabletypesize{\small}
\tablewidth{0pt}
\tablecaption{AIT photometric observations of HAT-P-41}\label{tab:ait_full}
\tablehead{
\colhead{Hel. Julian Date} & \colhead{Delta $R$} & \colhead{Sigma} \\
\colhead{(HJD $-$ 2,400,000)} & \colhead{(mag)} & \colhead{(mag)} \\
\colhead{(1)} & \colhead{(2)} & \colhead{(3)} \\
}
\startdata
58175.0255 & $-$0.56702 & 0.00122 \\
58176.0214 & $-$0.56986 & 0.00109 \\
58180.0102 & $-$0.56636 & 0.00052 \\
58181.0158 & $-$0.56444 & 0.00293 \\
58182.0022 & $-$0.56479 & 0.00185 \\
58184.9980 & $-$0.56390 & 0.00182 \\
\enddata
\tablecomments{Table~\ref{tab:ait_full} is presented in its entirety in the electronic 
edition of {\it ApJ}.  A portion is shown here for guidance regarding 
its form and content.}
\end{deluxetable}

\section{Stellar Properties}
\label{sec:properties}

\subsection{Analysis of Stellar Variability}\label{sec:activity}

\begin{deluxetable}{ccccc}
\tablewidth{0pt}
\tablecaption{Results of the analysis of photometric monitoring} observations for HAT-P-41\label{tab:ait}
\tablehead{
\colhead{Observing} & \colhead{} & \colhead{Date Range} & 
\colhead{Sigma} & \colhead{Seasonal Mean} \\
\colhead{Season} & \colhead{$N_{obs}$} & \colhead{(HJD $-$ 2,400,000)} & 
\colhead{(mag)} & \colhead{(mag)} \\
\colhead{(1)} & \colhead{(2)} & \colhead{(3)} & 
\colhead{(4)} & \colhead{(5)} 
}
\startdata
 2018 A & 110 & 58175--58295 & 0 00224 & $-0.56496\pm0.00021$ \\
 2018 B &  34 & 58386--58451 & 0.00291 & $-0.56979\pm0.00051$ \\
 2019 A &  42 & 58577--58657 & 0.00264 & $-0.56966\pm0.00041$ \\
 2019 B &  21 & 58756--58802 & 0.00394 & $-0.56602\pm0.00088$ \\
\enddata
\end{deluxetable}

The results of analysis of the AIT photometric observations (Sec.~\ref{sec:phot}) are given in Table~\ref{tab:ait}. The low numbers of observations
in 2018 B, 2019 A, and 2019 B are the result of the unusually cloudy weather at Fairborn for the past two years. This cloudy weather 
pattern continues to the present. Column~4 of the table gives the standard 
deviation of the individual observations with respect to their corresponding seasonal mean. The standard deviations range between 0.00224 and 0.00394~mag for the four observing intervals. This is near the limit of our nightly 
measurement precision with the C14, as determined from the constant comparison 
stars in the field. Periodogram analyses of the four intervals reveal no 
significant periodicities. The scatter in the seasonal means given in column 5 is consistent with the expected photometric precision considering the small number of observations in the last three intervals and the marginal photometric conditions prevalent at Fairborn Observatory over the past two years. Therefore, HAT-P-41 appears to be constant on night-to-night and year-to-year timescales to the limit of our precision.

Additionally, HAT-P-41 was not detected in any of the XMM Newton's EPIC detectors. Given the distance of the object we can set an upper limit of $L_{\rm X}=1\times 10^{29}$ erg\,s$^{-1}$ on the stellar X-ray luminosity. This implies a value of $\log L_{\rm X}/L_{\rm bol}<-5.2$, indicating that the star has a moderate activity level at most \citep{wright2011}.


The photometric observations of HAT-P-41 describe a relatively quiet star. Furthermore, the Ca II chromospheric activity index (S = 0.18) and the corresponding estimated parameter flux $\log{R'_{HK}}$ (-5.04) for HAT-P-41's spectral type (B-V = 0.29) are not indicative of high activity \citep{hartman2012, noyes1984} and may indicate instead a basal-level activity \citep{Isaacson2010}. \citet{rackham2019} show that for all but the most active F-dwarfs variability does not result in any detectable change to the transit spectrum. Specifically, the impact of potential complications such as false TiO/VO detections, false water detections, and optical offsets are all determined to be less than $\sim10$ppm. Therefore, we conclude that stellar variability is unlikely to contaminte HAT-P-41b's transit spectrum.

\subsection{Stellar Parameters}\label{sec:gaia}
Inferred atmospheric planetary parameters are directly dependent on host star parameters. For our analyses, we incorporate the stellar parameters from TESS Input Catalog --- version 8 \citep[TIC-8;][]{stassun2019b, stassun2019a}.  TIC-8 provides reliable stellar parameters for planetary host stars based primarily on Gaia Data Release 2 (GDR2) point sources \citep{gaia2016, gaia2018}. The algorithm for HAT-P-41's parameters is as follows: distance is first derived from Gaia DR2 parallax, using a correct inference procedure \citep{bj2018}. HAT-P-41's galactic longitude (-10.6) puts it in a region where uncertainty on reddening makes determining effective temperature from Gaia photometry difficult. As a result, a spectroscopically-derived effective temperature \citep[from the PASTEL catalog][]{soubiran2016} is preferred. The stellar radius and mass are then self-consistently derived from the distance and effective temperature \citep{andrae2018}. Finally, $\log{g_s}$ is calculated from the stellar radius and mass.

\begin{table*}
\centering       
\begin{threeparttable}
\caption{System parameters for HAT-P-41}

\begin{tabular}{l r r}     
\hline\hline      
    Parameter & TIC-8\tnote{a} & Discovery Paper \tnote{b}\\
\hline
    $R_s [R_{Sun}]$ & $1.65^{+0.08}_{-0.06}$ & $1.683^{+0.058}_{-0.036}$\\
    $M_s [M_{Sun}]$ & $1.32^{+0.25}_{-0.16}$ & $1.42\pm{0.047}$\\
    $\log{g_s}$ [cgs units] & $4.12^{+0.11}_{-0.06}$ & $4.14\pm{0.02}$\\
    $T_{s,\textrm{eff}}$ [K] & $6480^{+100}_{-100}$ & $6390\pm{100}$\\
    $R_p [R_{Jup}]$ & $1.65^{+0.08}_{-0.07}$ & $1.685^{+0.076}_{-0.051}$\\
    $M_p [M_{Jup}]$ & $0.76^{+0.14}_{-0.12}$ & $0.80\pm{0.10}$\\
    $\rho_p$ [g~cm$^{-3}$] & $0.21\pm{0.05}$ & $0.20\pm{0.03}$\\
    $T_{p,\textrm{eq}}$ [K] & $1960^{+40}_{-35}$ & $1940\pm{38}$\\
    $a$ [AU] & $0.0418^{+0.0021}_{-0.0019}$ & $0.0426\pm{0.0005}$\\
    Distance [pc] & $348\pm{4.5}$ & $344^{+12}_{-8}$\\
\hline
\end{tabular}
\begin{tablenotes}
\item[a] Provided by or derived from Tess Input Catalog \citet{stassun2019b}
\item[b]\citet{hartman2012}

\end{tablenotes}
\label{tab:sysparams}
\end{threeparttable}
\end{table*}

It is important to recalculate 
$R_{p}$, $M_p$, and semi-major axis $a$ based on the $R_s$ value from TIC-8, since those are derived in the discovery paper assuming a certain value for $R_s$. As a simple example, $R_p$ is derived by constraining $R_p/R_s$ in transit and multiplying by $R_s$. To re-derive the planetary parameters, we follow the methodology of \citet{stassun2017}. The resulting values and 1-$\sigma$ ranges are shown in Table~\ref{tab:sysparams} along with the values from the discovery paper \citep{hartman2012}. 

We favor the TIC-8 stellar parameters over the discovery paper values \citep[derived using isochrones and high-resolution spectroscopy;][]{hartman2012} since they are based on more recent and comprehensive data. We emphasize that the two sets of parameters are consistent to better than 1-$\sigma$, and using the discovery paper values has no impact on the conclusions of this paper.

A planet's composition is directly linked to its host star's composition. \citet{brewer2018} determined the stellar abundance of 15 different elements for HAT-P-41 as part of the Spectral Properties of Cool Stars (SPOCS) catalog. Table~\ref{tab:brewer} gives the abundances, relative to solar, for the relevant elements. \citet{brewer2018} find an effective temperature, metallicity, and $\log{g_s}$ consistent with both TIC-8 and the discovery paper. HAT-P-41 is a metal-enriched star, and notably has an elemental oxygen abundance of $2.3\times$ solar. Carbon is the only depleted element at $\sim0.8\times$ solar, resulting in a subsolar C/O ratio of 0.19 ($0.36\times$ solar).

\begin{table*}
\centering       
\begin{threeparttable}
\label{tab:brewer}
\caption{HAT-P-41 Elemental Abundances}  

\begin{tabular}{l r}     
\hline\hline      
    Elemental ratio & Abundance (log solar unit)\\
\hline
    $[$O/H$]$ & $0.37\pm{0.04}$ \\
    $[$C/H$]$ & $-0.08\pm{0.03}$ \\
    $[$Na/H$]$ & $0.17\pm{0.01}$ \\
    $[$Ti/H$]$ & $0.22\pm{0.01}$ \\
    $[$V/H$]$  & $0.09\pm{0.03}$ \\
    $[$Al/H$]$ & $0.07\pm{0.03}$ \\
    $[$M/H$]$  & $0.18\pm{0.01}$ \\
    $[$C/O$]$ & $-0.45\pm{0.05}$ \\
\hline
\end{tabular}
\tablecomments{All values from \citet{brewer2018b}}

\label{tab:sysparams}
\end{threeparttable}
\end{table*}

\section{Data Analysis} \label{sec:data}

\subsection{STIS}\label{sec:stis}
\subsubsection{Data Reduction}
Our data analysis procedures follow the general methodology detailed in \citet{nikolov2014, nikolov2015}. We commenced analysis from the flt.fits files, which were reduced (bias-, dark- and flat-corrected) using the latest version of the CALSTIS pipeline and the latest calibration frames. We used median combined difference images to identify and correct for cosmic-ray events in the images as described by \citet{nikolov2014}. We found $\sim4$ percent of the detector pixels were affected by cosmic ray events. We also corrected pixels identified by CALSTIS as bad with the same procedure, which together with the cosmic ray identified pixels resulted in a total of $\sim$14 percent interpolated pixels.

We performed spectral extraction with the IRAF procedure APALL using aperture sizes in the range from 6 to 18 pixels with a step of 0.5. The best aperture for each grating was selected based on the resulting lowest light-curve residual scatter after fitting the white light curves. We found that aperture sizes 13.5, 13.5 and 10.5 pixels satisfy this criterium for visits 83, 84 and 85, respectively. 

We cross-correlated and interpolated all spectra with respect to the first spectrum to prevent sub-pixel wavelength shifts in the dispersion direction. The STIS spectra were then used to extract both white-light spectrophotometric time series and custom wavelength bands after summing the appropriate flux from each bandpass.

The raw STIS light curves exhibit instrumental systematics on the spacecraft orbital time-scale, which are attributed to thermal contraction/expansion (referred to as the 'breathing effect') as the spacecraft warms up during its orbital day and cools down during orbital night. We take into account the systematics associated with the telescope temperature variations in the transit light-curve fits by fitting a baseline function depending on various parameters. 

\subsubsection{Light Curve Analysis}
White and spectroscopic light curves were created from the time series of each visit by summing the flux of each stellar spectrum along the dispersion axis. We fit each transit light curve using a two-component function that simultaneously models the transit and systematic effects. The transit model was computed using the analytical formulae given in \citet{mandel2002}, which are parametrized with the mid-transit times (T$_{{\rm{mid}}}$), orbital period (P) and inclination (i), normalized planet semimajor axis (a/R$_{\ast}$), and planet-to-star radius ratio (R$_{{\rm{p}}}$/R$_{\ast}$).

Stellar limb-darkening was accounted for by adopting the four parameter non-linear limb-darkening law with coefficients c1, c2, c3 and c4, computed using a three-dimensional stellar atmosphere model grid \citep{magic2015}. We adopted the closest match to the effective temperature, surface gravity, and metallicity values for HAT-P-41 determined by \citet{hartman2012}.

As in our past STIS studies, we applied orbit-to-orbit flux corrections by fitting for a fourth-order polynomial to the spectrophotometric time series phased on the HST orbital period and a linear time term. We also used a low-order polynomial (up to a third degree with no cross terms) of the spectral displacement in the dispersion and cross dispersion direction. The first exposures of each HST orbit exhibit lower fluxes and have been discarded in the analysis. Similar to our past HST STIS analyses \citep{nikolov2014,sing2016}, we intended to discard the entire first orbit to minimize the space craft thermal breathing trend, but found that for two of the three HST visits, a few of the exposures taken toward the end of the first orbit can be used in the analysis.  

We then generated systematics models that spanned all possible combinations of detrending variables and performed separate fits using each systematics model included in the two-component function. The Akaike information criterion \citep[AIC;][]{akaike1974} was calculated for each attempted function and used to marginalize over the entire set of functions following \citet{gibson2014}. Our choice to rely on the AIC instead of the Bayesian information criterion \citep[BIC;][]{bic1978} was determined by the fact that the BIC is more biased toward simple models than the AIC. The AIC therefore provides a more conservative model for the systematics and typically results in larger or more conservative error estimates, as demonstrated by \citet{gibson2014}. Marginalization over multiple systematics models assumes equal prior weights for each model tested.

For the white light curves, we fixed the orbital period, inclination and a/Rs to the values reported in Table\,\ref{tab:transit_params} and fit for the transit mid-time and planet-to-star radius ratio. We find central transit times Tc[MJD]$= 58000.6958\pm0.0029$ (visit 83), Tc[MJD]$= 58245.85414\pm0.00039$ (visit 84), Tc[MJD]$=  58280.87484\pm0.00036$ (visit 85). We derive the white light transit depths to be $10200\pm{104}\,$ppm and $10320\pm{85}\,$ppm for G430L and G750L, respectively.

For the spectroscopic light curves, a common-mode systematics model was established by simply dividing the white-light curve by a transit model \citep{berta2012, deming2013}. We computed the transit model using the orbital period, inclination and a/Rs from Table\,\ref{tab:transit_params} and the central times for each orbit from the whit-light analysis. The common-mode factors from each night were then removed from the corresponding spectroscopic light curves before model fitting.

We then performed fits to the spectroscopic light curves using the same set of systematics models as in the white-light curve analysis and marginalized over them as described above. For these fits, Rp/Rs was allowed to vary for each spectroscopic channel, while the central transit time and system parameters were fixed. We assumed the non-linear limb-darkening law with coefficients fixed to their theoretical values, determined in the same way as for the white-light curve. The detrended spectrophotometric light curves are shown in Figures\,\ref{fig:stis83_bins}, \ref{fig:stis84_bins}, \ref{fig:stis85_bins}. The derived STIS transit spectrum is shown along with the entire transit spectrum in Table~\ref{tab:spectrum}.

\begin{table*}
\centering       
\begin{threeparttable}
\caption{Transit parameters for HAT-P-41b}  

\begin{tabular}{l r}     
\hline\hline      
    Parameter & Value \\
\hline
    $R_p/R_s$ & $0.1028\pm{0.0016}$\\
    $a/R_s$ & $5.44^{+0.09}_{-0.15}$\\
    $i$ [Degrees] & $87.7\pm{1.0}$\\
    $T_c$ [BJD] & $2454983.86167\pm{0.00107}$\\
    $\log{g_p}$ [cgs units] & $2.84\pm{0.06}$ \\
    $P$ [days] & $2.694047\pm{4\times 10^{-6}}$ \\
\hline
\end{tabular}
\begin{tablenotes}
\item All values from discovery paper, \citet{hartman2012}

\end{tablenotes}
\label{tab:transit_params}
\end{threeparttable}
\end{table*}

\begin{table*}
\begin{threeparttable}  
\centering       
\caption{Transit Spectrum}  

\begin{tabular}{ l r r | l r r}     
\hline\hline      
         
    Instrument & Wavelength [$\mu$m] & Depth [ppm]\tnote{a} & Instrument & Wavelength [$\mu$m] & Depth [ppm]  \\
\hline
    STIS G430L\tnote{b} & 0.290--0.350 & 10091 $\pm$ 230 & STIS G750L & 0.711--0.731 & 10499 $\pm$ 364  \\  
    & 0.350--0.370 & 10006 $\pm$ 249  &  & 0.731--0.750 & 10038 $\pm$ 302  \\
    & 0.370--0.387 & 10397 $\pm$ 198 &  & 0.750--0.770 & 10142 $\pm$ 224 \\  
    & 0.387--0.404 & 10273 $\pm$ 163 &  & 0.770--0.799 & 10124 $\pm$ 281  \\ 
    & 0.404--0.415 & 9999 $\pm$ 137  &  & 0.799--0.819 & 10618 $\pm$ 317  \\
    & 0.415--0.426 & 9980 $\pm$ 185  &  & 0.819--0.838 & 9910 $\pm$ 239  \\   
    & 0.426--0.437 & 10324 $\pm$ 145 &  & 0.838--0.884 & 10252 $\pm$ 238  \\
    & 0.437--0.443 & 10428 $\pm$ 287  &  & 0.884--0.930 & 10217 $\pm$ 298  \\ 
    & 0.443--0.448 & 10245 $\pm$ 168  &  WFC3\tnote{e} & 1.122--1.141 & 10297 $\pm$ 107  \\   
    & 0.448--0.454 & 10411 $\pm$ 165  &  & 1.141--1.159 & 10620 $\pm$ 115  \\
    & 0.454--0.459 & 10367 $\pm$ 192  &  & 1.159--1.178 & 10347 $\pm$ 130  \\ 
    & 0.459--0.465 & 10227 $\pm$ 145  &  & 1.178--1.196 & 10479 $\pm$ 113  \\  
    & 0.465--0.470 & 10640 $\pm$ 164  &  & 1.196--1.215 & 10497 $\pm$ 111  \\ 
    & 0.470--0.476 & 10429 $\pm$ 165  & & 1.215--1.233 & 10644 $\pm$ 111  \\  
    & 0.476--0.481 & 10453 $\pm$ 144  &  & 1.233-1.252 & 10289 $\pm$ 100  \\ 
    & 0.481--0.492 & 10584 $\pm$ 125  &  & 1.252--1.271 & 10360 $\pm$ 107  \\
    & 0.492--0.498 & 10224 $\pm$ 192  &  & 1.271--1.289 & 10488 $\pm$ 122  \\ 
    & 0.498--0.503 & 10289 $\pm$ 166  &  & 1.289--1.308 & 10405 $\pm$ 97  \\ 
    & 0.503--0.509 & 10459 $\pm$ 149  &  & 1.308--1.326 & 10396 $\pm$ 94  \\  
    & 0.509--0.514 & 10519 $\pm$ 183  &  & 1.326--1.345 & 10581 $\pm$ 92  \\ 
    & 0.514--0.520 & 10506 $\pm$ 168  &  & 1.345--1.364 & 10684 $\pm$ 105  \\ 
    & 0.520--0.525 & 10429 $\pm$ 186  &  & 1.364--1.382 & 10622 $\pm$ 113  \\ 
    & 0.525--0.531 & 10558 $\pm$ 128  &  & 1.382--1.401 & 10477 $\pm$ 112  \\ 
    & 0.531--0.536 & 10247 $\pm$ 174  &  & 1.401--1.419 & 10689 $\pm$ 98  \\ 
    & 0.536--0.542 & 10451 $\pm$ 180  &  & 1.419--1.438 & 10535 $\pm$ 108  \\ 
    & 0.542--0.547 & 10476 $\pm$ 177  &  & 1.438--1.456 & 10626 $\pm$ 113  \\ 
    & 0.547--0.552 & 10422 $\pm$ 206  &  & 1.456--1.475 & 10564 $\pm$ 121  \\ 
    & 0.552--0.558 & 10794 $\pm$ 153  &  & 1.475--1.494 & 10686 $\pm$ 113  \\  
    & 0.558--0.563\tnote{c} & 9221 $\pm$ 279  & & 1.494--1.512 & 10799 $\pm$ 129  \\
    & 0.563--0.569 & 10444 $\pm$ 188  &  & 1.512--1.531 & 10551 $\pm$ 124  \\ 
    STIS G750L\tnote{d} & 0.526--0.555 & 10356 $\pm$ 220 & & 1.531--1.549 & 10566 $\pm$ 111  \\ 
    & 0.555--0.575 & 10497 $\pm$ 278  &  & 1.549--1.568 & 10515 $\pm$ 134  \\
    & 0.575--0.594 & 10317 $\pm$ 202  &  & 1.568--1.587 & 10436 $\pm$ 110  \\
    & 0.594--0.614 & 10061 $\pm$ 236  &  & 1.587--1.605 & 10492 $\pm$ 145  \\ 
    & 0.614--0.633 & 10386 $\pm$ 142  &  & 1.605--1.624 & 10340 $\pm$ 122  \\  
    & 0.633--0.653 & 10542 $\pm$ 219  &  & 1.624--1.642 & 10338 $\pm$ 142  \\ 
    & 0.653--0.672 & 10418 $\pm$ 276  &  & 1.642--1.661 & 10331 $\pm$ 138  \\ 
    & 0.672--0.692 & 10182 $\pm$ 225  & \textit{Spitzer} IRAC1 & 3.2--4.0 & 10191 $\pm$ 102  \\
    & 0.692--0.711 & 10168 $\pm$ 192  & \textit{Spitzer} IRAC2 & 4.0--5.0 & 10679 $\pm$ 145 \\
       \\ 
   
\hline
\end{tabular}
\begin{tablenotes}
\item[a] Transit depth $=R^2_{\textrm{planet}}/R^2_{\textrm{star}}$ 
\item[b] Typical STIS G430L bin size =  0.0055\,$\mu$m (median resolution $\sim 350$)
\item[c] Outlier bin strongly affected by systematics and ignored in retrieval analyses
\item[d]Typical STIS G750L bin size =  0.0196\,$\mu$m (median resolution  $\sim 130$)
\item[e]WFC3 bin size =  0.0186\,$\mu$m (median resolution $\sim75$)
\label{tab:spectrum}
\end{tablenotes}

\end{threeparttable}
\end{table*}

\subsection{WFC3}\label{sec:wfc3}
\subsubsection{Data Reduction}
The WFC3 data reduction generally follows the metholodogy of \citet{sheppard2017}, using programs adapted from IDL for Python. We download the ``ima'' data files from the HST archive, and remove background contamination following the ``difference reads'' methods of \citet{deming2013}, which allows us to easily resolve and remove potential contamination from the nearby companion \citep{evans2016}. We determine a wavelength solution by taking the zero-point from the F140W photometric observation and fitting for the wavelength coefficients that allow an out-of-transit spectrum to match the appropriate ATLAS stellar spectrum \citep{castelli2003}. 

We then divide the background-subtracted ``ima'' science frame by the WFC3 flat-field calibration file, and return the dark-, bias-, and flat-field-corrected flux array, in units of electrons. The uncertainty of the flux at each pixel is taken from the ``ima'' file's error frame, which accounts for gain-adjusted Poisson noise, read noise, and noise from dark current subtraction. This is further adjusted via error propagation for the added uncertainty from background removal and flat-field correction. We use the ``ima'' file's data quality frame to mask pixels (i.e., give zero weight) that are flagged as bad in every exposure in the time series. We then correct for cosmic rays using a conservative time series sigma-cut of 8-$\sigma$, while accounting for changes in flux that occur due to uneven scan rates and the transit itself, and set the affected pixels to the median value of that pixel in the time series. The average amount of pixels either impacted by cosmic ray events or flagged as bad pixels is $1.8\%$ of all pixels in a exposure. The reduced exposures are summed over the spatial scan direction to give a 1-D spectrum at each observation time. 

\subsubsection{Light Curve Analysis}\label{subsec:wfc3light}

We use a similar marginalization light curve analysis as \citet{sheppard2017}, applied to transit curves. This is a Bayesian model averaging method, first described by \citet{gibson2014} and applied by \citet{wakeford2016}, with further detrending by use of band-integrated (white light) residuals in spectral light curve fitting \citep{mandell2013, haynes2015}. We first analyzed the band-integrated light curve to simplify the spectral light curve analysis, then we fit the spectral light curves to derive the WFC3 transit spectrum.

We model the observed light curve as a \textit{BATMAN}\footnote{\url{https://github.com/lkreidberg/batman}} transit model \citep{kreidberg2015a} combined with a instrumental systematic model. For the transit model we assume nonlinear limb darkening and derive the coefficients by interpolating the 3-D values from \citet{magic2015} to the central wavelength of WFC3 (1.4\,$\mu$m). The limb darkening derivation is consistent with that used in the STIS analysis. We only fit for transit depth and central transit time, since the incomplete coverage of HST makes it difficult to improve constraints on other transit parameters, such as \textit{a/R$_{s}$} or inclination. We fix these values in the light curve analyses of each instrument (STIS, WFC3, and \textit{Spitzer}), which ensures consistent orbital parameters are used when analyzing different datasets. The transit and system parameters are shown in Table~\ref{tab:transit_params}. 

The systematic model grid comprises a set of 50 polynomial models which account for a visit-long linear slope, HST orbital phase-dependent systematics (i.e., ``breathing'', ramp), and sub-pixel wavelength shifts (for full model grid see \citep{wakeford2016}). It is computationally difficult to fully sample the parameter space of all 50 models using Markov Chain Monte Carlo (MCMC) samplers, so we instead fit each model using \textit{KMPFIT}\footnote{\url{https://github.com/kapteyn-astro/kapteyn/blob/master/doc/source/kmpfittutorial.rst}} \citep{kapetyn}, a Python implementation of the Levenberg-Markwardt least squares minimization algorithm, to more quickly determine parameter values and uncertainties. \citet{wakeford2016} found that uncertainties derived from these two methods typically agree within 10\%. We then weight each model by its Bayesian evidence --- approximated by the Akaike information criterion --- and marginalize over the model grid (assuming a prior that each model is equally likely) to derive the light curve parameters and uncertainties while inherently accounting for uncertainty in model choice.

\begin{figure}[t]
\centering
{
\includegraphics[width=0.5\textwidth,keepaspectratio]{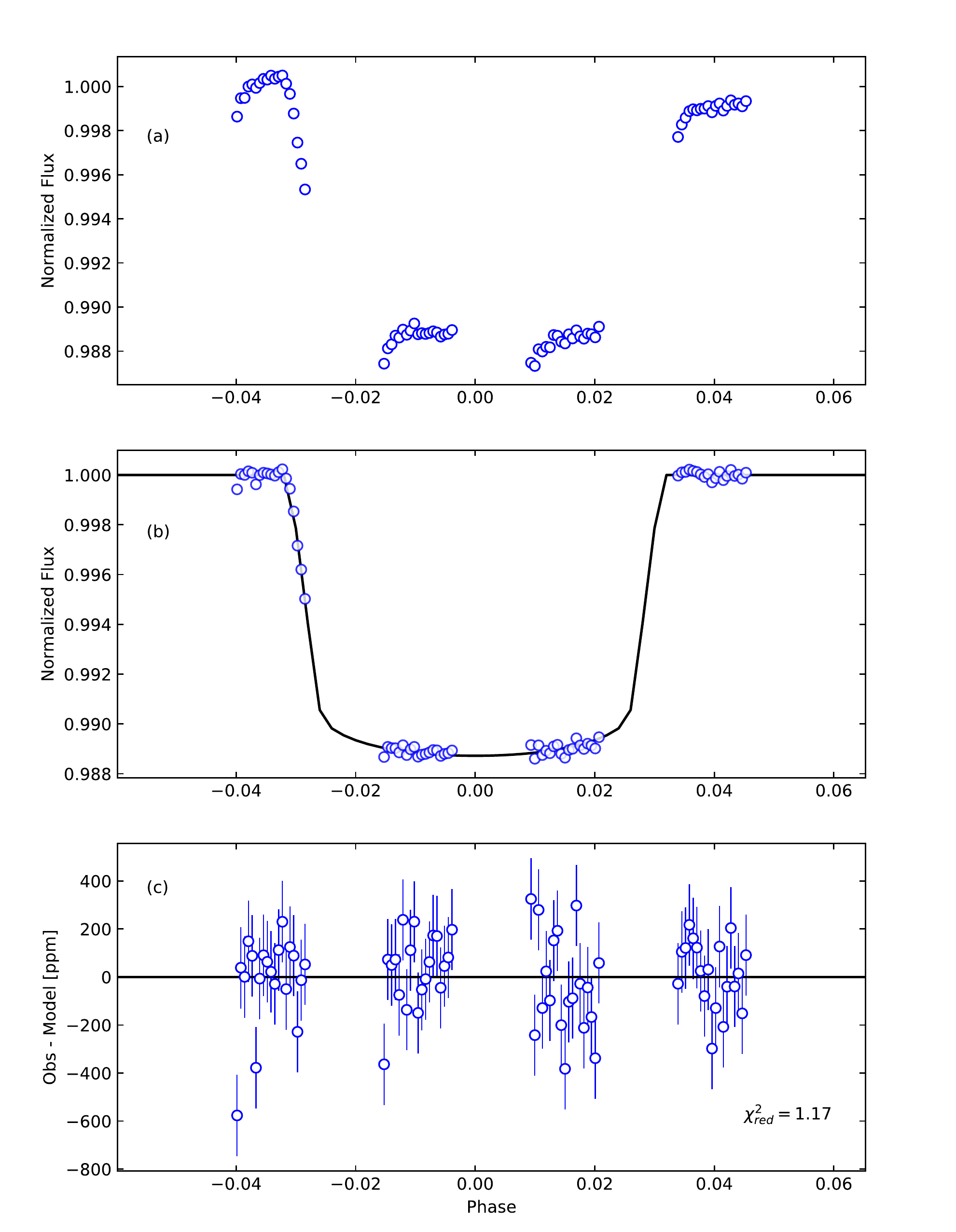}
\caption{Top panel: preprocessed band-integrated light curve for WFC3 observations. This is the band-integrated flux versus planet phase derived from the reduced data. The first orbit is excluded, as it is dominated by instrumental systematics. Middle panel: band-integrated light curve divided by the highest weighted systematic model (i.e., the detrended light curve). Bottom panel: residuals between light curve data and highest-weighted joint transit and systematic model. The reduced $\chi^2$ for the highest weighted model is 1.17, which is consistent with the model being a good fit for 67 degrees of freedom.
}
\label{fig:wl}
}
\end{figure}

As is common practice, we ignore the systematic-dominated first orbit in the white light analysis; however, the use of common-mode detrending allows us to include that orbit in the spectral light curve analysis. The raw light curve, the light curve with instrumental systematics removed, and the residuals from the highest-weight systematic model are shown in Figure~\ref{fig:wl}. We derive the white light depth to be $10490\pm{51}\,$ppm. 

To derive the transit spectrum, we bin the 1D spectra from each exposure between the steep edges of the grism response curve (1.12--1.66\,$\mu$m), deriving a flux time-series for each spectral bin. We use bins of width 0.0186\,$\mu$m (4 pixels) to maximize resolution without drowning the signal in noise. We note that the atmospheric retrieval is not sensitive to the choice of bin size. We fit each spectral light curve using a similar model as for the band-integrated light curve to derive the transit depth at each wavelength. 

The spectral light curve analysis mimics white light analysis with a few exceptions: all system parameters except transit depth are fixed to the white light values, limb-darkening coefficients are interpolated to the bin's central wavelength, and residuals from the band-integrated light curve are incorporated as a scalable parameter. The band-integrated residuals encode possible instrumental systematics that were not captured by the model grid, and so they can be used to further detrend the binned spectral light curves. The shape of the residuals are assumed to be constant with wavelength, though the amplitude is allowed to vary. This allows for removal of any wavelength-independent red noise from spectral bin curves at the penalty of slightly increasing the white noise. Note that the band-integrated uncertainty is sufficiently small relative to spectral light curve uncertainty such that the added noise has only a minor effect. The wavelength range, transit depth, an depth uncertainty for each WFC3 bin is shown in Table~\ref{tab:spectrum}.


The shape of our derived spectrum is in excellent agreement with the literature spectrum \citep{tsiaras2018}, though it is shifted to higher depths by $\sim90\,$ppm, indicating that we derive a larger  white light depth. This difference persists even if we derive the spectrum without using white light residuals. This could plausibly be due to different limb darkening treatments or different systematic modeling choices. This difference emphasizes the importance of considering offsets between instruments in retrieval analyses (Sec.~\ref{subsec:inst_offsets}). Further, the white light depth is subject to the choice of orbital parameters, which are typically fixed in spectroscopic light curve fits. We run sensitivity tests to determine that accounting for orbital parameter uncertainties increases the scatter between HST STIS and HST WFC3 by roughly 60\,ppm for HAT-P-41b. We capture this scatter by including it in WFC3's depth uncertainty, increasing it from 50$\,$ppm to $80\,$ppm.

As a check, we performed retrievals using the published spectrum from \citet{tsiaras2018} in combination with the derived STIS and \textit{Spitzer} depths and found differences well within the 1-$\sigma$ uncertainties for the retrieved parameters. The major results and conclusions in this paper are not sensitive to the WFC3 spectrum choice.




\subsubsection{WFC3 Transit Spectrum Verification}

Marginalization is only reliable if at least one model is a good representation of the data \citep{gibson2014, wakeford2016}. We therefore checked the goodness-of-fit of the highest-weighted systematic model for each light curve using both reduced $\chi^2$ and residual normality tests. Further, we explored if red noise is present in the light curve residuals, as that can bias inferred depth accuracy and precision \citep{cubillos2017} . 

Though $\chi^2$ cannot prove that a model is correct, it can demonstrate that the fit of a particular model is consistent with that of the ``true'' model with ``true'' parameter values \citep{andrae2010}. Therefore it is an informative goodness-of-fit diagnostic, and it is particularly useful due to its familiarity and simplicity. The ``true'' model with ``true'' parameters will have a reduced $\chi^2$ of one with uncertainty defined by the $\chi^2$ distribution. For both the band-integrated and spectral light curves (66 and $\sim88$ degrees of freedom, respectively), this results in an acceptable reduced $\chi^2$ range of roughly 0.7--1.3.

The band-integrated analysis ($\chi^2_{\nu}=1.17$) and all but one of the spectral bins (median $\chi^2_{\nu}=0.9)$ fall within this range. The exception is the 1.299\,$\mu$m light curve, which heighest-weighted model fit has a reduced $\chi^2$ of 0.56. This low value indicates that the uncertainties in this light curve are overestimated. This is likely due to incorporating white light residuals, which both inflate uncertainties and can potentially interpret random white noise as structure. However, it is not flagged by the normality or correlated noise analyses (described below), and fitting the light curve without incorporating white light residuals finds a consistent depth with a more reasonable $\chi^2_{\nu}=0.71$. Further, the derived depth and uncertainty exhibit good agreement with the published \citep{tsiaras2018} transit depth at this wavelength (accounting for the white light offset). Therefore, we include it in the transit spectrum. For the other 28 spectral bins, the reduced $\chi^2$ values provide no evidence against validity of the derived transit depths and uncertainties.

A residual normality test checks if the residuals for a model are Gaussian-distributed to determine goodness-of-fit, since this is expected for the ``correct'' model. Like reduced $\chi^2$, a normality test cannot prove that a model is correct, but can only diagnose incorrect models. We use the \textit{scipy} implementation of the common Shapiro-Wilk test for normality \citep{shapiro65}, and determine for which light curves the highest evidence model has normality ruled out at the $5\%$ significance level. At a sample size of around 90 this is by no means rigorous, but it is still a useful heuristic for flagging potentially problematic light curve models.

Normality is rejected at the $5\%$ significance level only for the band-integrated residuals and the 1.243\,$\mu$m spectral bin residuals. In both cases, normality is ruled out due to a single outlier in the time-series. Normality is rejected at $3\%$ significance for the band-integrated residuals, due entirely to the first exposure in the first orbit. When this exposure is ignored, we recover a consistent depth and uncertainty and the residuals are consistent with normality. We therefore keep this exposure in the analysis.

A possible cause of the spectral bin's outlier is a minor cosmic ray event or bad pixel that was small enough to both avoid the detection by the sigma-cut and not affect the band-integrated curve, but large enough to impact the much smaller bin flux. Removing this spectral bin from the retrieval had no noticeable effect on the results. Further, the derived depth and uncertainty are consistent with literature values \citep{tsiaras2018}. We therefore decide to include this bin in the retrieval.

\begin{figure}[t]
\centering
{
\includegraphics[width=0.75\textwidth,keepaspectratio]{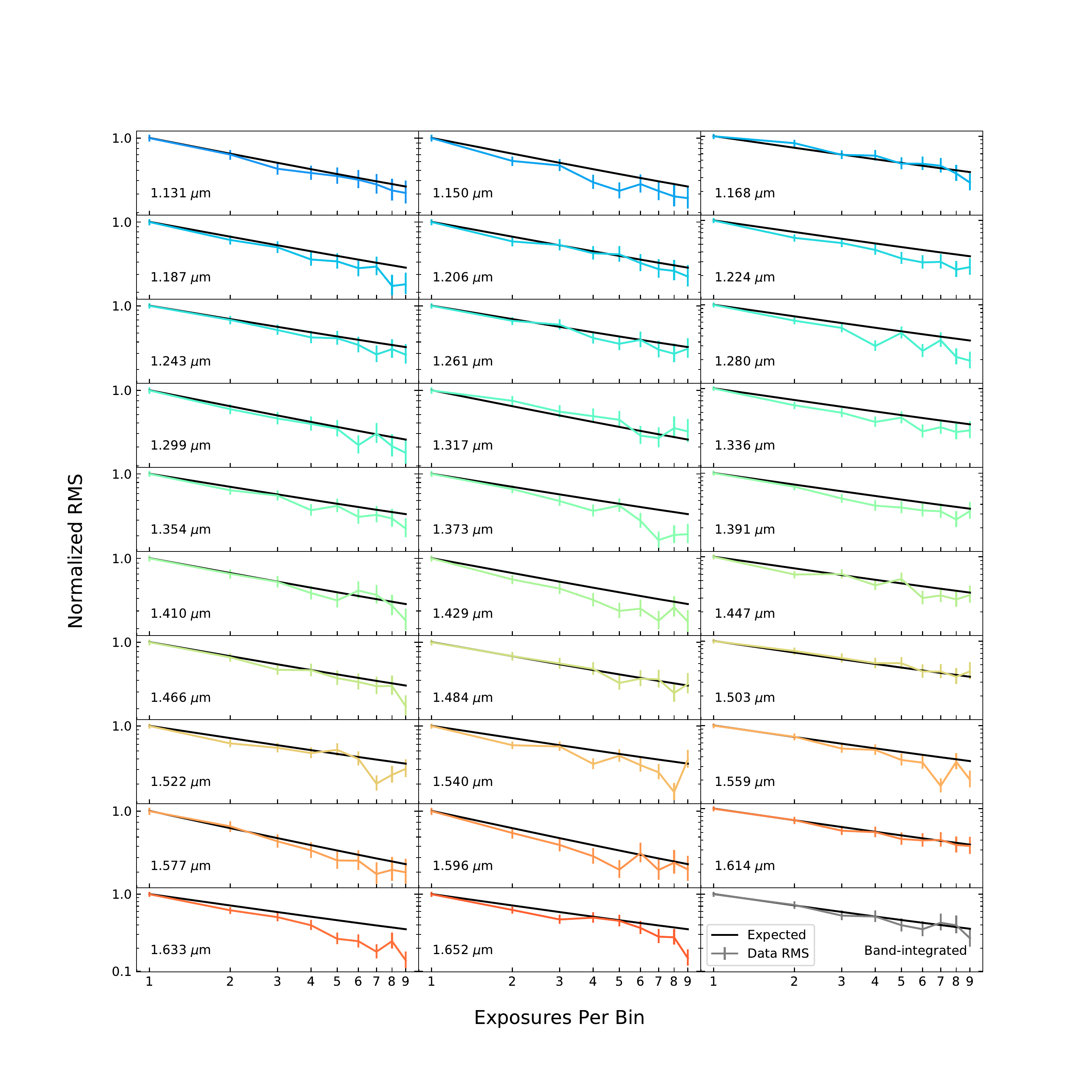}
\caption{Correlated noise analysis for each of the 29 WFC3 spectral bins and the band-integrated light curve. For each bin's light curve, we find the RMS of the residuals for an increasing number of exposures per bin. Pure white noise would scale with the black line, while correlated noise would increase with binning. Though not exact given the gaps between WFC3 data, this is a useful heuristic to search for correlated noise. See \citet{cubillos2017} and \citet{pont2006} for more details.
}
\label{fig:correlated}
}
\end{figure}

Finally, we test for correlated noise in the residuals following the methodology of \citet{cubillos2017} (also see \citet{pont2006}) and using MC$^3$\footnote{\url{https://github.com/pcubillos/mc3}
}. Though this method is not necessarily rigorous for HST due to the incomplete phase coverage and relatively small number of exposures, it is still a practical diagnostic. We find no evidence of correlated noise up to the timescale of half an HST orbit (Figure~\ref{fig:correlated}). Beyond this timescale (nine exposures per bin) the standard deviation is based on a light curve with fewer than eight points and so the relationship is less informative and subject to small-number statistics.

With the caveats noted above, marginalization does an excellent job in fitting the spectral light curves. We emphasize that removing any of the flagged bin spectra has no affect on the retrieval. Together, these tests support the validity of the derived transit depths and uncertainties in the WFC3 bandpass.



\subsection{Spitzer}\label{sec:spitzer}
\subsubsection{Data Reduction}
The Spitzer data consists of cubes of 64 subarray frames in each band, each of size 32$\times$32 pixels.  We extracted aperture photometry for each frame, totaling 21,632 frames at both 3.6 and 4.5\,$\mu$m.  To extract photometry, we used 11 numerical apertures with radii ranging from 1.6 to 3.5 pixels, and we centered those apertures on the position of the host star determined using both a 2-D Gaussian fit to the stellar point spread function, and also a center-of-light calculation.  Since HAT-P-41 has a companion star 3.6 arc-sec distant \citep{hartman2012, evans2016}, we measured the flux of the companion scattered into each of our numerical apertures, using the method described by \citet{garhart_etal2020}.  We adopted the magnitude difference in the Spitzer bands as deduced by \citet{garhart_etal2020}.  Accounting for our different aperture radii than was used by \citet{garhart_etal2020}, we derive dilution correction factors of 1.0171 and 1.0106 at 3.6- and 4.5\,$\mu$m, respectively.   The transit depths are then multiplied by those factors in order to correct for the presence of the companion star.

\subsubsection{Light Curve Analysis}
We fit transit curves to the 22 sets of photometry at each wavelength (eleven apertures, each with two centering methods). Our default fitting procedure fixes the orbital parameters at the values in the discovery paper by \citet{hartman2012}, fitting only for the central time and depth of the transit.  The shape of the Spitzer transits is well matched when fixing the orbital parameters to those values. However, we also explored including the orbital inclination and $a/R_s$ in the fit (see below), those being the orbital parameters that most strongly affect the shape of the transit.  We adopt quadratic limb darkening coefficients calculated by least squares for the \textit{Spitzer} bands by \citet{claret_etal2013}, using 2 km/sec microturbulence. We choose the values for $T_{eff}=6400$\,K and $log{\it g}=4.0$, without interpolation. We fix those coefficients in the fitting process, and we deem these choices to be appropriate given that the limb darkening is minimal at these infrared wavelengths.  Our fits to the transit account for the intra-pixel sensitivity variations of the Spitzer photometry using pixel-level decorrelation (PLD,  \citealp{deming_etal2015}), including a linear baseline (ramp) in time.  We use a Bayesian information criterion to decide between a linear versus quadratic ramp.  The details of the PLD fit are the same as described for secondary eclipses by \citet{garhart_etal2020}, except that we are fitting transits, so we include limb darkening.  Briefly, the fitting code bins the photometry and pixel basis vectors to various degrees, and selects the optimal bin size, aperture radius, and centering method, based on the smallest difference from an ideal Allan deviation relation \citep{allan_1966}.  The Allan deviation relation expresses that the standard deviation of the residuals should decrease as the square-root of the bin time.  Operating on binned data allows the PLD algorithm to concentrate on the longer time scales that characterize the red noise (and also the transit duration), as opposed to the 0.4-second cadence time of the raw photometry.

We determine the errors on our transit depths and times using an MCMC procedure, with a burn-in phase of 10,000 steps, followed by 800,000 steps to explore parameter space.  We calculate multiple chains for each transit, and verify convergence using a Gelman-Rubin (GR) statistic \citep{gelman&rubin_1992}.  Our GR values for our PLD fits are very close to unity, being 1.0027 at 3.6\,$\mu$m and 1.0004 at 4.5\,$\mu$m, indicating good convergence.  Our transit depths and times are listed in Table~\ref{spitzer}.

Our derived transit times are in excellent agreement with measurements of the same Spitzer transits by \citet{wakeford2020}.  Specifically, using our uncertainties, our fitted times differ by $1.1\sigma$ and $0.6\sigma$ at 3.6- and 4.5\,$\mu$m, respectively.  \citet{wakeford2020} use 8 HST transits as well as the discovery results and the Spitzer transits to derive a new ephemeris.  Our fitted times (Table~\ref{spitzer}) differ from that ephemeris by insignificant amounts (0.2 and 7.8 seconds).

\begin{figure}
    \centering
    \includegraphics{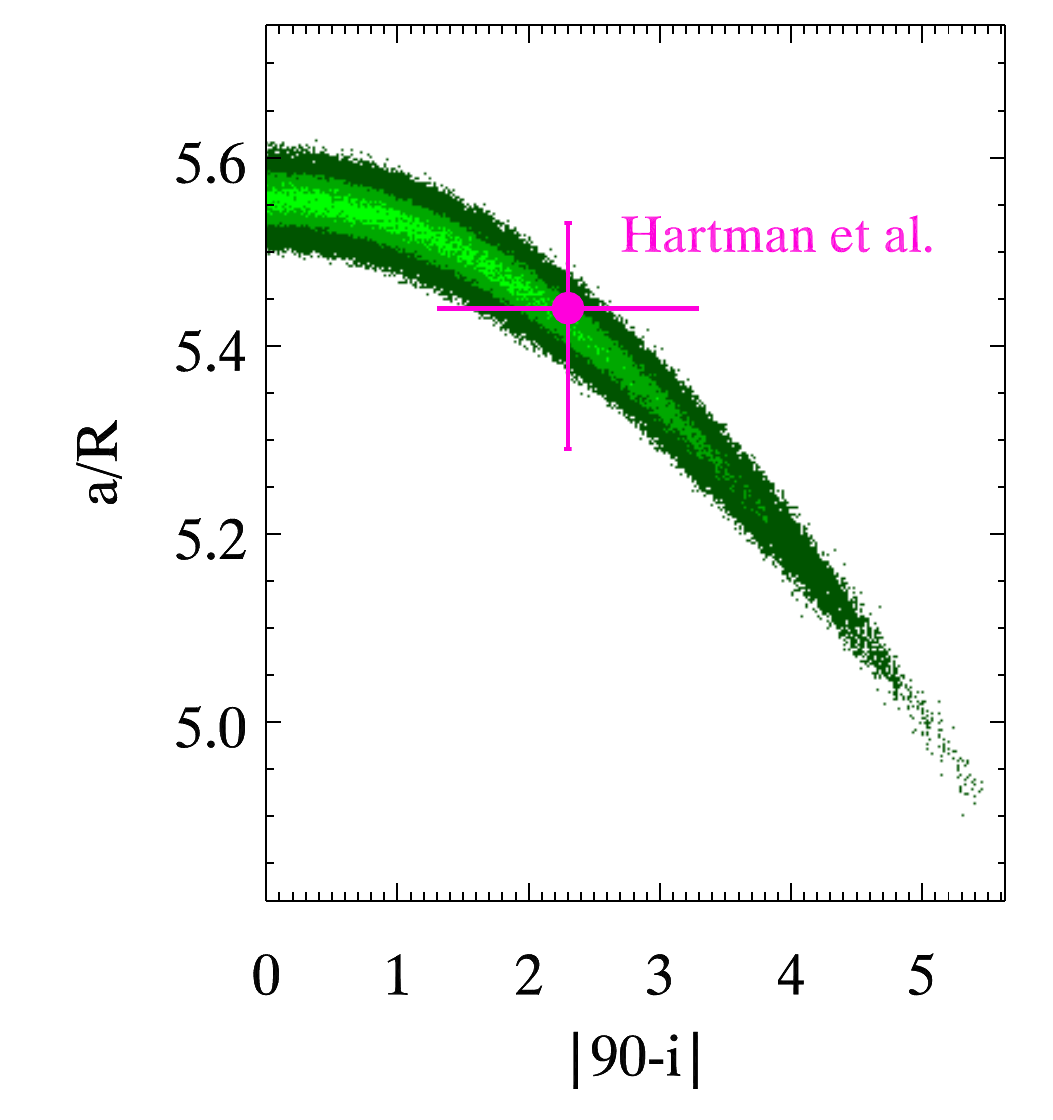}
    \caption{Likelihood distribution from the fit to the 3.6\,$\mu$m transit of HAT-P-41b, based on an MCMC using uniform priors, and shown versus $a/R_{s}$ and the orbital inclination. These two orbital parameters are degenerate when using only a Spitzer transit, and the values derived by \citet{hartman2012} are indicated by the point with error ranges.  The Spitzer transits at both 3.6- and 4.5\,$\mu$m are fully consistent with $a/R_{s}$ and $i$ from \citet{hartman2012}.}
\label{fig:transit_degen}
\end{figure}

We explored the effect of uncertainties in the orbital parameters, since those can affect the derived transit depth \citep{alexoudi_etal2018}. Adopting uniform priors for $a/R_s$ and inclination, we find that they are degenerate when fitting only the Spitzer transits.  That degeneracy is illustrated in Figure~\ref{fig:transit_degen}, where it is apparent that inclination and $a/R_s$ can trade off to maintain the observed transit duration, and the sharp ingress/egress that characterizes the Spitzer transits.  Changing the inclination changes the chord length across the stellar disk, and (when limb-darkening is minimal) that can be compensated by changing $a/R_s$ to maintain the same transit duration.  The orbital solution from \citet{hartman2012} is entirely consistent with our likelihood distribution for those parameters, as shown in Figure~\ref{fig:transit_degen} for 3.6\,$\mu$m (4.5\,$\mu$m is similar).  We therefore freeze the orbital parameters at the \citet{hartman2012} values when fitting our Spitzer transits. 

\begin{figure}
  \plottwo{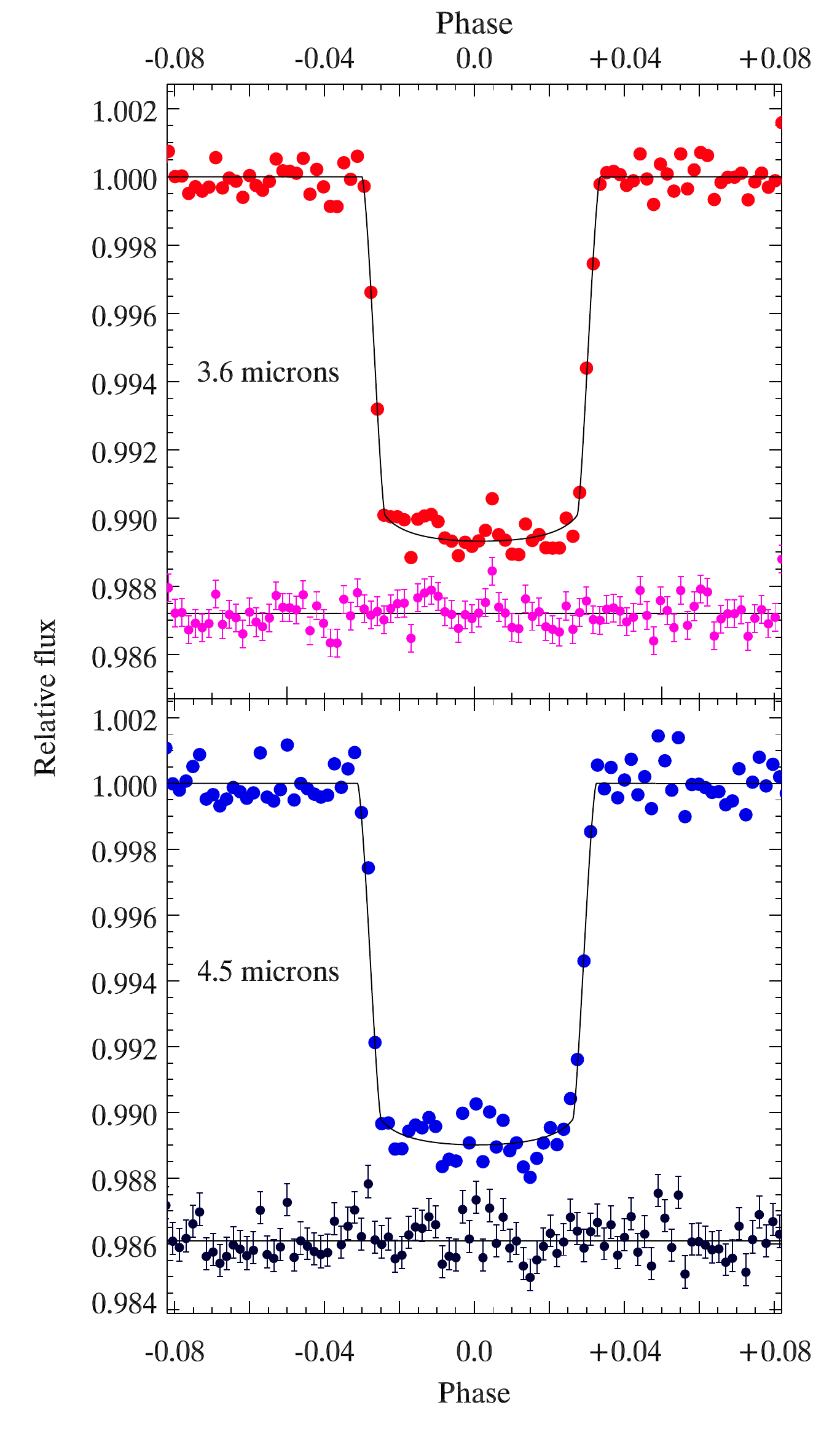}{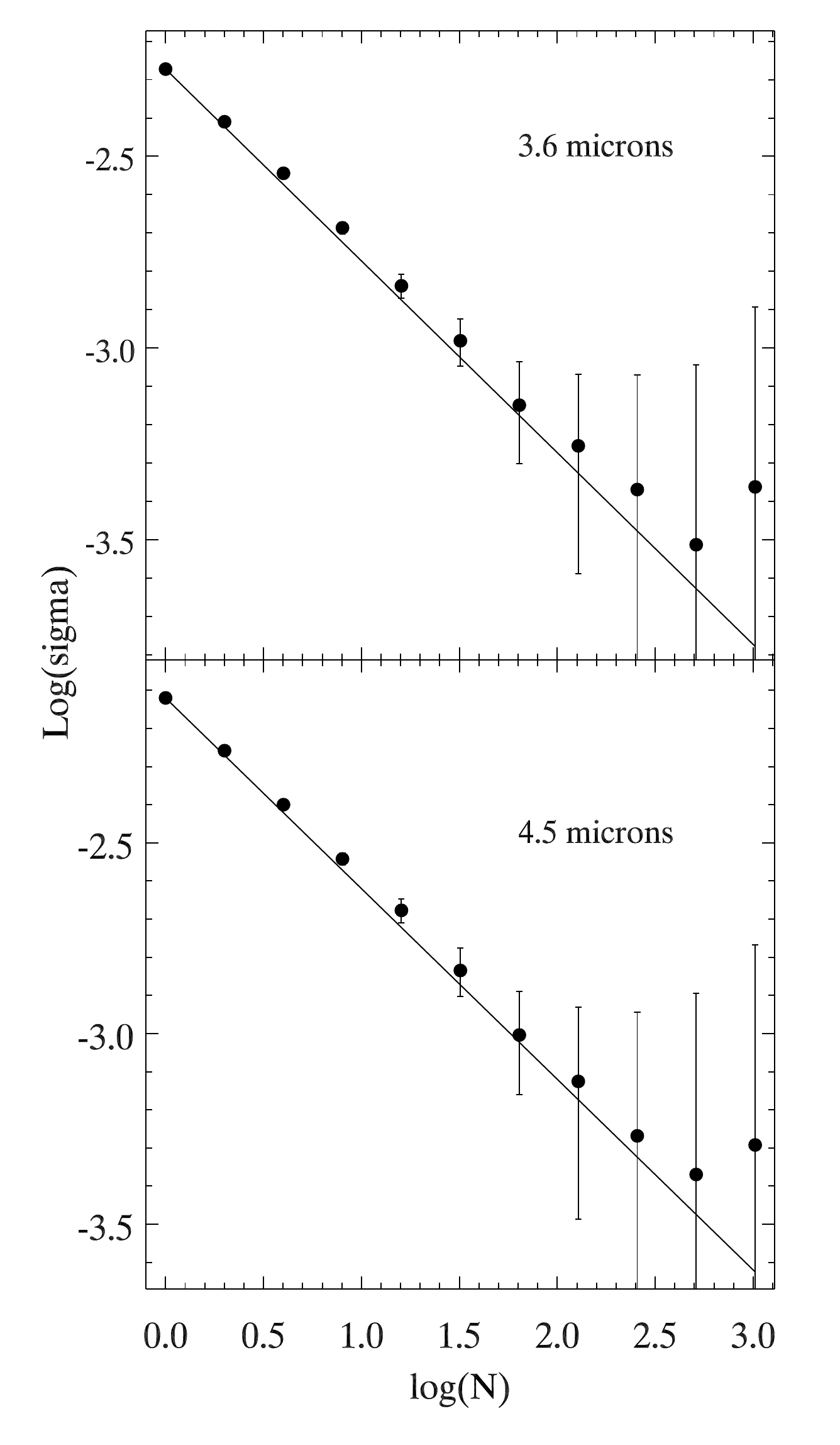}
  \caption{ Left: Spitzer transit light curves of HAT-P-41b at 3.6- and 4.5\,$\mu$m after correction of the intra-pixel effects of the detector and temporal ramps. The data are binned to 100 points per transit for clarity of illustration. The residuals (data minus fitted model) are shown below the transit curves, and have error bars added. Right: Allan deviation relations for the binned residuals, i.e. standard deviation of the residuals when the original data are binned over an increasing number of points, N.  The solid lines project the single-point ($N=1$) scatter to larger bin sizes with a slope of -0.5 as expected for photon noise.}
  \label{spitzer_figure}
\end{figure}

Figure~\ref{spitzer_figure} illustrates the transits at 3.6 and 4.5\,$\mu$m. The residuals from the best-fit model are included in the figure, and the right panel shows the residuals binned over increasing time scales, a so-called Allan deviation relation \citep{allan_1966}.  The slopes of those relations are close to the -0.5 value expected for photon noise. 
 
 \begin{table*}
\centering       
 \begin{threeparttable}  
    \caption{Transit times and depths for HAT-P-41b in the Spitzer bands.
    } 
    \begin{tabular}{lll}
    Wavelength  &  BJD(TDB)  & $R_p^2/R_s^2$ (ppm)  \\
    \hline \hline
    3.6\,$\mu$m & 2457772.20440$\pm$0.00033  &  10020$\pm$100    \\
    4.5\,$\mu$m & 2457788.36860$\pm$0.00032  &  10568$\pm$135    \\
    \end{tabular}
    \tablecomments{These are ``as observed'' transit depths, not corrected for dilution by the companion star.  To correct for dilution, multiply the values of $R_p^2/R_s^2$  by 1.0171 at 3.6\,$\mu$m, and by 1.0106 at 4.5\,$\mu$m. The corrected values are shown with the rest of the transit spectrum in Table~\ref{tab:spectrum}.}
\label{spitzer} 
\end{threeparttable}  
    \end{table*}

\section{Atmospheric Retrieval}\label{sec:retrieve}

There are two common frameworks to retrieve physical parameters from transmission spectra. The first is by assuming chemical equilibrium, where the abundance of a molecule is only dependent on local temperature, local pressure, and global elemental abundances such as O/H and C/H \citep[e.g.,][]{kreidberg2015b}. The second is by instead fitting for molecular abundances based on observed spectral features, then determining global elemental abundances from the molecular abundance values \citep[e.g.,][]{Madhusudhan2012}. Since carbon and oxygen-based molecules are typically the most spectroscopically active species over the wavelengths covered by HST and \textit{Spitzer}, the elemental abundances are commonly parameterized by metallicity --- defined as the enhancement of metal elements relative to hydrogen compared to solar values (see Sec~\ref{sec:metalco} for more detail) --- and carbon-to-oxygen ratio (C/O). Some retrievals improve flexibility by allowing other elements --- such as sodium or vanadium --- to also vary from their solar ratios \citep{amundsen2014, tremblin2015, tremblin2016, sing2016}.

The open source code PLATON\footnote{\url{https://github.com/ideasrule/platon}} \citep{zhang2019} is able to perform quick retrievals which assume chemical equilibrium, whereas AURA \citep{pinhas2018} is able to capture possible disequilibrium chemistry by not assuming chemical equilibrium. 
We retrieve the atmospheric parameters with both frameworks to see how interpretations compare, and to explore how sensitive the conclusions are to retrieval assumptions. 

\subsection{PLATON}\label{subsec:platon}
PLATON is a fast, open-source retrieval code developed by \citet{zhang2019}. Like many retrieval codes, it comprises a forward model and an algorithm for Bayesian inference. Though there are some differences, it essentially uses the same forward model as Exo-Transmit \citep{kempton2017}. 
Here, we summarize the forward model: To calculate a spectrum, it first determines the abundances of 34 potentially relevant chemical species for a given atmospheric metallicity and C/O. These include Na and K as well as molecules CH$_4$, CO, CO$_2$, HCN, H$_2$O, MgH, NH$_3$, TiO, and VO (see \citet{kempton2017} for complete list). The metallicity and C/O provide elemental abundances, which are combined with a temperature-pressure grid as input into GGchem \citep{woitke2018} to compute equilibrium molecular and atomic abundances at each pressure layer in the atmosphere, accounting for the effects of condensation on equilibrium abundances. PLATON allows for a grey cloud deck, below which no light can penetrate, and the abundance-temperature-pressure grid facilitates the determination of total opacity at each pressure layer in the atmosphere which lies above this cloud top. PLATON includes the same opacity sources as Exo-Transmit, and accounts for opacity from gas absorption, CIA, and scattering (either parametric Rayleigh scattering or Mie scattering). The forward model converts the opacity-pressure grid to a opacity-height grid using hydrostatic equilibrium with $P_{ref}=1$ bar, which is then used as an input to a radiative transfer code to determine the uncorrected transit depths. After correcting for possible stellar activity, due to either unocculted spots or faculae, PLATON's forward model outputs the corrected transit spectrum. The largest source of computational uncertainty is opacity sampling error, which is a source of white noise from using a relatively low resolution (R=1000) that cannot resolve individual lines \citep{zhang2019}. Accounting for opacity sampling for the transit spectrum of HAT-P-41b typically increases the depth uncertainty by 1.5\% (2.5ppm), which is sufficiently small such that it does not affect intepretation. For more details on PLATON, see \citet{zhang2019}. We note that the version of PLATON we describe and use in this paper is Platon 3.1. A newer version, PLATON 5.1, has since been released with additional features as described in \citet{platon2}.


Our PLATON analysis does not retrieve individual abundances. Instead, it fits for the isothermal temperature, atmospheric metallicity as a multiple of solar values for atomic species, and C/O ratio; the retrieved equilibrium abundances for atomic and molecular absorbers are a natural consequence of those values. This is in contrast with the AURA analysis (Sec.~\ref{subsec:AURA}), which retrieves individual molecular and atomic abundances.

The algorithm PLATON uses for Bayesian inference is nested sampling \citep{skilling2004}. Specifically, PLATON uses multimodal nested sampling from the Python implementation \textit{nestle.}\footnote{\url{https://github.com/kbarbary/nestle}} Like MCMC samplers, nested sampling efficiently samples posterior distributions with dimensionalities typical of atmospheric retrievals (n=5--20), and so it is effective at atmospheric parameter estimation. Unlike MCMC routines, it automatically calculates the Bayesian evidence for a model, which is necessary for model comparison. The Bayesian evidence  intrinsically accounts for overfitting by punishing too much model structure and thus determines if extra parameters are warranted. We use this to justify excluding parameters which add structure and do not significantly improve the fit. Nested sampling also has a well-defined stopping criteria, so there is no need to check for convergence. For an excellent write up on this algorithm, especially about using it in practice, see the documentation of \textit{Dynesty}\footnote{https://dynesty.readthedocs.io/en/latest/} \citep{higson2019}. 

In addition to the standard set included in PLATON's forward model, we added three new fittable parameters: a partial cloud parametrization and three instrumental transit depth bias parameters (henceforth referred to as instrumental offsets). 

\subsubsection{Partial Clouds}
The partial cloud parameter is motivated by work by \citet{line2016} and \citet{MacDonald2017}, which showed that if the grey cloud deck were inhomogenous, then the spectrum we observe (D) would be a weighted average of the clear atmosphere transit spectrum ($D_{clear}$) and the cloudy atmosphere spectrum ($D_{cloudy}$) with weights given by the cloud fraction ($f_{c}$). We implement this as $D = f_{c}*D_{cloudy} + (1 - f_{c})*D_{clear}$. Since high altitude grey clouds are seen in spectra as flat lines, averaging a spectrum with features with this line will shrink the features and can mimic the effect of a high mean-molecular mass, small scale height atmosphere. Thus, including this parameter allows us to account for this possible degeneracy and prevents us from overconfidently claiming a high-metallicity atmosphere. 

\subsubsection{Instrumental Offsets}\label{subsec:inst_offsets}
The instrumental offsets are nuisance parameters that can capture the extent to which transit depths from STIS G430L, STIS G750L, or WFC3 are biased relative to the depths from the other instruments. This is motivated by the use of common-mode corrections in the light curve analysis of each instrument, which can potentially introduce a uniform bias for the depth at each spectral bin for that instrument. Offsets are also able to account for inter-instrumental transit depth scatter introduced by uncertainty in orbital parameters a/R$_s$ and inclination (Section~\ref{subsec:wfc3light}). 

We explore three offset scenarios. The first is physically-motivated. In this scenario, we use Gaussian priors with sigmas determined by the uncertainties in the band-integrated transit depths to try to reflect the correlated uncertainty that exists between spectral bins for each instrument, whether due to common mode corrections or orbital parameter uncertainties. This essentially propagates the white light depth uncertainty into the retrieval. Derived in Section~\ref{sec:data}, these uncertainties are 105\,ppm, 85\,ppm, and 80\,ppm for STIS G430L, STIS G750L, and WCF3, respectively. The second scenario investigates the potential impact of unknown sources of bias by setting a large, uniform offset prior for each instrument's offset. The third scenario extends this, by setting two large, uniform priors: a WFC3 offset and a single offset for both STIS instruments. The third scenario allows the absolute depths at STIS to vary while preserving the optical spectral shape.

We caution that offsets --- especially penalty-free, uniform prior offsets --- can cloak missing physics in a model. We do not think they should act as a safety net to achieve a good fit to a spectrum, and the inferred atmospheric properties should be unsderstood in context. However, offsets offer a way to both investigate potential instrumental biases and account for absolute depth uncertainty for each instrument. It is valuable to include offsets as model parameters and marginalize over these possible values in order to understand how the uncertainty in the absolute transit depth for each instrument affects the marginalized posterior distributions of the other model parameters.



\subsection{AURA}\label{subsec:AURA}

We complement our analysis of HAT-P-41b by performing retrievals on the STIS, WFC3 and \textit{Spitzer} observations without the assumption of chemical equilibrium. We employ an adaptation of the retrieval code AURA \citep{pinhas2018}, as described in \citep[e.g.,][]{Welbanks2019}. 

The code computes line by line radiative transfer in a transmission geometry and assumes hydrostatic equilibrium. We consider a one-dimensional model atmosphere consisting of 100 layers uniformly spaced in $\log_{10}$(P) from 10$^{-6}$-10$^2$ bar. The pressure-temperature (P-T) profile in the atmosphere is retrieved using the P-T parameterization of \citet{Madhusudhan2009}. The measured radius of the planet R$_p$ is assigned to a reference pressure level in the atmosphere through a free parameter P$_{ref}$. 

The model atmosphere assumes uniform mixing ratios for the chemical species and treats them as free parameters. We consider sources of opacity expected to be present in hot Jupiter atmospheres \citep[e.g.,][]{Madhusudhan2012} and include H$_2$O \citep{Rothman2010}, Na \citep{Allard2019}, K \citep{Allard2016}, CH$_4$ \citep{Yurchenko2014}, NH$_3$ \citep{Yurchenko2011}, HCN \citep{Barber2014}, CO \citep{Rothman2010}, CO$_2$ \citep{Rothman2010}, TiO \citep{Schwenke1998}, AlO \citep{Patrascu2015}, VO \citep{McKemmish2016}, and H$_2$-H$_2$ and H$_2$-He collision induced absorption \citep[CIA;][]{Richard2012}. The opacities for the chemical species are computed following the methods of \citet{Gandhi2017}. The CO$_2$ abundance is restricted to remain below the H$_2$O and CO abundances as expected at these temperatures for H-rich atmospheres \citep{Madhusudhan2012}.

We allow for the presence of clouds and/or hazes following the parametrization in \citet{line2016, MacDonald2017}. Non-homogenous cloud coverage is considered through the parameter $\bar{\phi}$, corresponding to the fraction of cloud cover at the terminator. Hazes are incorporated as $\sigma=a\sigma_0(\lambda/\lambda_0)^\gamma$, where $\gamma$ is the scattering slope, $a$ is the Rayleigh-enhancement factor, and $\sigma_0$ is the H$_2$ Rayleigh scattering cross-section ($5.31\times10^{-31}$~m$^2$) at the reference wavelength $\lambda_0=350$~nm. Opaque regions of the atmosphere due to clouds are included through an opaque (gray) cloud deck with cloud-top pressure P$_{\text{cloud}}$.

Lastly, we allow for the same three instrumental offset scenarios as described in Section~\ref{subsec:inst_offsets}. In these model runs, a constant offset in transit depth is applied to the data set of choice. The offset priors for each scenario are given in Table~\ref{table:AURA_priors}.


In summary, our retrievals consist of up to 25 parameters: 11 chemical species, 6 parameters for the P-T profile, 1 for the reference pressure, 4 for clouds and hazes, and up to 3 extra parameters for instrumental shifts. Table~\ref{table:AURA_priors} shows the parameters and priors used in our retrievals. 

\begin{deluxetable*}{c|c|c}
\tablecaption{Parameters and priors used in the AURA retrievals.  \label{table:AURA_priors} }
\tablecolumns{3}

\tablewidth{0pt}
\tablehead{
\colhead{Parameter} & \colhead{Prior distribution} & \colhead{Prior range}
}
\startdata
X$_i$ & Log-uniform & 10$^{-12}$--10$^{-1}$ \\ \hline
T$_0$ & Uniform &  800--2000~K \\ \hline
$\alpha_{1,2}$  &  Uniform & 0.02--2.00~K$^{-1/2}$\\ \hline
P$_{1,2}$ & Log-uniform & 10$^{-6}$--10$^{2}$ bar\\ \hline
P$_{3}$ & Log-uniform & 10$^{-2}$--10$^{2}$ bar\\ \hline
a & Log-uniform  & 10$^{-4}$--10$^{10}$\\ \hline
$\gamma$ & Uniform & -20-- 2\\ \hline
P$_{\text{cloud}}$ &Log-uniform  &10$^{-6}$--10$^{2}$ bar \\ \hline
$\bar{\phi}$ & Uniform & 0--1\\\hline
\makecell{STIS${_\textrm{shift}}$} & \makecell{Uniform} & \makecell{-500--500\,ppm} \\\hline
\makecell{STIS G430L${_\textrm{shift}}$} & \makecell{ Gaussian \\ Uniform} & \makecell{105\,ppm \\ -500--500\,ppm} \\\hline
\makecell{STIS G750L${_\textrm{shift}}$} & \makecell{ Gaussian \\ Uniform} & \makecell{85\,ppm \\ -500--500\,ppm} \\\hline
WFC3${_\textrm{shift}}$ & \makecell{Gaussian \\ Uniform} & \makecell{80\,ppm \\ -500--500\,ppm}
\enddata 
\tablecomments{Instrument shifts were employed in a subset of the retrievals and had either Gaussian or Uniform priors as explained in section~\ref{subsec:AURA_shifts}. For Gaussian priors a mean of 0\,ppm was assumed.}
\end{deluxetable*}

\subsection{A note on metallicity and C/O }\label{sec:metalco}
Atmospheric metallicity is a broadly used term that does not always have the same definition or assumptions built into its derivation \citep{Madhusudhan2014a, Kreidberg2014}. Here we define C/O and global atmospheric metallicity explicitly. For AURA, the abundances of different elements are derived independently from the corresponding gaseous absorbers - O/H from oxygen-bearing molecules such as H$_2$O, Na from gaseous Na, and so on. This retrieval approach allows for different elements to be enhanced or depleted in different quantities. As such, there is no single metric for metallicity in this approach. Nonetheless, as described below, we use the O/H ratio from the retrieved H$_2$O abundance using AURA as a proxy for metallicity in order to facilitate comparisons with PLATON retrievals in which all the elements are enhanced by a single metallicity parameter. 

In PLATON, metallicity is a factor that scales the solar elemental abundances, denoted M/H. The ratios between metals (e.g., Fe/O, Ti/V)  are fixed to solar metal ratios, but the solar metal-to-hydrogen ratio 
is allowed to vary. Thus, all metals are scaled by the same factor. Then, the elemental carbon abundance is determined by  C/O $\times$ metallicity. Therefore, only carbon is allowed to differ from its solar ratio compared to other metals. While allowing carbon instead of oxygen to vary is arbitrary, C/O variation is motivated: it is the only metal-to-metal ratio for which predicted molecular abundances are typically sensitive to in the wavelength range covered by HST/\textit{Spitzer} transit spectra. In this paradigm, metallicity is effectively a heuristic for O/H, since that dominates the retrieval both because of molecular opacity (e.g., H$_2$O, TiO, VO) and because over 99\% of the mean molecular weight is due to C, O, and H. In reality we retrieve O/H and C/O, then set all other elements to $X = X_{\odot} * \frac{O/H}{O/H_{\odot}}$. Therefore, PLATON's derived metallicity is reasonably comparable to AURA's derived O/H.


\section{PLATON Retrieval Analysis} \label{sec:platon_retrieval}
 
The relative importance of each physical process that affects an observed transit spectrum is not clear ahead of time. PLATON, though less flexible than free-chemistry retrievals or retrievals that allow elemental abundances to vary from solar ratios, is able to quickly perform chemically-constrained retrievals ($\sim$30 minute runtime for fiducial model retrieval on full data set). This makes it well suited for testing an array of models, which is important in order to determine how different model assumptions impact the conclusions of a retrieval. To explore this, we choose the fiducial model to be the set of parameters necessary to fully describe the simplest physical processes that we know affect the spectrum: opacity from gas absorption, CIA, and Rayleigh scattering. We then detail the effect of incorporating more complicated physics. In Section~\ref{subsec:model_selection}, we use both physical and statistical arguments to determine the ``best'' model. However, it is important to show the sensitivity of the results to each model assumption to not provide overconfident constraints and to be able to predict how new observations might affect the conclusions.

\subsection{Fiducial Model}\label{subsec:fid}
The priors for the fiducial model are shown in Table~\ref{tab:priors}. The metallicity of the atmosphere, the temperature of the limb, and the C/O ratio are necessary to include in order to calculate molecular and atomic abundances, which determine the gas absorption, CIA, and Rayleigh scattering opacities. While we cannot improve constraints on $M_p$ and $R_s$, it is best practice to include them as parameters with Gaussian priors in order to propagate the uncertainties on those measurements \citep{zhang2019}. Otherwise, we would mistakenly assume that $M_p$ and $R_s$ are precisely known. We also include the cloud top pressure of a grey cloud deck in the fiducial model. We fix the reference pressure to 1 bar and retrieve the planetary radius at that pressure. Note that \citet{Welbanks2019} demonstrated that it is justified to assume a reference pressure and retrieve the planetary radius without affecting the ability to constrain atmospheric composition.

Although $R_p$ and $M_p$ are both allowed to vary independently in PLATON retrievals, their uncertainties are not independent: the uncertainties for $M_p$ are derived from $\log(g_p)$ (from transit observations) and $R_p$ (derived from $R_p/R_s$ from transit and $R_s$ from TIC-8). PLATON does not constrain $M_p$ and $R_p^2$ to match $\log(g_p)$ a priori, however this is only an issue if regions of high likelihood extend to combinations of values that should not be allowed (i.e., more than 3-$\sigma$ from observed $\log(g_p)$). We re-derive $\log(g_p)$ using values at $R_p$ and $M_p$ at the edge of significant likelihood and find good agreement with the prior, well within 3-$\sigma$). 

\begin{table*}
\centering       
\begin{threeparttable}
\caption{Prior distributions for fiducial model}  

\begin{tabular}{l l l l r}     
\hline\hline      
         
    Parameter & Symbol & Distribution & Range/Width\tnote{a} & Default Value \\
\hline
    Planet Radius & $R_p$ & Uniform & 0.83--2.48~R$_{Jup}$\tnote{b}& $1.65~R_{Jup}$ \\  
    Limb Temperature & $T$ & Uniform & 850--2550K\tnote{b}& 1700K\\  
    Carbon-oxygen ratio & C/O & Uniform & 0.2--2.0& 0.53\tnote{d}\\ 
    Metallicity & $Z$ & Log-uniform & 0.1--1000 $Z_{\odot}$ & 1 $Z_{\odot}$\\  
    Planet Mass & $M_p$ & Gaussian & $0.14~M_{Jup}$ & $0.76M_{Jup}$ \\
    Stellar Radius & $R_s$ & Gaussian & $0.08~R_{\odot}$ & $1.65~R_{\odot}$ \\
    Cloudtop Pressure & $P_{cloud}$ & Log-uniform & $10^{-3}$--$10^8$ Pa& 1 Pa\\ 
   
\hline
\end{tabular}
\begin{tablenotes}

\item[a] Range for uniform or log-uniform; width is sigma of a Gaussian
\item[b] Range is 50--150\% of the default value.
\item[c]Solar C/O
\end{tablenotes}
\label{tab:priors}
\end{threeparttable}
\end{table*}

We use uninformative priors where appropriate in order to fully explore the possible parameter space. For quantities that can range over many orders of magnitude, such as the cloud top pressure or the metallicity, this means a log-uniform prior is necessary to avoid bias towards higher values. Otherwise --- for $T_{limb}$, $R_p$, and C/O --- we use uniform priors with limits either set by the functionality of the code (e.g., C/O) or conservatively derived from previous observations. Widening the prior for any parameter in the fiducial model has no significant effect on the result of the retrieval.

\begin{figure}[t]
\centering
{
\includegraphics[width=\textwidth,keepaspectratio]{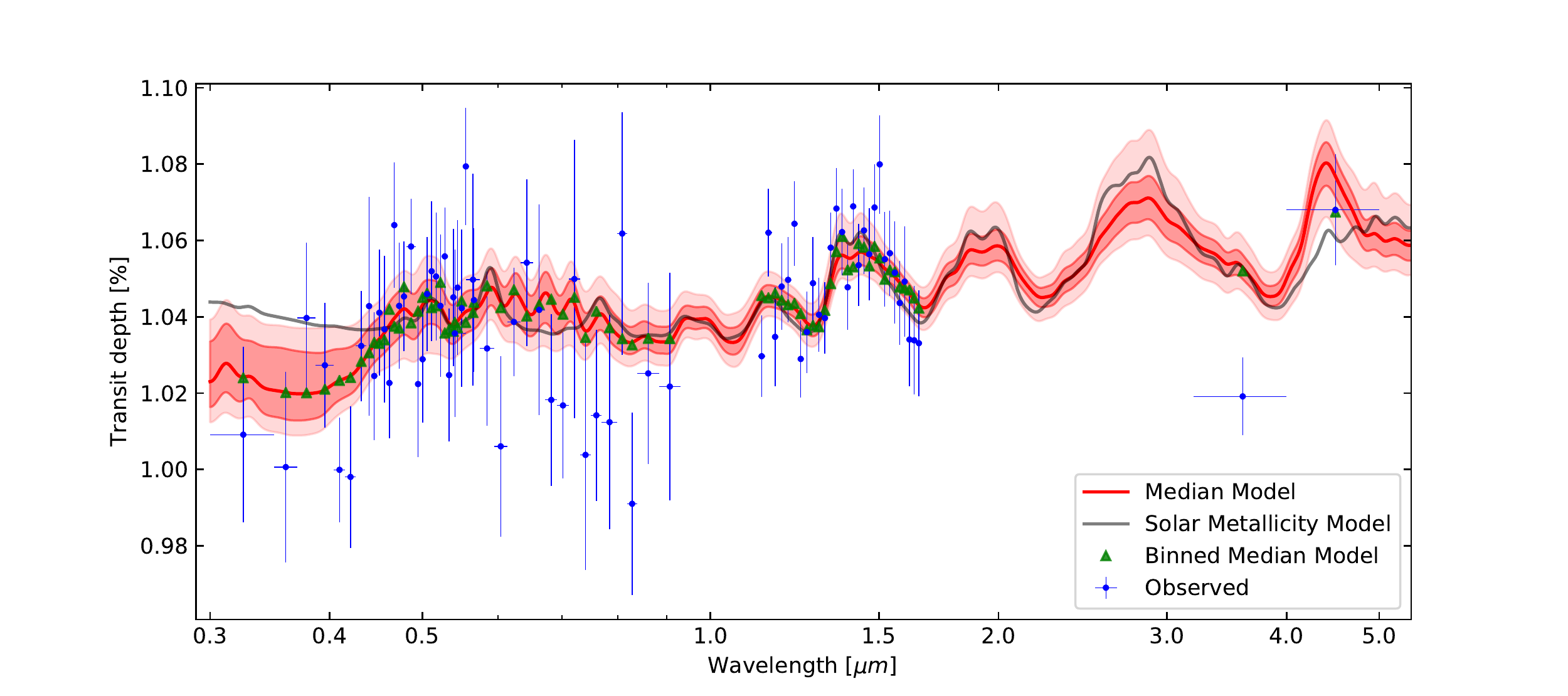}
\caption{Median retrieved model with 1-$\sigma$ and 2-$\sigma$ uncertainty contours for the fiducial model. Also shown is the median retrieved model when metallicity is fixed to solar.
The median model and uncertainties are derived by generating 100 samples from the correctly-weighted posterior and calculating the depth at each bin for each sample. The contours are given by the 2nd, 16th, 50th, 84th, and 98th-percentile depths at each bin. The continuous model is smoothed with a Gaussian filter with $\sigma=15$, which approximates the resolution of HST WFC3 \citep{zhang2019}.}
\label{fig:spectra}
}
\end{figure} 

The retrieved median fiducial model with uncertainty contours is shown with the observed spectrum in Fig~\ref{fig:spectra}. The model is an excellent fit (reduced $\chi^2$=1.09; consistent with the $\chi^2$ of the true model for 70 degrees of freedom to 1-$\sigma$). We clearly detect water vapor ($>$5-$\sigma$ significance) via the $1.4\mu$m water band in the WFC3 data. The bump in the STIS data is indicative of TiO, and the lack of any optical slope or flat-line indicates that grey clouds and scattering haze do not contribute significant opacity in the planet's spectroscopically active region. The difference between the two \textit{Spitzer} points is attributed to CO$_2$, though since they are photometric observations we do not resolve any feature. 

\begin{figure}
\centering
{
\includegraphics[width=0.75\textwidth,keepaspectratio]{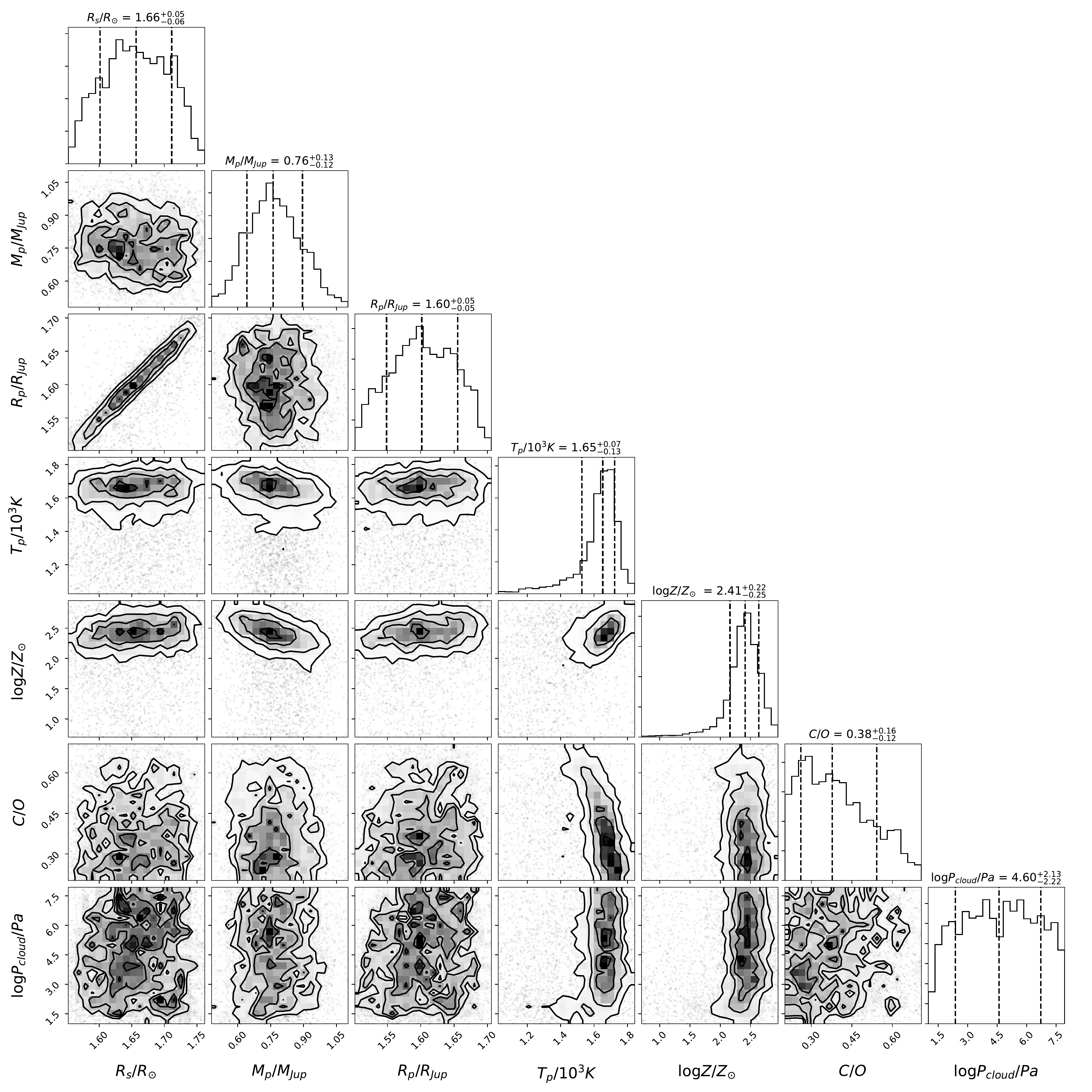}
\caption{Corner plot illustrating posterior probability distributions from PLATON for the fiducial model. The 16th, 50th, and 84th percentile values are indicated by vertical dashed lines and stated in the title of each parameter's 1D marginalized posterior distribution. The contours indicate the joint 0.5-, 1-,1.5-, and $2-\sigma$ levels for each 2D distribution. The 1-$\sigma$ metallicity range is $\log_{10}{Z/Z_{\odot} = 2.41^{+0.23}_{-0.25}}$ (145--437$\times$ solar metallicity), the isothermal limb temperature is well constrained around 1650\,K, and the C/O ratio is likely subsolar. The spectrum is consistent with a cloud-free atmosphere, and the marginalized posterior distributions of $M_p$ and $R_s$ are dominated by their priors. The slight correlations between $T$--$\log{Z}$ and $M_p$--$\log{Z}$ are due to their relation in the scale height equation.}
\label{fig:fiducial_corner}
}
\end{figure} 

The posterior probability distribution is represented by the corner plot \citep{foremanmackey2016}\footnote{\url{https://github.com/dfm/corner.py}} in Fig~\ref{fig:fiducial_corner}. This figure provides the marginalized posterior distribution for each parameter (with median, 16th, and 84th percentile values indicated by vertical dashed lines), as well as every two-dimensional projection of the posterior (with 0.5-, 1-, 1.5, and 2-$\sigma$ contours) to reveal covariances. Well-constrained parameters have narrow distributions with clear peaks, and slanted or diagonal shapes are indicative of correlated sets of parameters (e.g., the $R_p$-$R_s$ shape). We find a super-solar metallicity ($259^{+174}_{-114}\times$ solar metallicity, henceforth Z$_{\odot}$), a likely sub-solar C/O (C/O $<$ 0.6) that is consistent with stellar C/O (0.19), a clear atmosphere ($P_{cloud}>0.5$mBar), and $T_{limb}=1650^{+70}_{-120}$K. The temperature is driven primarily by the STIS data, mostly because PLATON interprets the bump in the STIS data as a metallic oxide, which is only the dominant opacity source above $\sim$1500K in chemical equilibrium. Below $\sim1500$K the optical spectrum would be dominated by an atomic sodium line, and this is not seen in the data. The upper limit on C/O is related mainly to the H$_2$O: in equilibrium chemistry for T$\sim$1650K and P$\sim$1 bar, H$_2$O opacity decreases exponentially when C/O $>$ 0.6 \citep{Madhusudhan2012}. Any model with C/O $>$ 0.6 struggles to capture the water feature and relatively high infrared baseline opacity (compared to the optical) and is thus a poor fit to the data. 

The high metallicity is constrained by the size of both the STIS and WFC3 features, as well as the lack of a significant Rayleigh scattering slope. While the metallicity affects chemistry, it is primarily constrained via its effect on the mean molecular mass of the atmosphere. Increasing metallicity increases the ratio of metals to hydrogen by definition, which increases the atmosphere's mean molecular mass. This lowers the atmospheric scale height and consequently decreases the predicted feature size. The equation for approximate feature size \citep[$\delta_\lambda$;][]{kreidberg2018}, where $\mu$ is the mean molecular mass, clarifies its dependencies: 




\begin{equation}
\delta_{\lambda}\propto \frac{TR_p}{\mu g_pR_s^2}\propto \frac{TR^3_p}{\mu M_pR_s^2}
\label{eq:prop}
\end{equation}

PLATON can decrease the feature size by changing $R_p$ or $R_s$, but both are well-constrained by the continuum baseline as well as priors and thus are relatively fixed. It can also be lowered by decreasing the temperature, increasing the metallicity (and thus the mean molecular weight), or by increasing $M_p$. The temperature is strongly constrained by chemistry, 
and $M_p$ is constrained by previous observations, so only metallicity can vary enough to explain the observed feature sizes. 
Note that this relationship explains the correlations between $M_p$, $T_p$, and metallicity seen in Fig~\ref{fig:fiducial_corner}: as mass increases or temperature decreases, metallicity decreases since a lower value is necessary to achieve the scale height that predicts the observed feature sizes. For reference, the median derived mean molecular weight is 5.8 AMU and the derived scale height is roughly 322km. 

At solar metallicity, the model predicts features that are much larger than what the data shows. Consequently, solar metallicity atmospheres in the fiducial model can only explain the observed feature by invoking a cloud to mute the troughs of the features. The median retrieved model for metallicity fixed to solar is shown in Figure~\ref{fig:spectra}. Note that this model is dispreferred by 3.5-$\sigma$, since the cloud leads to a poor fit to the bluest transit depths.

The same metallicity value is an excellent fit for both the water feature in the WFC3 data and the TiO feature in the STIS data. Retrieving only on the STIS data or only on the WFC3 data recovers supersolar atmospheric metallicities. Additionally, due to predicting a greater abundance of CO$_2$, it is better than low-metallicity solutions at explaining the large change in depth between the \textit{Spitzer} points. 
Observations from all three instruments support the high metallicity solution.

\subsection{More Complex Models}\label{subsec:complex}

In this section we incorporate additional model parameters to explore if more complex physics impacts the inferred atmospheric parameters. We demonstrate the insensitivity of our results to model assumptions. 

\subsubsection{Partial Cloud Coverage}
\citet{line2016} showed that partial cloud coverage (i.e., clouds at the same height but not uniformly covering the limb azimuthally) could mimic the effect of a high mean molecular mass atmosphere for WFC3 spectra. When partial clouds are present, the observed spectrum would be the weighted average of the cloudy and clear spectra. The transit depth of a grey-cloud dominated atmosphere does not vary with wavelength, and so the cloudy spectrum is a straight line. Averaging a clear spectrum with molecular features and a cloudy, flat spectrum reduces the size of the features by an amount proportional to the cloud fraction. Given that we find a significantly supersolar metallicity, we investigate if this possible mean molecular mass-cloud fraction degeneracy affects the results from the fiducial case.

When fit independently and allowing the cloud fraction to vary, both WFC3 and STIS spectra retrievals no longer constrain the metallicity to be supersolar. However, fitting the infrared and optical data jointly breaks this degeneracy, as predicted by \citet{line2016}. Effectively, a low-mean molecular mass and non-uniform cloud solution should be impacted by Rayleigh scattering in the optical data, especially the bluest six wavelength bins. The dominance of gas absorption opacity over Rayleigh scattering opacity in the STIS data disallows this solution, breaking the degeneracy in favor of the high mean molecular mass solution.


However, it is possible that removing assumptions made in the fiducal model --- such as fixed Rayleigh scattering or no instrumental offsets --- could muddle this decisive degeneracy break and allow for a low-metallicity solution. We investigate this below.

\begin{table*}
\centering       
\begin{threeparttable}
\caption{Prior distributions for more complicated models}  

\begin{tabular}{l l l l r}     
\hline\hline      
         
    Parameter & Symbol & Distribution & Range/Width\tnote{a} & Default Value \\
\hline
 Cloud fraction & $f_{c}$ & Uniform & 0--1 & 1 \\
 Scattering slope &$\gamma$ & Uniform & -2--20 & 4 \\
 Scattering factor & $a_0$ & Log-uniform & $10^{-4}$--$10^8$& 1\\
 WFC3 offset & - & Uniform/Gaussian & -500--500 ppm/80 ppm & 0 ppm\\
 STIS G430L offset & - & Uniform/Gaussian & -500--500 ppm/105 ppm & 0 ppm\\
 STIS G750L offset & - & Uniform/Gaussian & -500--500 ppm/85 ppm & 0 ppm\\
 STIS offset & - & Uniform & -500--500 ppm & 0 ppm\\
 Stellar Effective Temperature& $T_{star}$ & Fixed & 0 & 6480K\\
 Faculae Temperature & $T_{fac}$ & Fixed\tnote{b} & 0 & 6580K\\
 Faculae covering fraction & $f_{fac}$& Uniform & 0--0.10 & 0\\
 Mie Particle Size & $r_{part}$ & Log-uniform & 0.01--1\,$\mu$m & 0.1\,$\mu$m\\
 Mie Number Density & $n$ & Log-uniform & $10$--$10^{15}$ m$^{-3}$ & $10^5$ m$^{-3}$\\
 Fractional Scale Height & $H_{cloud}/H_{gas}$ & Log-uniform & 0.1--10 & 1.0\\
   
\hline
\end{tabular}
\begin{tablenotes}
\item[a]Range for uniform distribution, width is sigma of a Gaussian
\item[b] $T_{fac}=T_{star} + 100K$ \citep{rackham2019}
\end{tablenotes}
\label{tab:priors2}
\end{threeparttable}
\end{table*}

\subsubsection{Parametric Rayleigh Scattering}
The fiducial model assumes Rayleigh scattering. In lieu of complicated microphysics, PLATON allows parametric scattering, in which the slope and the magnitude of Rayleigh scattering vary in order to capture the possible signature of many hazes. For a more detailed explanation, see \citet{zhang2019}. Though there is no obvious signature of haze in the optical data (i.e., no linear slope decreasing with increasing wavelength), it is worth exploring if loosening the assumption of exact Rayleigh scattering affects the results.

\begin{figure}
\centering
{
\includegraphics[width=0.75\textwidth,keepaspectratio]{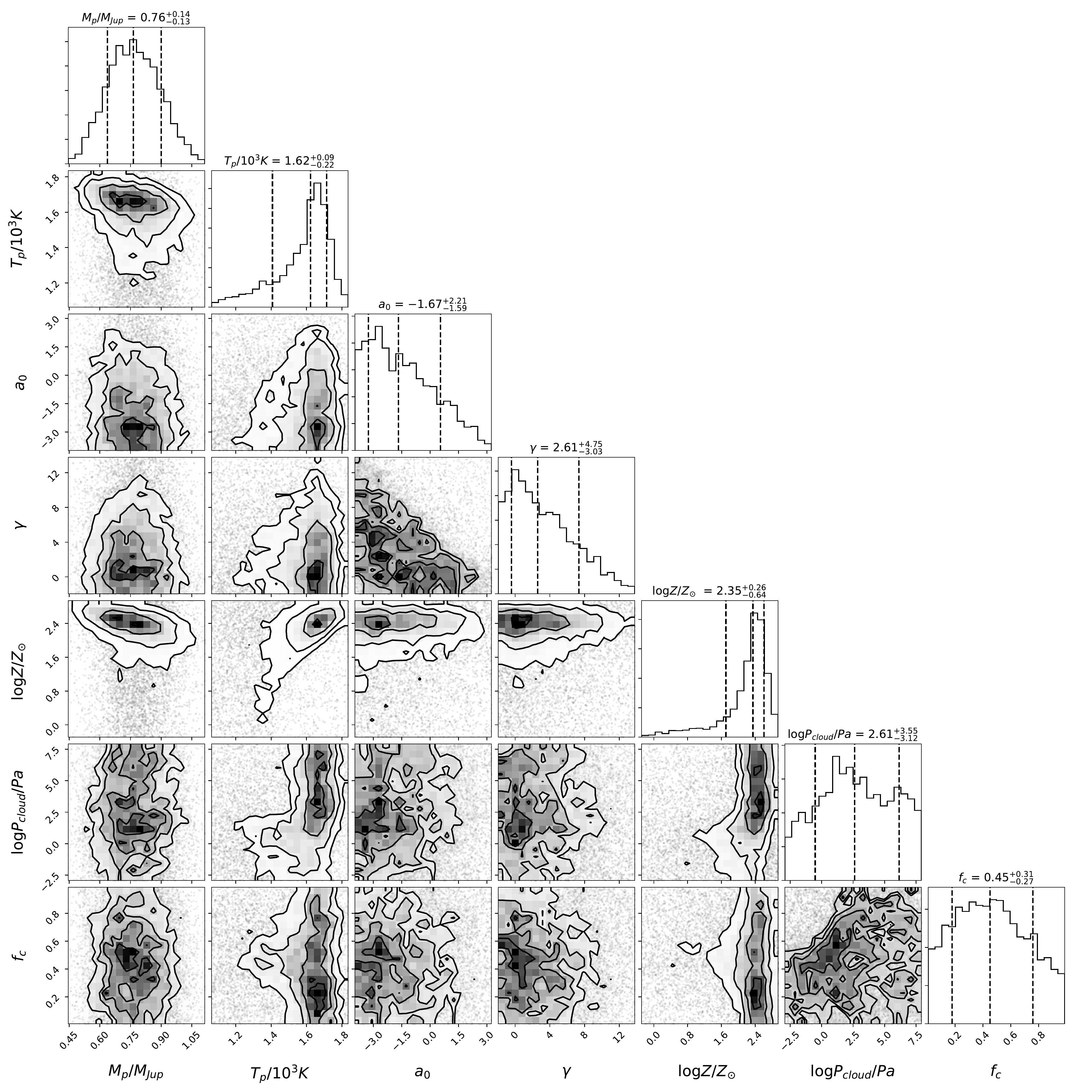}
\caption{Corner plot illustrating the posterior probability distribution from PLATON for the partial cloud and parametric scattering case. $R_p$, $R_s$, and C/O are not shown for clarity, since their marginalized posterior distributions are the same as in the fiducial case. The 1-$\sigma$ metallicity range is shifted down to $\log_{10}{Z/Z_{\odot}} = 2.34^{+0.27}_{-0.64}$. 
Note that at $f_c=0$ or $\log{P_{cloud}}>10^{2.5}$, we recover the fiducial marginalized posteriors.}
\label{fig:scat}
}
\end{figure}

Allowing the full scattering parameter space (see Table~\ref{tab:priors2}) has little effect: the clear lack of slope in the STIS data conclusively leads to a haze-free atmosphere. Further, the median scattering factor is 0.01, implying that the data is easiest to fit when opacity from Rayleigh scattering is muted. This complicates the mean molecular weight-cloud fraction ($\mu-f_c$) degeneracy. Lower values of $\mu$ are now possible, since the model no longer expects scattering opacity to be important at optical wavelengths. The lower the magnitude of Rayleigh scattering --- and the shallower the scattering slope --- the lower $\mu$ can be. This is because decreases in Rayleigh opacity allow for gas absorption to still be dominant at larger scale heights. As a result, a patchy cloud and low metallicity solution is viable. Though possible, the low $\mu$ solution requires a specific combination of cloud top pressure, scattering slope, scattering factor, and cloud fraction, and does not improve the fit. Therefore, it is much less likely than the high metallicity solution. The marginalized posterior probability distribution for metallicity has the same maximum likelihood value as the fiducial model. The difference is that the distribution has a tail extending to lower metallicities (Fig~\ref{fig:scat}). The resulting median log metallicity and 1-$\sigma$ range (as determined by the 16th and 84th percentile values) is $\log_{10}{Z/Z_{\odot}}=2.34^{+0.27}_{-0.64}$.

Though all cloud fractions and cloud pressures are allowed, the posterior is consistent with a clear atmosphere due to the likelihood-desert in the upper-left corner of the cloud fraction-cloud top pressure pairs plot: clouds are only seen above the altitude corresponding to the $\sim10$~Pa pressure level at fractions below 0.50.



\subsubsection{Instrumental Offsets}\label{subsec:offsets}

We model an instrumental offset as a constant value added to the forward model's binned transit depth in the wavelength range of the instrument of interest. For the physically motivated scenario (Scenario 1 from Section~\ref{subsec:inst_offsets}) , we set the priors for STIS G430L, STIS 750L, and WFC3 to be Gaussians centered on zero\,ppm with widths set to the uncertainty on the transit depth from their white light curves (105, 85 and 80\,ppm, respectively; see Table~\ref{tab:priors2}).

The retrieved median WFC3 offset is non-trivial, with a median of about 1.5$\times$ the white light uncertainty ($130\pm{50}$\,ppm). The offsets in the STIS G430L and G750L are less significant, at $58\pm{58}$\,ppm and $-65\pm{55}$\,ppm, both well within their white light uncertainties. However, there is a significant median offset of $\sim120$\,ppm between the two instruments. This is driven primarily by the retrieval attempting to align transit depths in the overlapping wavelength region between the instruments.

The \textit{Spitzer} 3.6\,$\mu$m point drives the WFC3 offset: shifting the WFC3 depths down necessitates a smaller radius ratio, which better captures the relatively low transit depth at 3.6\,$\mu$m. The ability to better capture the Spitzer 3.6\,$\mu$m results in a higher evidence, indicating that this model is strongly preferred over the fiducial model (for a more detailed discussion, see Section~\ref{subsec:model_selection}).


When combined with partial clouds, the instrumental offsets cause a small decrease in the retrieved median metallicity ($\log_{10}{Z/Z_{\odot}} = 2.33^{+0.23}_{-0.25}$). This is for similar reasons as explained in the parametric scattering section; whereas parametric scattering justified the absence of an optical scattering slope in the low-metallicity solution by effectively removing Rayleigh scattering opacity, the instrumental offset model can decrease the WFC3 depths relative to the STIS depths to artificially allow for it. 

Note that increasing STIS depths (instead of decreasing WFC3 depths) has the same affect on Rayleigh opacity and thus metallicity. However, it is not a viable solution since --- unlike decreasing WFC3 depths --- it does not improve the forward model's ability to capture the low Spitzer 3.6\,$\mu$m point.

Allowing Gaussian-prior instrumental offsets had no significant effect on the results. However, it is not impossible that there is some unknown wavelength-independent systematic that biases the absolute transit depths of the instruments relative to one another. Though unlikely, to explore this we allowed offsets in the STIS G430L, STIS G750L, and WFC3 data to vary by about 5\% (500 ppm) in either direction (Scenario 2 from Section~\ref{subsec:inst_offsets}). Due to the model preferring a lower radius ratio to best explain the Spitzer 3.6\,$\mu$m point, the median WFC3 offset is a 250\,ppm decrease, about 3$\times$ the white light uncertainty. Surprisingly, this large offset does not significantly change the 1-$\sigma$ ranges for metallicity ($\log_{10}{Z/Z_{\odot}} = 2.26^{+0.24}_{-0.40}$). The size of the molecular features and the large differential between \textit{Spitzer} photometric points drive the supersolar metallicity in this case. Regardless of the magnitude of the offset, we retrieve a high metallicity. The two large, uniform offset case, where both STIS instruments are offset by the same amount (Scenario 3 from Section~\ref{subsec:inst_offsets}), retrieves effectively identical posterior distributions as Scenario 2.


While it is worthwhile to understand the effect on the retrieval, there is no reason to expect such large instrumental offsets for HAT-P-41b. A transit depth offset can be caused by the necessity of analyzing each instrument differently. For example, not handling limb darkening consistently and not using consistent orbital parameters (i.e., inclination) for each analysis might cause an offset, but this is easily fixed and is not an issue for our dataset. Since the instruments' observations are from different dates, it is also possible that stellar variability could cause an offset. However, we have long-term photometry (Sec~\ref{sec:phot}) that shows no such variability. Additionally, the STIS depths are in good agreement with HST UVIS observations \citep{wakeford2020}. There is no indication that this particular observation is biased in any way, and unresolved companions are confidently ruled out \citep{evans2018}. The most plausible source is unaccounted for uncertainties or bias from the spectral analysis, as the WFC3 spectrum derived in this paper is shifted up $\sim90$\,ppm relative to the literature spectrum \citep{tsiaras2018}, as noted in Section~\ref{subsec:wfc3light}. However, this is still well below the 250\,ppm value preferred by the large, uniform offsets models. We determine that offsets beyond the physically motivated values are unlikely.




\subsubsection{Stellar Activity}
Section~\ref{sec:activity} demonstrated that HAT-P-41 is consistent with a quiet star and stellar activity is not expected to impact the transit spectrum. However, to be conservative, we investigated if allowing for greater stellar variability impacted our conclusions.

The typical signature of un-occulted, cool starspots is to mimic a haze-like slope in the transit spectrum, and such a signature is clearly absent in the derived transit spectrum of HAT-P-41b. On the other hand, the signature of hot faculae is a steep optical drop-off towards shorter wavelengths \citep{rackham2019}. Given that we see a drop in transit depths in the optical, the retrieval could plausibly be affected if faculae dominate over star spots, and so we focus on a faculae overabundance.


We assumed that the temperature of the stellar photosphere equals the stellar effective temperature. We modeled the faculae following the prescription from \citet{rackham2019} and, accordingly, fixed the faculae temperature to $T_{phot} + 100$K. PLATON weights the contributions from the different temperature regimes via the fractional coverage parameter, which represents the overabundance of faculae in the unocculted regions. \citet{rackham2019} states that moderately active F5V-dwarfs will have around 1\% faculae coverage, and up to about 7\% on the more active end. This is the faculae fraction, which is much higher than the faculae overabundance. However, we set a conservative uniform prior on the fractional coverage of 0-10\% in order to determine if high activity would significant alter our conclusions. 

We find that including stellar activity has no effect on the posterior probability distribution. It may seem that the STIS data could be explained by a featureless flat line and stellar activity instead of a TiO feature. However, the overabundance of faculae necessary to explain the drop in bluest six points (0.32--0.42\,$\mu$m) produces a poor fit to the rest of the STIS data. Therefore, even when including stellar variability, TiO is necessary to explain the STIS depths. Allowing a wider range of faculae temperatures also had no effect. 

In summary: there is no evidence of stellar variability from prior observations, and allowing for activity does not affect the retrieval.



\subsubsection{Mie Scattering}
\citet{benneke2019} recently invoked Mie scattering to explain anomalously low \textit{Spitzer} transit depths. Given the relatively low value of HAT-P-41b's Spitzer 3.6\,$\mu$m depth relative to the rest of the spectrum --- the fiducial model's predicted depth at 3.6\,$\mu$m is about 3.3-$\sigma$ away from the observed depth --- we included Mie scattering in our analysis. In PLATON, Mie scattering can be used in lieu of parametric Rayleigh scattering. 


Each condensate is described by a wavelength dependent complex refractive index, \textit{n-ik}, where \textit{n} is the real part and \textit{k} is the imaginary part of the index. This index explains how that particular condensate interacts with light with wavelengths similar to the particle size. PLATON assumes log-normal distribution in particle size with geometric standard deviation 0.5 to determine the abundance of different radii condensates for a given mean particle size \citep{zhang2019}. The other relevant factors are cloud height (condensates are only relevant above that pressure; below it gray cloud opacity dominates), particle density at the cloud top pressure, and condensate scale height. The condensate scale height is parameterized as a fraction of the gas scale height, and it describes how the abundance of Mie scattering particles decreases with height. The refractive index is fixed for a given condensate, and the other four parameters are fit for in the retrieval (see \ref{tab:priors2}).

PLATON tests one Mie scattering species at a time. Only a few species expected to form clouds in hot Jupiter atmospheres have condensation temperatures above HAT-P-41b's limb temperature ($T\sim1600$K) \citep{wakeford2015}. Those five (SiO$_2$, Al$_2$O$_3$, CaTiO$_3$, FeO, and Fe$_2$O$_3$) fall into two phenotypes: "low-\textit{n}" with real refractive index $n\approx1.5$ and "high-\textit{n}" with $n\approx2.5$. 
Though the \textit{k} values vary more significantly, we find that they do not have a significant impact on the absorption cross section of the condensates. 
We use \textit{n} and \textit{k} values from \citet{kitzmann2018}, which we average over our wavelength range (0.3--5\,$\mu$m). The \textit{n} values are flat over this range, and so the average is an excellent approximation. 
We tested retrievals with both the low-\textit{n} (corundum; Al$_2$O$_3$) and high-\textit{n} (hematite; Fe$_2$O$_3$) phenotypes. 

The priors for the fittable parameters are shown in Table~\ref{tab:priors2}. The prior for cloudtop pressure is the same as the fiducial model. Since the condensate radii must be such that they cause a relative drop in opacity around 4\,$\mu$m (i.e., increase opacity more in the near-UV by more than around 4$\,\mu$m) , we can constrain the mean particle size reasonably well. We set the prior to be log-uniform with a range that contains all plausible values. The number density is not known ahead of time, so we set an uninformative log-uniform prior; widening the prior further did not affect the retrieval. Finally, it is unclear what physical constraints there are on condensate scale height. \citet{fortney2005} finds that condensate scale heights can be one third of the gaseous scale height for hot Jupiters, and \citet{benneke2019} found $H_{part}/H_{gas}\approx3$ for a sub-Neptune. Using these values as guides, we set a conservative uniform prior on the fractional scale height and constrain it to be in the range 0.1--10.

\begin{figure}
\centering
{
\includegraphics[width=170mm]{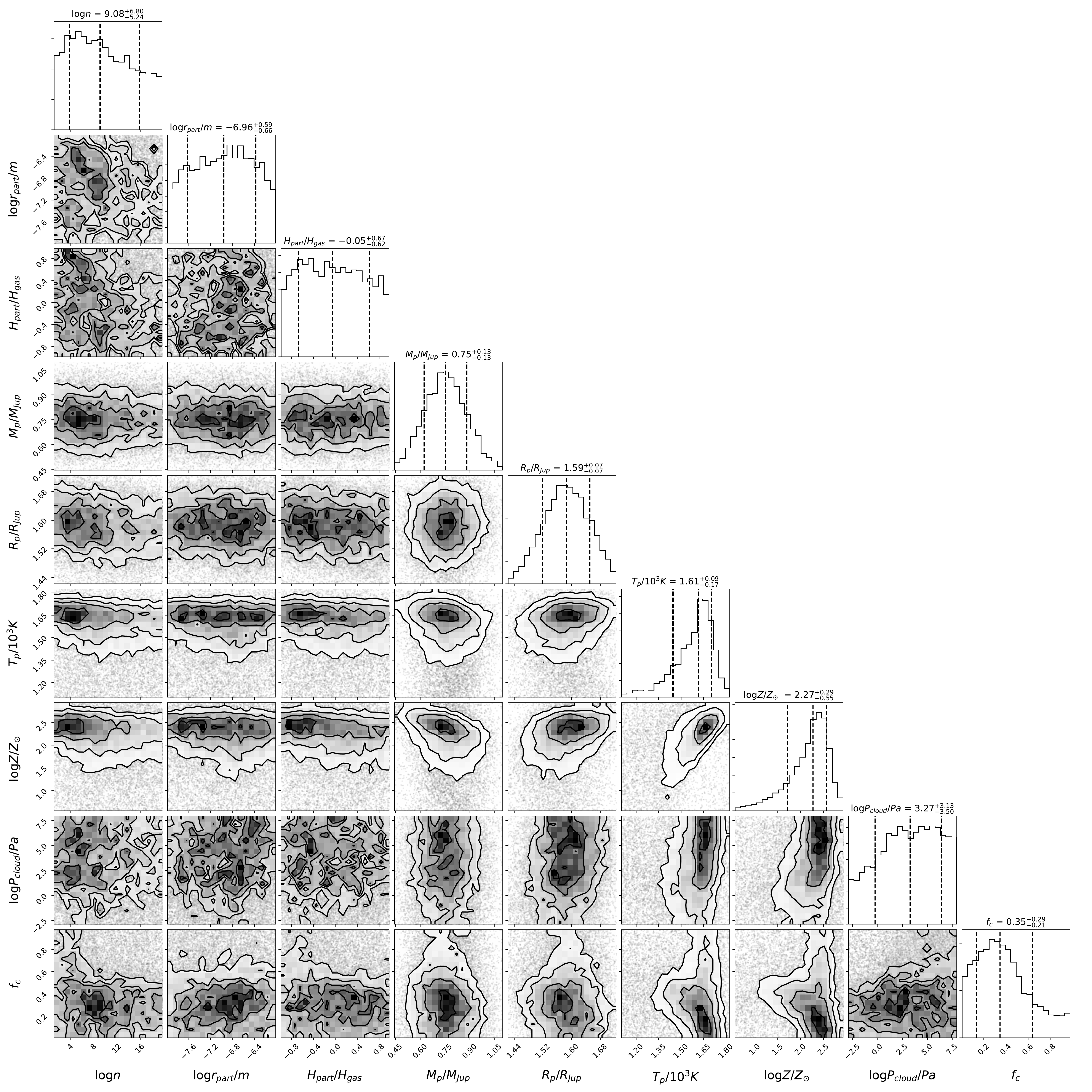}
\caption{Corner plot illustrating posterior probability distribution from PLATON for the fiducial plus Mie scattering and partial clouds model. $R_s$, and C/O are not shown for clarity, since their marginalized posterior distributions are the same as in the fiducial case. The 1$\sigma$ metallicity range is shifted down a bit to $\log_{10}{Z/Z_{\odot}} = 2.27^{+0.30}_{-0.55}$. Note that we recover the fiducial marginalized posteriors when any of the mean particle size, the particle number density, or the condensate scale height are too small.}
\label{fig:mie_corner}
}
\end{figure} 

Including Mie scattering opacity does not noticeably affect the results of the retrieval, and the Mie scattering parameters are not constrained by the retrieval. The inferred small gaseous scale height --- which dampens features and is necessary to explain STIS and WFC3 feature sizes --- makes it difficult to explain the large variations in the radius ratio. Combining Mie scattering with partial clouds --- physically, an atmosphere with patchy clouds and Mie scattering particles distributed only above those clouds --- alleviates the issue of explaining the large transit depth variation. Since partial clouds allow for higher scale heights, Mie scattering could then cause a larger drop in transit depth near \textit{Spitzer} without needing to invoke an unreasonably high fractional scale height.

Figure~\ref{fig:mie_corner} shows the corner plot for this model. Though the Mie scattering parameters are not constrained, at number densities above $\sim10^{8}$\,m$^{-3}$, particle radii around 0.15\,$\mu$m, and condensate scale heights greater than the gaseous scale height, lower metallicity and temperature values are possible. This is because added Mie opacity tends to mute features near its peak opacity. This provides a physical reason to expect smaller spectral features, and so a less small scale height is necessary to fit the features. The net impact is a decreased --- but still supersolar --- median metallicity of $\log_{10}{Z/Z_{\odot}} = 2.27^{+0.30}_{-0.55}$

\subsection{Model Selection}\label{subsec:model_selection}

Section~\ref{subsec:complex} stepped through the PLATON retrieval for increasingly complex models, examining both how each additional parameter affected the posterior and why it affected it in that way. While knowing the effect of each model assumption is useful, it is important to determine a preferred model in order to effectively convey the results. In this section, we use Bayesian model comparison to select the best model.

Model selection is as important as parameter estimation in atmospheric retrievals. We determine the preferred model by a combination of physical arguments and Bayesian statistics. Specifically, we check if it is necessary to consider more complicated physics using the odds ratio, which is the Bayes factor between models (defined as the ratio of their evidences) multiplied by their prior probability ratio. The prior probability ratio is typically assumed to be one (i.e., the models are assumed to be equally likely). The odds ratio determines if one model should be preferred over another by intrinsically rewarding better fits while punishing overcomplicated structure \citep{trotta2008}. This is entirely data-and-model defined, assuming appropriately uninformative priors are used. We compare the Bayesian evidences of each model in order to determine which should be favored. 

The Bayesian evidences and 1-$\sigma$ metallicity ranges for every model discussed in Section~\ref{sec:platon_retrieval} are shown in Table~\ref{tab:evidence}. The 1-$\sigma$ range is represented by both the median metallicity with quantiles (i.e., the central 68\% of metallicity values) 
The 1-$\sigma$ metallicity ranges are included to illustrate the uncertainty caused by model choice. The retrieved atmospheric metallicities are remarkably consistent across the models, and a supersolar metallicity is ubiquitous. This demonstrates that under PLATON's assumptions, supersolar metallicity is a robust conclusion.

\begin{figure}[t]
\centering
{
\includegraphics[width=0.5\textwidth,keepaspectratio]{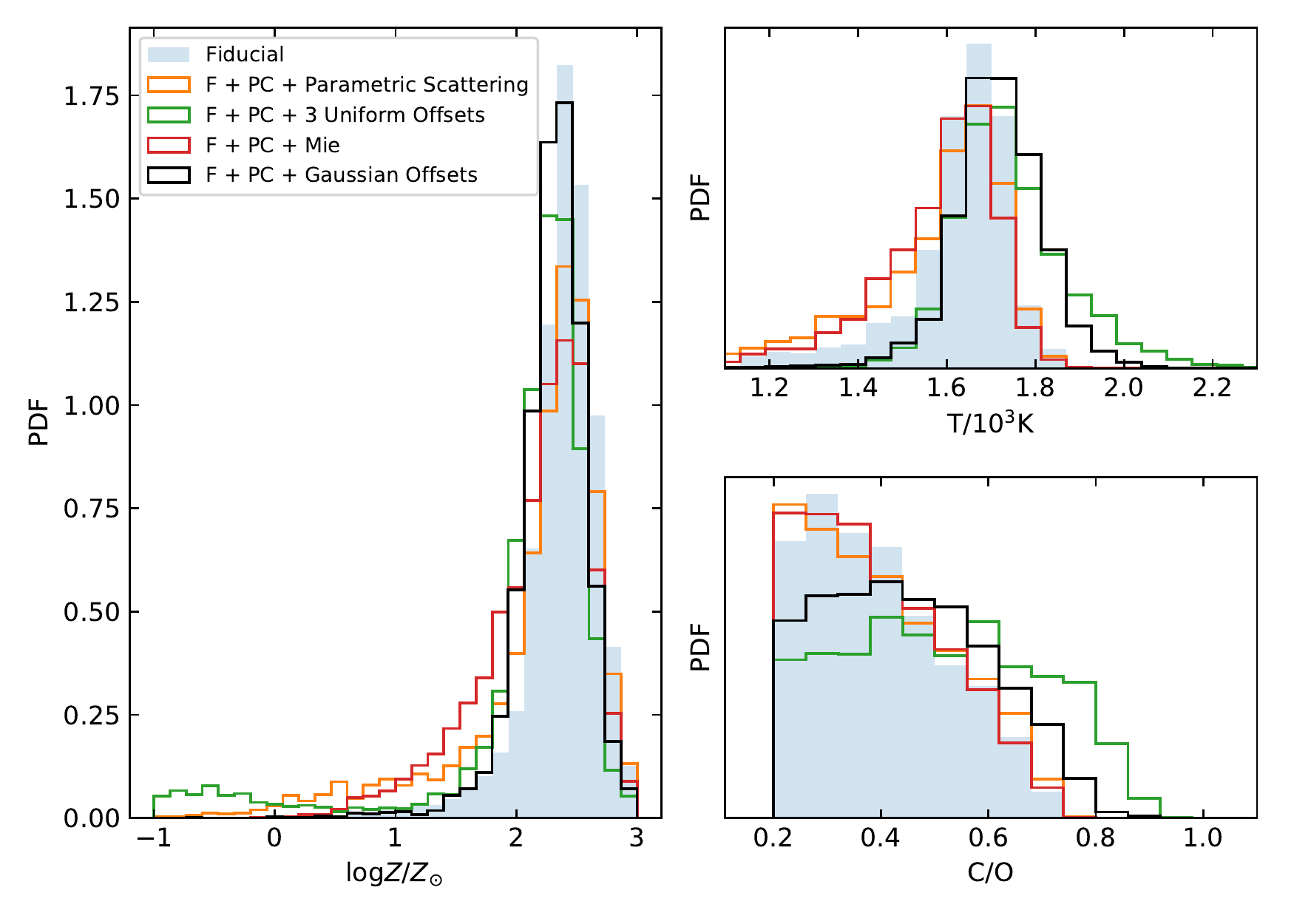}
\caption{Marginalized posterior distributions for metallicity, temperature, and C/O from select models compared to solar values. Note that stellar C/O = 0.19 (Table~\ref{tab:brewer}). Though some models have low metallicity tails, the 68\% credible interval for metallicity is robust (Table~\ref{tab:evidence}). Offsets allow for higher temperatures and C/O ratios, while both parametric and Mie scattering allow for lower temperatures.}
\label{fig:posteriors}
}
\end{figure}

Figure~\ref{fig:posteriors} emphasizes the insensitivity of the atmospheric parameters to model assumptions. This shows the one dimensional marginalized posterior distributions for metallicity, temperature, and C/O for five of the models we examined. These specific models are shown because they are ``interesting'' in that they differ from the fiducial model's posteriors the most. We emphasize that these are these are the models that most differ from the fiducial case. While including instrumental offsets tends to flatten the distributions, the peaks of all of the models are incredibly consistent. 

We define the model selection-relevant columns here:
\begin{itemize}
\item $\ln{\mathcal{Z}}$ is the natural log of the Bayesian evidence. A higher value indicates the model is better able to describe the data without overfitting.
\item \textbf{Weight} Weight assigned to each model for Bayesian model averaging based on Bayesian evidence and prior probability (Section~\ref{subsec:bma}). This can be roughly interpreted as the probability of each model.
\item {$\mathcal{O}$} is the odds ratio in favor of a model over the fiducial model. It is the product of their Bayes factor and their prior probability ratio. The prior probability ratio is often assumed to be one, as is the case here. This can be directly interpreted: an odds ratio of 100 indicates 100:1 odds in favor of the more complex model. Values less than one indicate evidence against the corresponding model.
\item \textbf{Interpretation} This is the empirically derived interpretation of odds ratio based on the Jeffreys' scale \citep{trotta2008}. 

\end{itemize}

\begin{table*}[t]

\caption{Evidence and metallicity ranges for each plausible model. Only the models including instrumental offsets are preferred over the fiducial model. No model assumption changes the conclusion of a supersolar atmospheric metallicity.}

\begin{threeparttable}
\begin{tabular*}{\textwidth}{l @{\extracolsep{\fill}} r r r r r l}

\hline\hline      
    Model & $\log_{10}$Z\tnote{a} & Z\tnote{b} & $\ln{\mathcal{Z}}$\tnote{c} & Weight & $\mathcal{O}$\tnote{d} & Interpretation\tnote{e}  \\
\hline
Fiducial (F) & $2.41^{+0.23}_{-0.25}$ & 145--437 & 551.6 & 0.0 & Ref & Default Model \\
F + Partial Clouds (PC) & $2.38^{+0.22}_{-0.37}$ & 102--398 & 551.7 & 0.0 & 1.1 & Inconclusive\\
{F + PC + Parametric Scattering} & $2.34^{+0.27}_{-0.64}$ & 50--407 & 550.7 & 0.0 & 0.4 & Inconclusive\\
F + PC + Stellar Actvity & $2.41^{+0.27}_{-0.41}$ & 100--479 & 551.8 & {0.0} & {1.3} & Inconclusive \\
F + Mie Scattering & ${2.41^{+0.24}_{-0.29}}$ & {132--447} & {551.0} & {0.0} & {0.6} & Inconclusive\\
F + PC + Mie & ${2.27^{+0.3}_{-0.55}}$ & {52--372} & {551.8} & {0.0} & {1.3} & Inconclusive\\ 
F + PC + {3} Gaussian Offsets & ${2.33^{+0.23}_{-0.25}}$ & {120--363} & {556.4} & {0.33} & {122} & {Strongly preferred}\\
F + PC + {3} Uniform Offsets & ${2.26^{+0.24}_{-0.4}}$ & {72--316} & {556.9} & {0.58} & {213} & {Strongly preferred} \\
{F + PC + 2 Uniform Offsets} & ${2.30^{+0.26}_{-0.39}}$ & {81--363} & {554.9} & {0.08} & {29} & {Moderately preferred} \\ 

\hline
\end{tabular*}
\begin{tablenotes}
\item[a] Median {log} metallicity with 16\% and 84\% quantiles, in units of log solar metallicity
\item[b] 68\% credible interval for metallicity, in units of solar metallicity
\item[c] Natural log of Bayesian Evidence
\item[d] Odds ratio between model and the fiducial model
\item[e] According to Jeffreys' scale \citep{trotta2008}
\end{tablenotes}

\label{tab:evidence}
\end{threeparttable}
\end{table*}

Table~\ref{tab:evidence} contains every notable model we considered. We did not do an iterative combination of every model scenario, for two primary reasons. Most importantly, we are weary of overfitting the data. The fiducal model is already an excellent fit to the data ($\chi^2_{\nu} = 1.09$), so we must be careful about adding complications. Layering multiple parameter physical processes, such as Mie scattering and stellar activity, involves an extra five parameters and significantly overcomplicates the fit. Instead, we only combine complications when there is a physically motivated reason to do so, e.g. partial clouds. The second reason is computational difficulty. Some model combinations have enough free parameters to describe the data with many different combinations, and so the retrieval does not converge on the timescale of weeks. Mie scattering combined with offsets falls in this category. However, given the overfitting concerns and the lack of evidence for just Mie scattering, we do not think this is a worrying omission.


It is generally best practice to assume the simplest model unless there is evidence in favor of extra parameters. That is why we list the fiducial model as the reference, and determine the evidence of the more complicated models. If the evidence of the model with extra parameters is not significantly greater, it means that the ability to explain to data was not improved enough to justify the added complexity. This essentially quantifies Occam’s razor.

Following this logic, we determine the ``fiducial + partial clouds + {3} Gaussian offsets'' model to be the best model. Only the Gaussian offset and uniform offset models are preferred over the fiducial model. {While the three uniform offsets model has the highest evidence/weight, the odds ratio between that and the Gaussian offsets model is 1.75. This is inconclusive on the Jeffrey's scale, meaning we are unable to distinguish between the models by evidence alone. Instead, we favor the Gaussian offsets model as more plausible, since {its Gaussian priors are physically motivated by common-mode corrections.}}

The evidence for partial clouds is inconclusive, however partial clouds are more plausible than assuming 100\% cloud coverage, as argued by \citet{MacDonald2017, Welbanks2019}. Therefore, to be conservative, we choose the model which account for partial clouds as the ``best'' model. 

The odds ratio only works in a direct comparison of two models and is not a statement on the absolute goodness-of-fit. The reduced chi-squared test statistic is a useful sanity check to ensure that the model is able to explain the variance in the data. The value for the best model is {an ideal} ${\chi^2_{\nu} = 1.0}$. The results section (Section~\ref{sec:results}) --- and the abstract values --- are based on parameter estimation from the ``fiducial + partial cloud + Gaussian offsets'' model.

\subsubsection{Bayesian Model Averaging}\label{subsec:bma} Instead of model selection, it is possible to take a weighted average of the results from each model and therefore automatically take their respective evidences into account \citep{gibson2014, wakeford2016, wakeford2018}. The benefit of Bayesian model averaging is the ability to quantify uncertainty in model selection, as well as avoiding having to arbitrarily choose between models with slightly different evidences. However, it requires a few assumptions: it is only valid if the set of models comprises the full model space, i.e, at least one model is a good description of the data. The weight-averaged uncertainties assume Gaussian-distributed posteriors, which is not strictly correct. However, it is useful in combining information from every model. 


Here, we show the assumptions we make to use Bayesian model averaging. The $\chi^2_{\nu}$ values for the models we tested are clustered around one, so it is fair to assume that a ``correct'' model is contained in the set. {Figure~\ref{fig:posteriors} shows that although the posteriors are not perfectly Gaussian, they have sharp, unimodal peaks, and so the uncertainty derived from marginalization is informative. }



The model weights are defined by Eq~\ref{eq:marg} (adapted from \citet{gibson2014}). $W_q$ is the weight assigned to model \textit{q}, $P(M_q|D)$ is the the likelihood of model \textit{q} given the data, and $P(D|M_q)$ is the likelihood of the data given model \textit{q}, which is equivalent to the Bayesian evidence of model \textit{q}, $E_q$. The denominator is a normalization term, summed over \textit{N} models. {We assume a conservative prior that each model is equally likely ($P(M_i) = 1$ for all $i$).
}
\begin{equation}
W_q = \frac{P(M_q|D)}{\sum_{i=1}^{N}P(M_i|D)} = \frac{P(D|M_q)P(M_q)}{\sum_{i=1}^{N}P(D|M_i)P(M_i)} = \frac{E_qP(M_q)}{\sum_{i=1}^{N}E_iP(M_i)}
\label{eq:marg}
\end{equation}

The marginalized log metallicity with 1-$\sigma$ uncertainties is calculated from equations 15 and 16 from \citet{wakeford2016}. The result is
${\log_{10}{Z/Z_{\odot}} = 2.29^{+0.24}_{-0.36}}$ (${194^{+144}_{-109}\times}$~Z$_{\odot}$). As expected, the highest weighted models are the offset models.
Bayesian model averaging demonstrates that in PLATON's chemical equilibrium framework, a supersolar metallicity is the most likely result even after accounting for uncertainty in model selection. 

The marginalized metallicity is useful as a reference, but it is valuable to give the metallicity distribution for each specific model assumption. Marginalization is most appropriate when the specific model parameters are unimportant, however we are interested in the impact that modeling assumptions have on the atmospheric parameters. We emphasize that even for apparently ``data-defined'' methods, many assumptions have to be made and those should be explicitly stated for an appropriate interpretation.




\section{Results for the Favored PLATON Model} \label{sec:results}

\begin{figure}[t]
\centering
{
\includegraphics[width=1.0\textwidth,keepaspectratio]{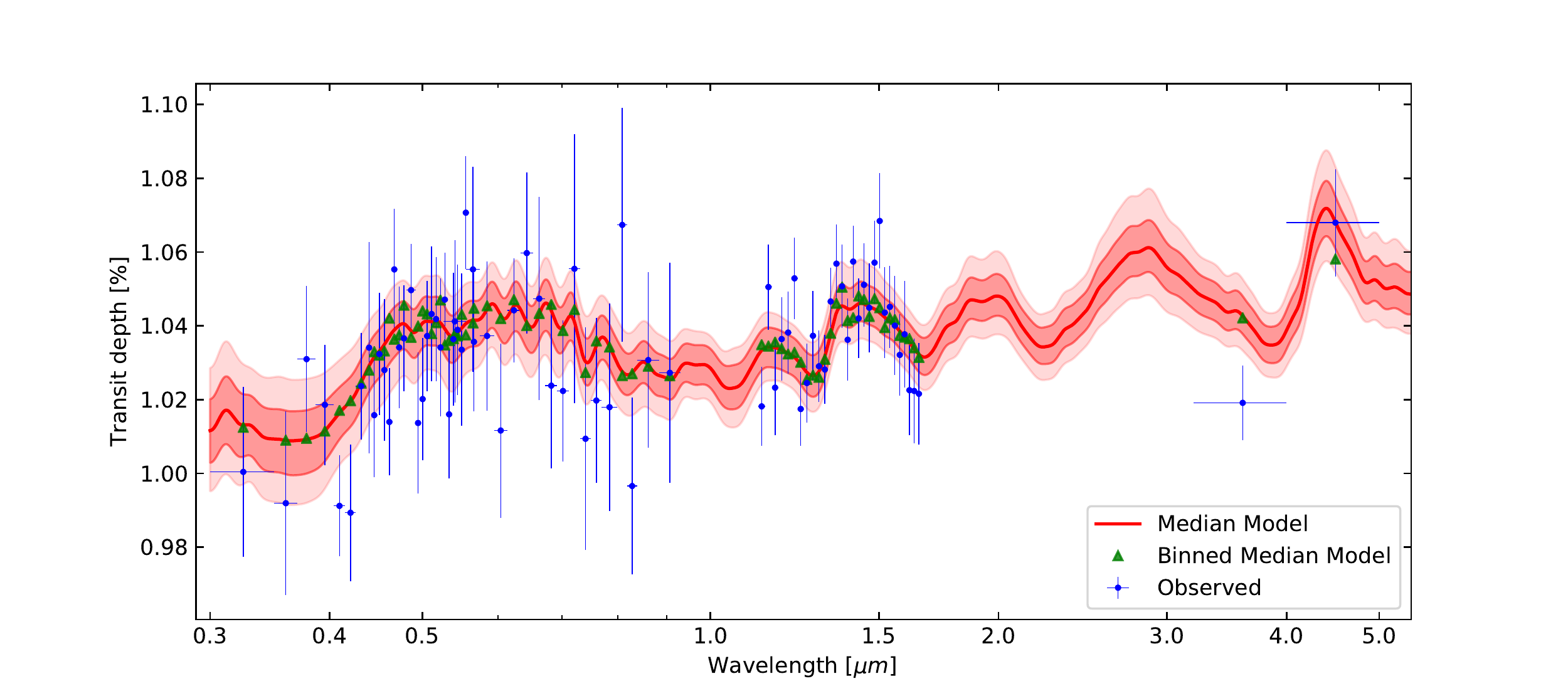}
\caption{Median retrieved model with 1-$\sigma$ and 2-$\sigma$ uncertainty contours for the {favored} PLATON model (fiducial model with partial clouds and {instrumental shifts with physically-motivated Gaussian priors}).}
\label{fig:fpcoff_spectrum}
}
\end{figure} 
In Section~\ref{subsec:model_selection} we argued that the best PLATON model scenario is the fiducial model with partial clouds and {physically motivated, Gaussian prior }instrumental offsets added. The retrieved median spectrum with uncertainty contours is shown in Figure~\ref{fig:fpcoff_spectrum}. {It is an excellent fit to the data, with $\chi^2_{\nu} = 1.0$}. In this section we discuss the details of the retrieved atmospheric parameter values.

\begin{figure}[t]
\centering
{
\includegraphics[width=1.0\textwidth,keepaspectratio]{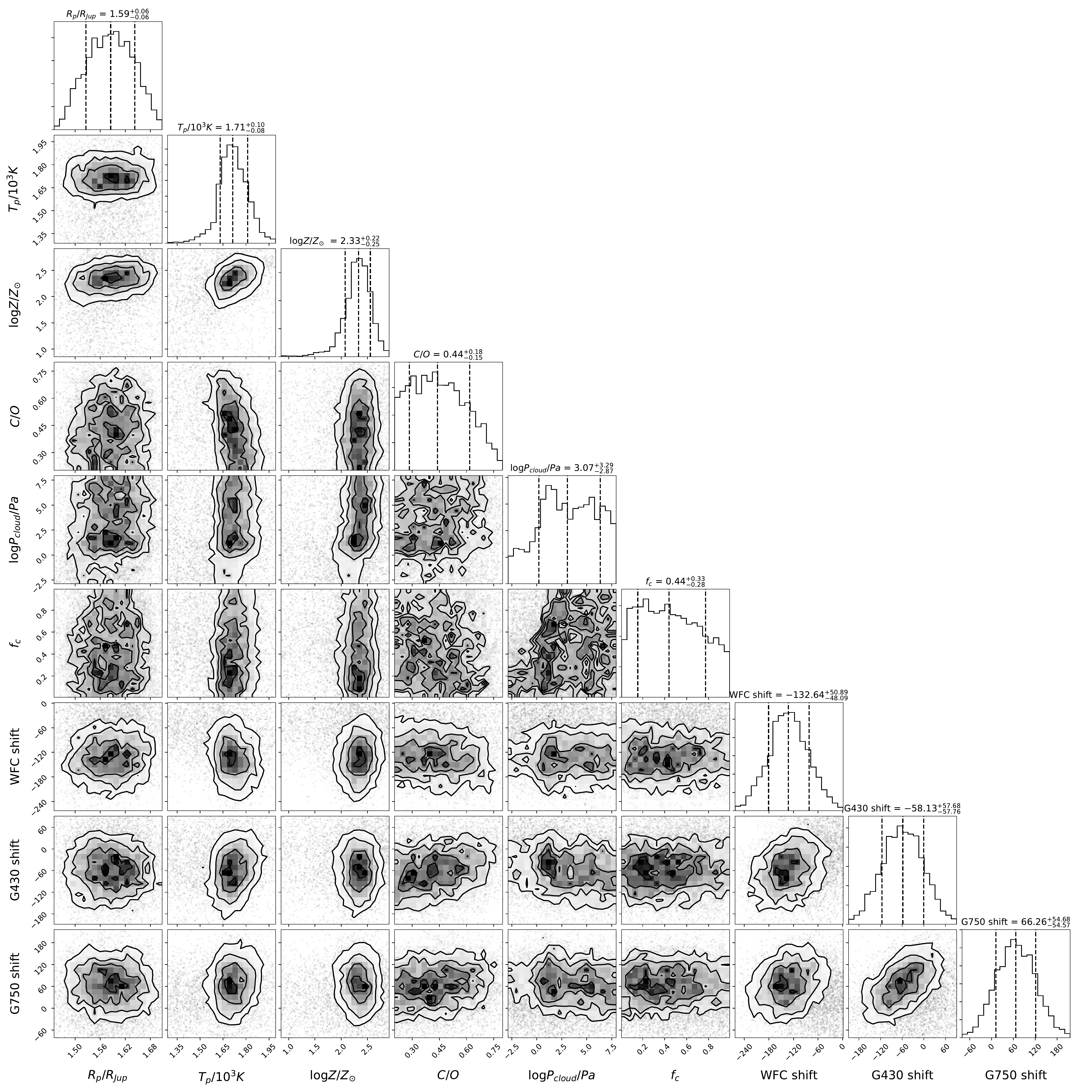}
\caption{Corner plot for the best PLATON model (fiducial with partial clouds and Gaussian offsets). {R$_s$ and M$_p$ are prior dominated and are excluded for clarity.} The offsets are given in parts per million; for example, the median WFC3 offset {indicates the retrieval favors shifting} the WFC3 depths down by ${\sim132}$\,ppm.}
\label{fig:fpcoff_corner}
}
\end{figure} 

\subsection{Summary of Retrieved Parameters}
The posterior distribution corner plot is shown in Figure~\ref{fig:fpcoff_corner}. Both atmospheric metallicity and temperature are well constrained, and the C/O ratio, though relatively flat, has a strict upper limit. The median retrieved metallicity is super-solar (${\log_{10}{Z/Z_{\odot}} = 2.33^{+0.23}_{-0.25}}$), and solar metallicity is inconsistent to 3-$\sigma$ (lower limit {4.8}$\times$~Z$_{\odot}$). {As noted in Section~\ref{sec:metalco}, PLATON's derived metallicity is a proxy for [O/H], enabling comparison with its host star's oxygen abundance ([O/H]=0.37; Table~\ref{tab:brewer}). PLATON determines HAT-P-41b to be metal-enriched relative to its host star ((${\log_{10}{Z/Z_{star}} = 1.97^{+0.23}_{-0.25}}$), and it is inconsistent with the stellar metallicity to  to 3-$\sigma$ (lower limit 2.1$\times$~Z$_{star}$). The planetary C/O ($0.44^{+0.18}_{-0.15}$) has a 3-$\sigma$ upper limit of 0.83. Though the planetary C/O is technically inconsistent with the stellar C/O to 1-$\sigma$ (0.19; Table~\ref{tab:brewer}), the comparison is not valid as the planetary C/O prior had a computational lower limit of 0.20, and the posterior has significant likelihood at that limit. This ``piling'' at the prior boundary implies that the planetary C/O is consistent with the stellar C/O.} The median isothermal limb temperature ($T_{\textrm{limb}}={1710^{+100}_{-80}}$~K) is close to the equilibirum temperature of the planet ($T_{eq}=1960$~K), which implies an {efficient heat} recirculation.  These parameters lead to a high mean molecular weight (${\mu\sim5.5}$~AMU) atmosphere with a scale height of about {320~km}.

The retrieved results are consistent with a clear atmosphere. Though cloud top pressure and cloud fraction are unconstrained, their joint marginalized posterior is constrained. A uniform grey cloud is only allowed deeper than $\sim10$~Pa (0.1~mBar), and clouds above that pressure are only possible if they cover less than about 40\% of the limb. Hazes are dispreferred by model selection, and the median scattering opacity was {50}$\times$ weaker than Rayleigh scattering in the model which allowed parametric scattering.

{The retrieved relative shift between the STIS G430L and G750L instruments is 120ppm, due in part to the model attempting to align their overlapping regions.} A downshift for the WFC3 data is preferred (${\textrm{WFC3 offset} =-132\pm{50}}$\,ppm). The stellar radius, the planetary mass, and the planetary radius are consistent with the prior values. The planetary mass and stellar radius are, as expected, dominated by their priors. The planetary radius (${R_p=1.59\pm{0.06}}$) is at the reference pressure of 1~Bar, and when calculated at the planet's photosphere it is consistent with the planetary radius {derived based on stellar parameters from TIC-8.}

\subsection{{Evidence of Water and Optical-Wavelength Absorbers}}\label{subsec:platon_oxides}
While the spectral features in STIS, WFC3, and \textit{Spitzer} are attributed by PLATON to TiO, H$_2$O, and CO$_2$, respectively, the retrieval only robustly detects H$_2$O - the H$_2$O abundance is constrained by observations, while the abundances of other species are primarily constrained by the assumption of chemical equilibrium. {We note that while CO is more abundant than CO$_2$, CO$_2$ has a much larger cross section at 4.5\,$\mu$m, such that even with a smaller abundance its opacity dominates over that of CO at the temperatures and C/O ratios inferred by the retrieval.}  

We determine if a species is detected by finding the odds ratio between the best model with and without opacity from a particular species. This breaks the assumption of chemical equilibrium, so it is not strictly correct, but it is a useful heuristic nonetheless. A species is considered detected only when the odds ratio significantly favors the model with the species' opacity. {Table}~\ref{tab:detection} {shows the odds ratios --- and their more familiar frequentist analog, the detection significances \citep{Benneke2013} --- for several relevant spectroscopically active species.}

\begin{table*}
\centering
\begin{threeparttable}
\caption{{PLATON species detection evidences.}}
\begin{tabular}{l c c}
\hline\hline      
      &   & {Detection} \\
    {Species} & ${\mathcal{O}}$\tnote{a} & {Significance}\tnote{b}\\
\hline
{H}${_2}${O} & {46630} & {5.0}$\sigma$\\
{TiO} & {2.1} & {1.9}$\sigma$ \\
{VO} & {2.3} & {1.9}$\sigma$ \\
{TiO/VO} & {9.4} & {2.7}$\sigma$\\
{Na} & {1.1 }& {1.2}$\sigma$ \\
{CO$_2$} & {3.3} & {2.1}$\sigma$ \\
{CO} &{ 0.4} & N/A\\

\hline
\end{tabular}
\begin{tablenotes}
\item[a] {Odds ratio between model and the preferred PLATON model (fiducial model with partial clouds and Gaussian-prior instrumental shifts included)}
\item[b] {\citet{Benneke2013}}
\end{tablenotes}

\label{tab:detection}
\end{threeparttable}
\end{table*}

The odds ratio in favor of H$_2$O is ${\sim46630}$, indicating that the model with water is ${46630\times}$ more likely than the model without water opacity. This is equivalent to a {5.0}-$\sigma$ detection in frequentist terms. {The odds ratio in favor of CO$_2$ is 3.3, which is barely enough evidence to claim a weak detection. PLATON finds no evidence of Na, and CO is dispreferred.} 
The odds ratios for TiO {and VO }are 2.1 {and 2.3}, respectively, and these are not favored enough to claim detections (less than 2-$\sigma$). However, TiO and VO are only seen as non-detections because they have similar cross-sections. When TiO opacity is ignored, the retrieval can compensate because VO opacity is able to describe the STIS feature just as well as TiO. If we ignore both VO and TiO then the model cannot describe the STIS data as well, and so the odds ratio in favor of TiO/VO is {9.4} ({2.7}-$\sigma$). {Therefore, we find suggestive evidence of metallic oxide opacity, but we are} unable to discern if it is due to TiO or VO. Based on the assumption of chemical equilibrium {at the retrieved temperatures}, PLATON attributes the STIS feature to TiO because it is more abundant {and opaque in the spectrscopically active region} for a solar Ti/V ratio. 

\section{AURA Retrieval Analysis and Results}\label{sec:AURA_retrieval}

We perform a {second, complementary atmospheric retrieval analysis: a} series of {free-chemistry} retrievals on HAT-P-41b {using AURA (\ref{subsec:AURA})} to constrain the atmospheric properties at the day-night terminator of the planet {while allowing for deviations from chemical equilibrium}. First, we consider the presence of different chemical species in the atmosphere of HAT-P-41b using its full broadband spectrum. {Then}, we consider the presence of possible transit depth offsets between data sets and their possible impact in the derived chemical abundances and associated metallicities.

\subsection{{Evidence of Water and Optical-Wavelength Absorbers}}
\label{subsec:AURA_main}

\begin{figure}
\includegraphics[width=1.0\textwidth]{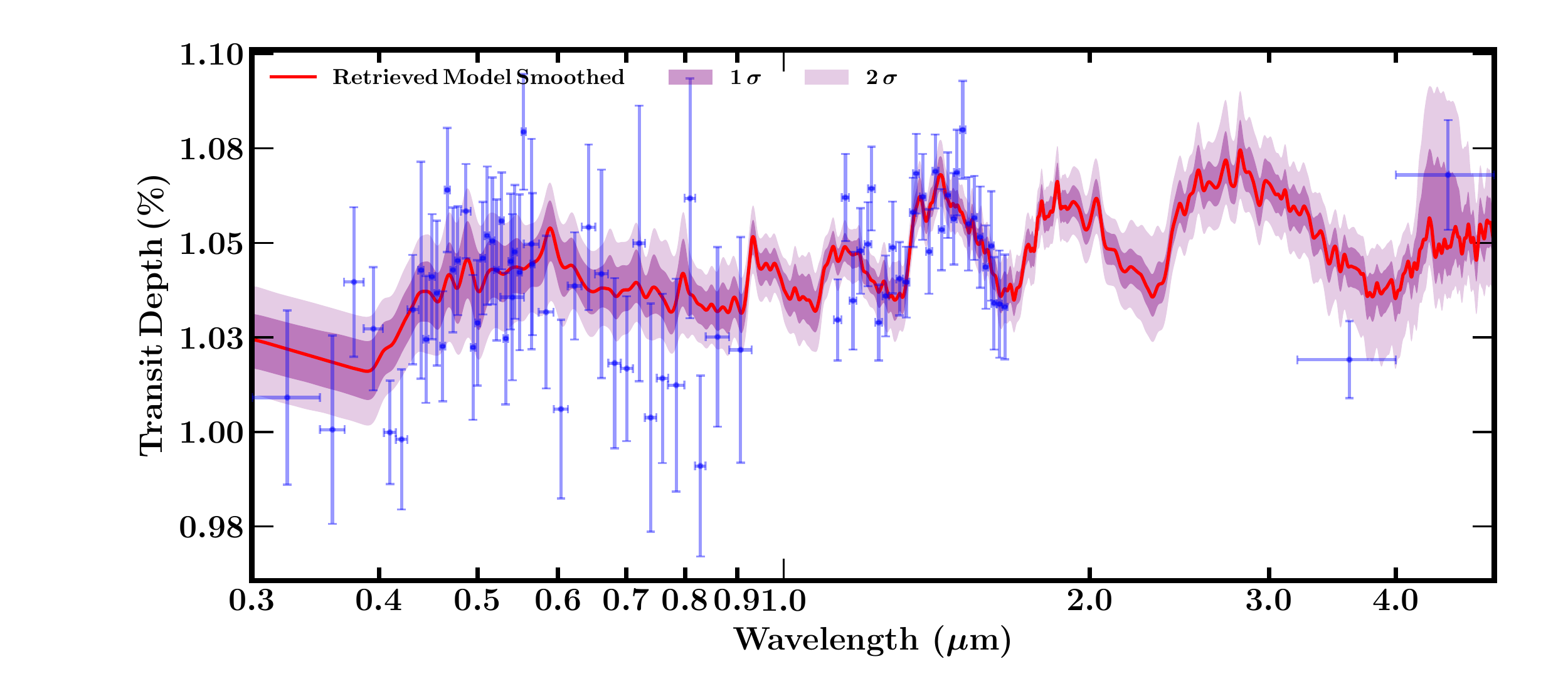}
\centering
  \caption{Retrieved spectrum of HAT-P-41b using STIS, WFC3 and \textit{Spitzer} data. Observations are shown using blue markers. The retrieved median spectrum is shown in red while the 1-$\sigma$ and 2-$\sigma$ regions are shown using the shaded purple areas. }
\label{fig:AURA_full_spectrum}
\end{figure}

\begin{figure}
\includegraphics[width=1.0\textwidth]{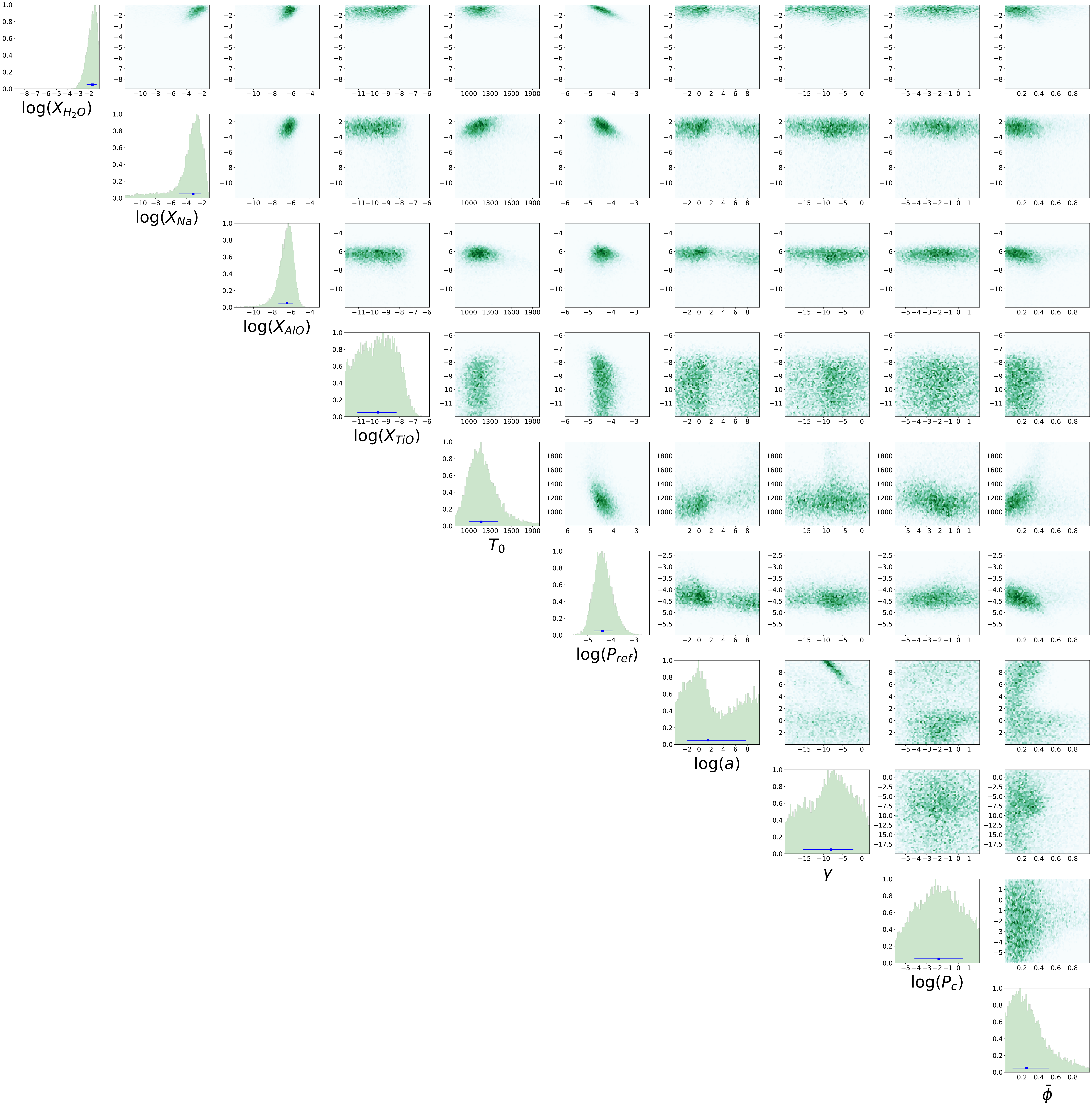}
\centering
  \caption{Posterior distributions of the relevant parameters for the full retrieval (Model 1 in Table~\ref{table:AURA_models}) using STIS, WFC3 and \textit{Spitzer} data. The abundances of H$_2$O, Na and AlO are constrained, while the cloud and haze parameters are not constrained. { The parameter T$_0$, the temperature at the top of the atmosphere (10$^{-6}$bar) is shown as a subset of the P-T parameters used in the model. }}
\label{fig:AURA_full_posterior}
\end{figure}

We perform a full retrieval on the broadband spectrum of HAT-P-41b and present the observations and retrieved median spectrum in Figure~\ref{fig:AURA_full_spectrum}. The full retrieval provides constraints on the presence of H$_2$O, {and provides indications for the presence of { Na and/or AlO} in the optical}. The full retrieval finds  $\log_{10}$(X$_{\text{H}_2\text{O}})={-1.65 ^{+ 0.39 }_{- 0.55 } }$, $\log_{10}$(X$_{\text{Na}})={ -3.09 ^{+ 1.03 }_{- 1.83 } }$ and  $\log_{10}$(X$_{\text{AlO}})={ -6.44 ^{+ 0.66 }_{- 0.91 } }$. While the retrieval with PLATON prefers {TiO/VO} to explain the STIS observations, the retrieval with AURA does not, {and instead prefers a combination of Na and AlO. The retrieved TiO abundance is low and unconstrained} ($\log_{10}$(X$_{\text{TiO}})={ -9.58 ^{+ 1.37 }_{- 1.50 } }$). Neither the CO nor CO$_2$ abundances are constrained by the retrieval. While the cloud/haze parameters are not tightly constrained, our retrieval indicates a coverage fraction of $\bar{\phi}= { 0.25 ^{+ 0.26 }_{- 0.16 }  }$ consistent with a mostly clear atmosphere. The temperature profile of the atmosphere is mostly unconstrained. We infer the temperature near the photosphere, at 100mbar, to be $T= {1345 ^{+349}_{-206} }$~K. The posterior distributions for the relevant parameters are shown in Figure~\ref{fig:AURA_full_posterior}.

{We utilise this full retrieval as a reference model to perform a Bayesian analysis and assess the impact of not considering some of these parameters in the models. This change in model evidence is then converted to its more familiar frequentist counterpart, a detection significance (DS) following \citet{Benneke2013}. Table~\ref{table:AURA_models} shows the different models considered, their model evidence, DS, and $\bar{\chi}^{2}$. We find a robust detection of H$_2$O at a  { 4.89}-$\sigma$ confidence. There is suggestive evidence of Na and/or AlO with confidence levels of { 2.09-$\sigma$ and 2.58-$\sigma$}, respectively. The removal of TiO from the models results in an increase in the model evidence, indicating a disfavour for this molecule to be present in our models. VO is similarly undetected. However, removing opacity from the three primary metal oxides (TiO, VO, and AlO), finds a moderate-to-strong ``detection'', with 3.59-$\sigma$ confidence. This is similar to PLATON, which did not find evidence of TiO or VO individually, but found weak-to-moderate evidence of their combined presence (Sec.~\ref{subsec:platon_oxides}). This can be interpreted as follows: AURA is confident (to 3.6-$\sigma$) that the sharp dip in the blue STIS data (0.4--0.5\,$\mu$m) is a real molecular feature due to a metallic oxide. The retrieval finds that the most likely candidate for the metallic oxide is AlO, as shown by it's 2.6-$\sigma$ preference, whereas TiO and VO are individually dispreferred.}

We assess the retrieved H$_2$O abundance relative to expectations from thermochemical equilibrium for solar elemental compositions \citep{Asplund2009}. Assuming a solar composition and 50\% of the available oxygen in H$_2$O, the retrieved H$_2$O abundance corresponds to a log metallicity ([O/H]) of ${\log_{10}{Z/Z_{\odot}} = 1.72^{+0.39}_{-0.55}}$ (metallicity of ${53^{+82}_{-38}\times}$~Z$_{\odot}$). { We also compare the retrieved  H$_2$O abundance to the stellar metallicity of the host star ([O/H]=0.37, Table~\ref{tab:brewer}) and obtain a value of ${\log_{10}{Z/Z_{star}} = 1.35^{+0.39}_{-0.55}}$ (metallicity ${23^{+33}_{-17}\times}$~Z$_{star}$})

\begin{deluxetable*}{cc|ccccccccc}
\tabletypesize{\scriptsize}
\tablecaption{Retrieved models \label{table:AURA_models}}
\tablecolumns{11}

\tablewidth{0pt}
\tablehead{
\colhead{$\#$} & \colhead{Model} &  \colhead{$\log_{10}$(X$_{\text{H}_2\text{O}}$)} &  \colhead{$\log_{10}$(X$_{\text{Na}}$)}  & \colhead{$\log_{10}$(X$_{\text{AlO}}$)} & \colhead{$\log_{10}{Z/Z_{\odot}}$} & \colhead{STIS$_{\text{Shift}}$} & \colhead{WFC3$_{\text{Shift}}$}  & \colhead{$\ln$($\mathcal{Z}$)} & \colhead{$\bar{\chi}^2$} & \colhead{DS} 
}
\startdata
1   &   Full model   & ${-1.65 ^{+ 0.39 }_{- 0.55 }}$ & ${-3.09 ^{+ 1.03 }_{- 1.83 }}$   &  ${-6.44 ^{+ 0.66 }_{- 0.91 }}$  & ${1.72 ^{+ 0.39 }_{- 0.55 }}$ & N/A & N/A &  { 559.1} & {0.93} & Ref.\\
2   &   No H$_2$O & N/A & ${-2.41 ^{+ 0.99 }_{- 2.99 }}$& ${-5.71 ^{+ 0.99 }_{- 1.39 }}$ &   N/A & N/A & N/A & {548.9} & {1.37} & {4.89}\\
3   &   No Na    &   ${-1.62 ^{+ 0.42 }_{- 0.67 }}$ & N/A & ${-6.90 ^{+ 0.84 }_{- 1.05 }}$  & N/A & N/A & N/A & {558.0} & {0.95} & {2.09} \\
4   &   No AlO    &${-1.49 ^{+ 0.35 }_{- 0.70 }}$&${-4.32 ^{+ 1.88 }_{- 4.31 }}$ &   N/A   & N/A & N/A & N/A &  { 557.0} &  { 1.03} & { 2.58}\\
5   &   No TiO    &${-1.70 ^{+ 0.41 }_{- 0.56 }}$ &${-2.97 ^{+ 0.95 }_{- 1.25 }}$& ${-6.39 ^{+ 0.66 }_{- 0.88 }}$  & N/A & N/A & N/A &  { 559.7} & { 0.92}& N/A\\
6   &   No Metal Oxides   &${-1.52 ^{+ 0.38 }_{- 0.91 }}$ &${-3.59 ^{+ 1.28 }_{- 1.47 }}$&  N/A & N/A & N/A & N/A &  { 554.2} & { 1.21}& {3.59}\\
7   &   Simpler model   &  ${-1.65 ^{+ 0.40 }_{- 0.63 }}$ &${-2.60 ^{+ 0.94 }_{- 1.10 }}$&${-5.81 ^{+ 0.51 }_{- 0.66 }}$&  ${1.72 ^{+ 0.40 }_{- 0.63 }}$ & N/A & N/A &   { 560.0 }& { 0.89} & N/A\\[10pt]
8  &  \makecell{{Gaussian shifts}\\ {G430L, G750L, \& WFC3}}   & ${-1.91 ^{+ 0.53 }_{- 0.68 }}$ & ${-2.38 ^{+ 0.81 }_{- 1.33 }}$ & ${-6.64 ^{+ 0.70 }_{- 0.96 }}$ & ${1.46^{+0.53}_{-0.68}}$ & \makecell{{G430L} ${-51 ^{+60}_{-62}}$ \vspace{1mm} \\ {G750L} ${79^{+59}_{-56}}$} & ${-91^{+59}_{-56}}$ & { 562.0} & { 0.90}& { N/A}\\
9  &  \makecell{{Uniform shifts}\\ {G430L, G750L, \& WFC3}} & ${-2.96 ^{+ 0.98 }_{- 0.88 }}$ & ${-2.43 ^{+ 0.84 }_{- 1.34 }}$& ${-7.05 ^{+ 0.75 }_{- 0.94 }}$& ${0.40 ^{+ 0.98 }_{- 0.88 }}$ & \makecell{{G430L} ${1 ^{+144}_{-156}}$ \vspace{1mm}\\ {G750L}  ${175^{+150}_{-160}}$} & ${-189^{+90}_{-94}}$ &  { 561.8} & { 0.88}& { N/A}\\[10pt]
10  &  \makecell{{Uniform shifts}\\ {STIS \& WFC3}} &${-3.34 ^{+ 1.00 }_{- 0.86 }}$ & ${-3.43 ^{+ 1.35 }_{- 2.19 }}$ &${-6.98 ^{+ 0.77 }_{- 0.78 }}$&${0.03 ^{+ 1.00 }_{- 0.86 }}$ & ${89 ^{+166}_{-156}}$ & ${-203^{+97}_{-97} }$ & { 560.7} &{ 0.89}&{ N/A}\\[10pt]
\enddata
 \tablecomments{ {The metallicity is approximated from water abundance (see Section~\ref{subsec:AURA_main} for details)}. N/A means that the parameter was not considered in the retrieval or that it is not possible to estimate the detection significance (DS) as the model has a larger evidence than the reference model (model 1). Shifts are given in ppm. }
\end{deluxetable*}

\begin{figure}
\includegraphics[width=1.0\textwidth]{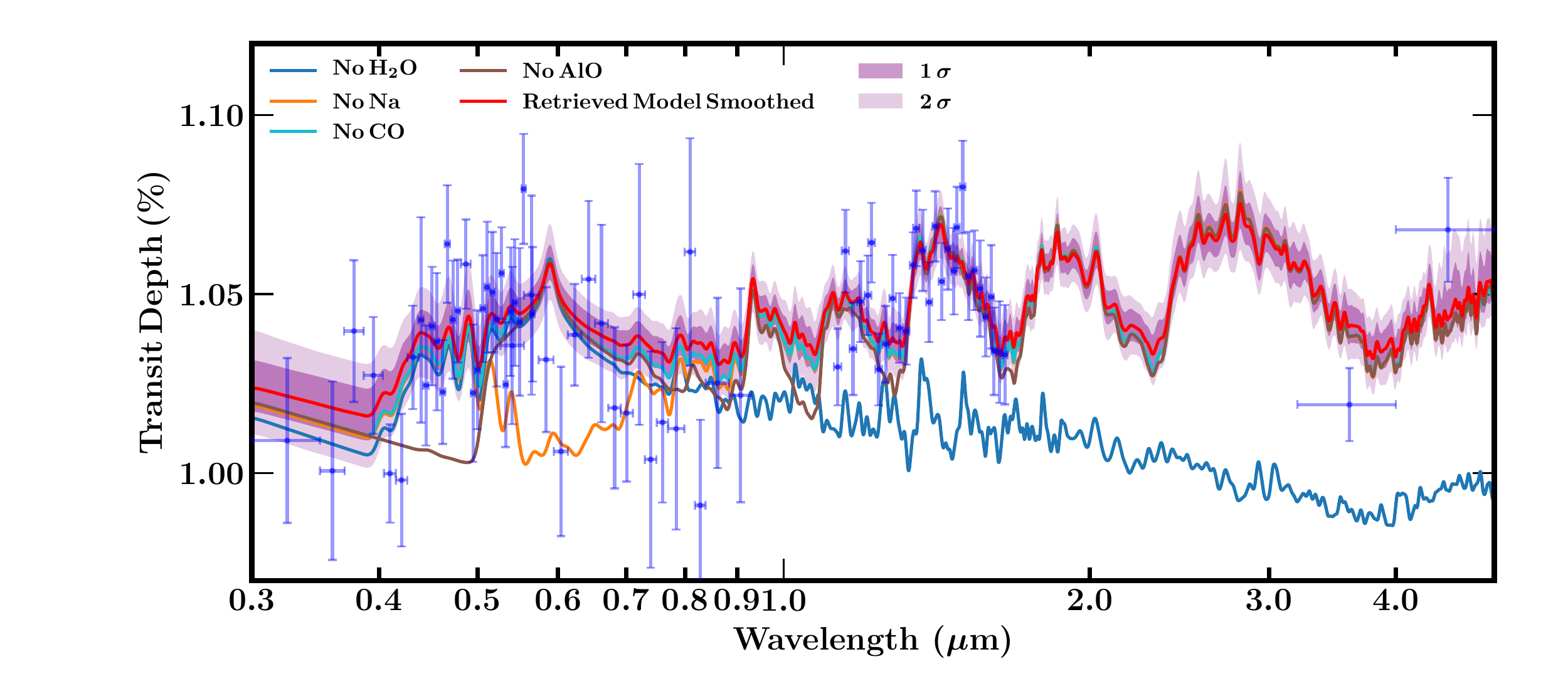}
\centering
  \caption{Retrieval of HAT-P-41b using a simplified model compared with the fiducial parameter set (see Section~\ref{subsec:AURA_main}). Observations are shown using blue markers. The retrieved median spectrum is shown in red while the 1-$\sigma$ and 2-$\sigma$ regions are shown using the shaded purple areas. Forward models using the retrieved { median} parameters show the contributions to the spectra due to individual chemical species. The forward models shown exclude absorption due to H$_2$O (blue), Na (orange), {CO} (cyan), and AlO (brown).}
\label{figure:AURA_contribution_plot}
\end{figure}

\begin{figure}
\includegraphics[width=1.0\textwidth]{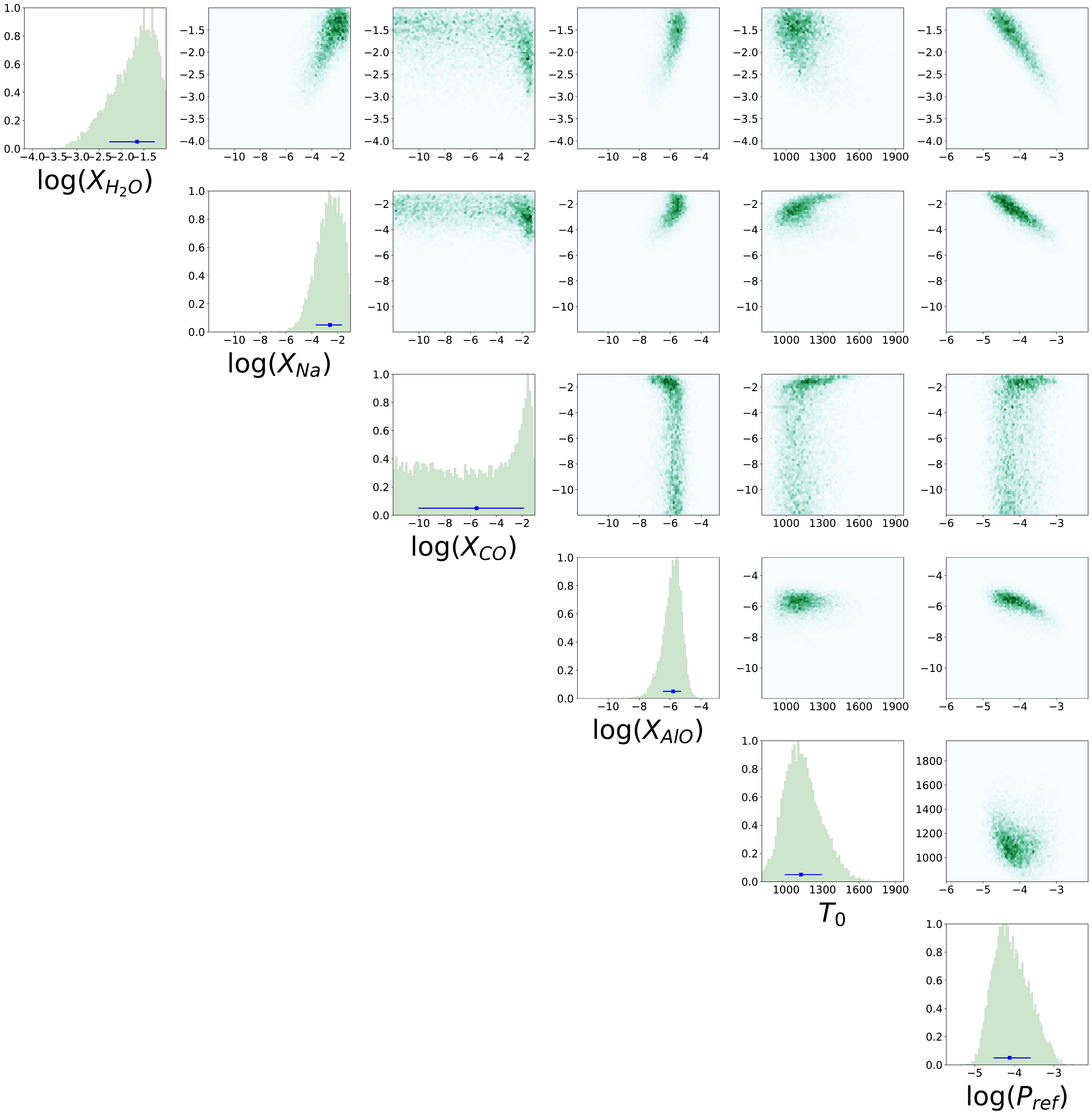}
\centering
  \caption{Full posterior distributions for the simpler model, Model 7 in Table~\ref{table:AURA_models}.}
\label{fig:AURA_simple_posterior}
\end{figure}

We consider the possibility of fitting the data using a simpler model consisting {mainly} of the parameters that are reasonably constrained by the full model. The simpler model considers the chemical abundances of H$_2$O, Na, {CO}, AlO, an isothermal {pressure-temperature profile}, and a clear atmosphere. The retrieved median fit and confidence contours are shown in Figure~\ref{figure:AURA_contribution_plot}. The simplified model retrieves values consistent with the full model. The retrieved values are $\log_{10}$(X$_{\text{H}_2\text{O}})={-1.65 ^{+ 0.40 }_{- 0.63 }}$, $\log_{10}$(X$_{\text{Na}})={-2.60 ^{+ 0.94 }_{- 1.10 } }$, $\log_{10}$(X$_{\text{AlO}})={ -5.81 ^{+ 0.51 }_{- 0.66 }}$, and ${\log_{10}{Z/Z_{\odot}} = 1.72^{+0.40}_{-0.63}}$ . The retrieved isothermal temperature is T$={ 1120 ^{+ 170 }_{- 140 } }$ and consistent with the inferred temperature at 100~mbar from the full retrieval. The posterior distribution for the retrieved parameters is shown in Figure~\ref{fig:AURA_simple_posterior}.

We use these retrieved parameters to generate a set of forward models to assess the spectroscopic contribution from each chemical species. Figure~\ref{figure:AURA_contribution_plot} shows that the WFC3 observations are better explained by the H$_2$O absorption feature at $\sim$1.4\,$\mu$m driving its strong detection in the spectrum of HAT-P-41b. On the other hand, a series of chemical species in the optical can provide some degree of fit to the STIS observations. In the optical, between $\sim$0.5--0.7\,$\mu$m, the broadened wings of Na along with its absorption peak provide a fit to observations. { AlO provides some fit to the substructure present in the STIS observations, particularly the increased transit depth between 0.4--0.5\,$\mu$m. Lastly the abundance of CO is not constrained and its contribution to the spectrum is minimal. CO is responsible for small changes in the optical and infrared that are well within the error bar of the observations.}

{ Our retrieval analysis of the broadband transmission spectrum of HAT-P-41b provides excellent fits to the data; using our fiducial model (Model 1) we obtain a best-fit $\bar{\chi}^2$ of 0.93 and $\ln$($\mathcal{Z})=559.1$. We note that we do not require additional continuum opacity sources (e.g., H$^-$) in our models in order to explain the data, as recently claimed by \citet{Lewis2020}.}



\subsection{Possible offsets in the data}
\label{subsec:AURA_shifts}

{ Lastly, we consider the presence of offsets in the data and their effect on the retrieved atmospheric properties. { We consider the three scenarios from Section~\ref{subsec:inst_offsets}. We note that these retrieved offsets are relative to the atmospheric model and that in all scenarios the \textit{Spitzer} observations remain unchanged.} We consider both Gaussian and uniform priors, as seen in Table~\ref{table:AURA_priors}.}


{We present the results of considering the presence of three offsets with Gaussian priors informed by the analysis of the white light transit curves (Model 8; Scenario 1 from Section~\ref{subsec:inst_offsets}). These priors are shown in Table~\ref{table:AURA_priors}. The retrieved shifts are ${-52 ^{+61}_{-63}}$\,ppm  for G430L, ${80 ^{+59}_{-56}}$\,ppm for G750L and ${-91 ^{+48}_{-50}}$\,ppm for WFC3. Similar to PLATON, the retrieval generally prefers to increase the G750L depths, primarily motivated by aligning the transit depths in the overlapping wavelength region between G430L and G750L. The retrieval also prefers to decrease WFC3 depths in order to better capture the \textit{Spitzer} 3.6\,$\mu$m depth. The retrieved abundances are $\log_{10}$(X$_{\text{H}_2\text{O}})= { -1.91 ^{+ 0.53 }_{- 0.68 } }$, $\log_{10}$(X$_{\text{Na}})= { -2.38 ^{+ 0.81 }_{- 1.33 } }$, and  $\log_{10}$(X$_{\text{AlO}})={ -6.64 ^{+ 0.70 }_{- 0.96 } }$. Although the retrieved H$_2$O abundance corresponds to a lower metallicity estimate, the derived range ${\log_{10}{Z/Z_{\odot}} = 1.46^{+0.53}_{-0.68}}$ is consistent with the fiducial model and describes a metal-rich atmosphere. The median metallicity is superstellar (${\log_{10}{Z/Z_{star}} = 1.09^{+0.40}_{-0.63}}$ ), though it is consistent with stellar metallicity to within 2-$\sigma$.}

Second, {we present the results for the case with three uniform shifts between HST-STIS G430L, HST-STIS G750L, and HST-WFC3 observations (Scenario 2 from Section~\ref{subsec:inst_offsets}). The retrieved G430L shift is consistent with 0 (${1 ^{+144}_{-156}}$\,ppm), but the retrieval prefers a large positive G750L offset (${176 ^{+151}_{-160}}$\,ppm) and a large negative WFC3 offset (${-189 ^{+91}_{-94}}$\,ppm). In addition to aligning the overlapping G430L-G750L region, it is possible that this large G750L shift is due to the model forcing the data to match features it finds easier to explain. This uncertainty is the danger in using uniform prior offsets, especially in an already-flexible free-chemistry retrieval.} The retrieved abundances are shown in Table~\ref{table:AURA_models} as Model {9} and are $\log_{10}$(X$_{\text{H}_2\text{O}})= { -2.96 ^{+ 0.98 }_{- 0.88 } }$, $\log_{10}$(X$_{\text{Na}})= { -2.43 ^{+ 0.84 }_{- 1.34 } }$, and  $\log_{10}$(X$_{\text{AlO}})={ -7.05 ^{+ 0.75 }_{- 0.94 } }$. While the retrieved abundances { for these three species are consistent within 1-$\sigma$ with the full unshifted model, the retrieved H$_2$O abundance corresponds to a lower metallicity estimate consistent with solar, sub-solar, and sub-stellar values ${\log_{10}{Z/Z_{\odot}} = 0.40^{+0.98}_{-0.88}}$} .

{ Lastly, we present the results accounting for offsets} in the STIS and WFC3 observations using a uniform prior, while keeping the \textit{Spitzer} observations unshifted ({Scenario 3 from Section~\ref{subsec:inst_offsets})}. The retrieval results in a shift in the STIS data of ${90 ^{+ 167}_{-157}}$\,ppm and a shift in the WFC3 data of ${ -204 ^{+ 97 }_{- 98 }}$\,ppm. { While the retrieved value for the STIS observations is consistent with no shift, the WFC3 observations preferentially retrieve a negative offset}. The derived abundances, shown as Model {10} in Table~\ref{table:AURA_models}, are $\log_{10}$(X$_{\text{H}_2\text{O}})={ -3.34 ^{+ 1.00 }_{- 0.86 } }$, $\log_{10}$(X$_{\text{Na}})={  -3.43 ^{+ 1.35 }_{- 2.19 } }$,   $\log_{10}$(X$_{\text{AlO}})={ -6.98 ^{+ 0.77 }_{- 0.78 } }$. {The H$_2$O abundance, like Model 9, corresponds to a metallicity consistent with solar and sub-solar values: ${\log_{10}{Z/Z_{\odot}} = 0.03^{+1.00}_{-0.86}}$.}

\begin{figure}
\includegraphics[width=0.82\textwidth, keepaspectratio]{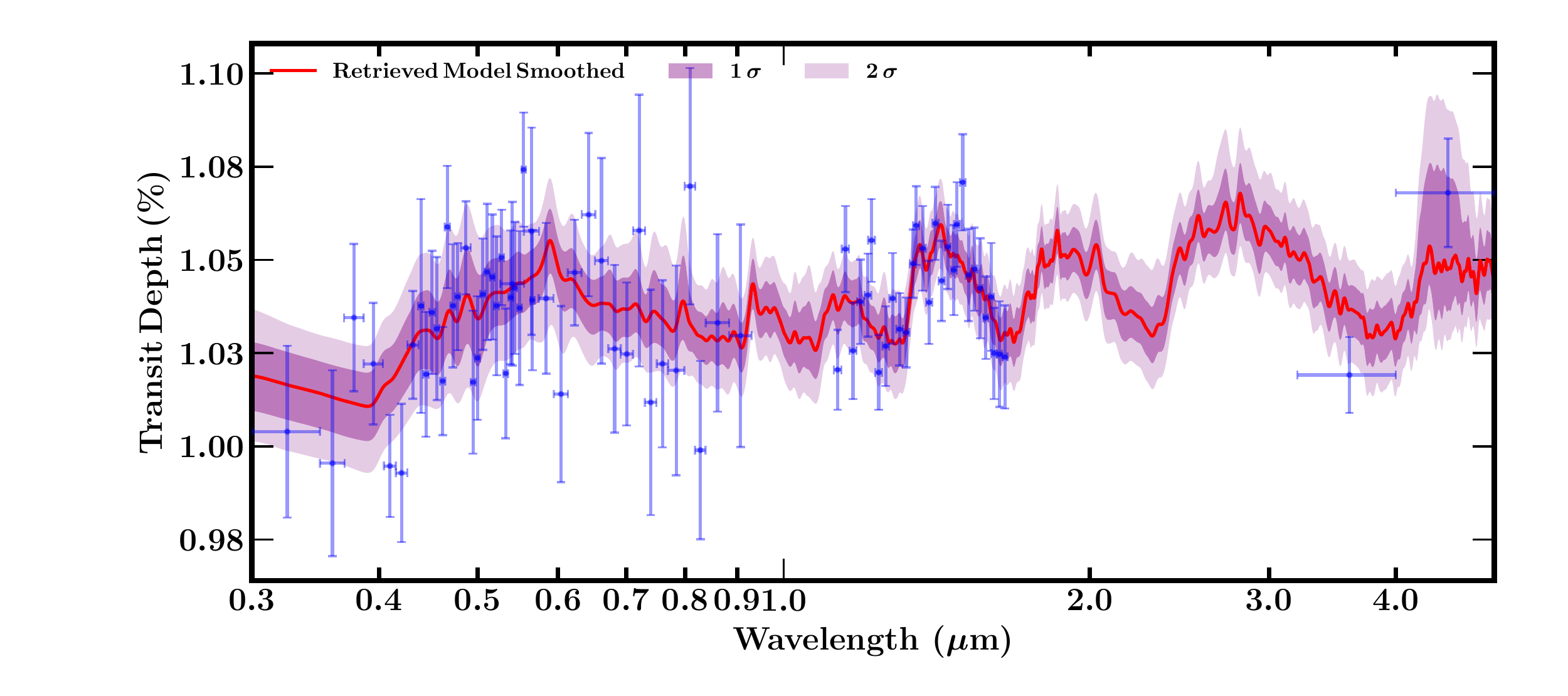}
\includegraphics[width=0.82\textwidth,keepaspectratio]{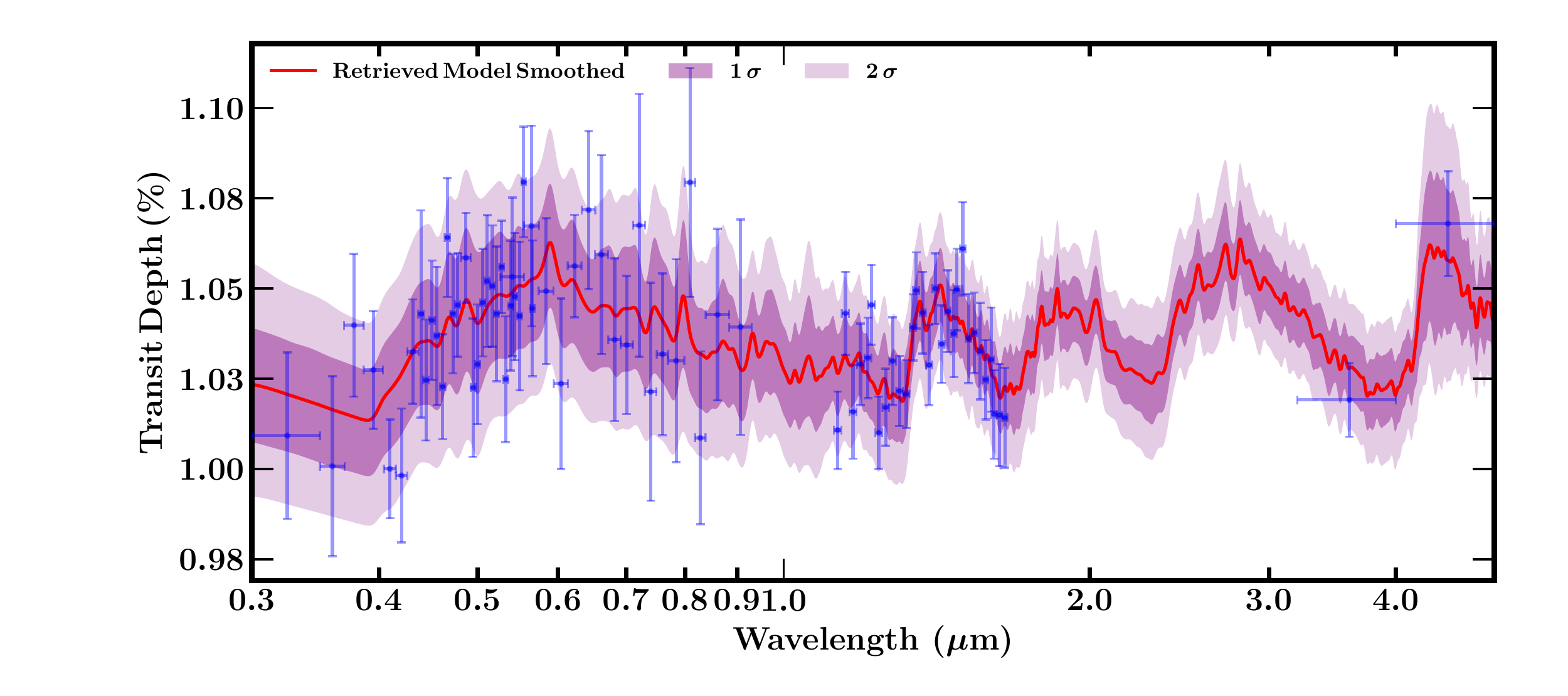}
\includegraphics[width=0.82\textwidth,keepaspectratio]{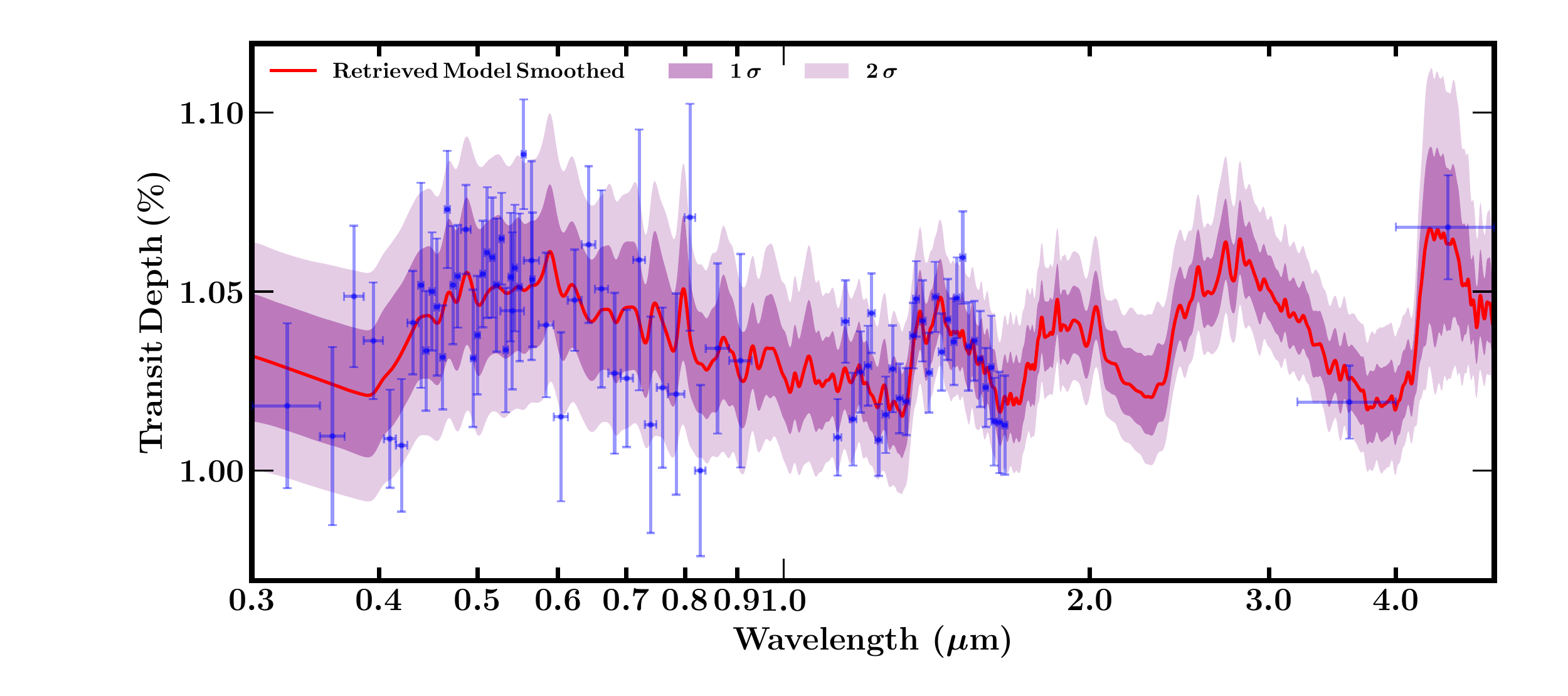}
\centering
  \caption{Retrieved spectrum of HAT-P-41b allowing for offsets in the STIS and WFC3 data sets. Observations are shown using blue markers and are shifted according to the models' retrieved median shifts. The retrieved median spectrum is shown in red while the 1-$\sigma$ and 2-$\sigma$ regions are shown using the shaded purple areas. \textit{Top:} Three shifts with Gaussian priors (Model 8) and retrieved median offsets of ${\sim -50}$\,ppm for STIS G430L,${\sim 80}$\,ppm for G750L, and ${\sim-90}$ for WFC3. \textit{Middle:} Three shifts with uniform priors (Model 9) and retrieved median offsets of ${\sim 0}$\,ppm for STIS G430L,${\sim 180}$\,ppm for G750L, and ${\sim-190}$ for WFC3.  \textit{Bottom:} {Two shifts with uniform priors (Model 10) and retrieved median offsets of ${\sim90}$\,ppm for STIS and ${\sim-200}$\,ppm for WFC3.
  }}
\label{fig:AURA_shifts}
\end{figure}

Figure~\ref{fig:AURA_shifts} shows the retrieved median models and confidence contours along with their respectively shifted observations for {the cases described in this Section (Models 8, 9, and 10)}.

{The models considering instrumental shifts are all prefered over the fiducial model at above the 2-$\sigma$ level. The model with Gaussian priors has a preference at the 2.9-$\sigma$ level, followed by the model with three uniform shifts at a 2.8-$\sigma$ level. The model with two uniform shifts is preferred over the fiducial model at 2.3-$\sigma$. We note that while both models with three offsets are similarly preferred over our fiducial model, the associated metallicity ranges are different. The model with three uniform shifts retrieves an H$_2$O abundance corresponding to a metallicity estimate consistent with substellar and stellar values. On the other hand, the model with Gaussian priors retrieves an associated metallicity range mostly superstellar and in agreement with the fiducial model. These results highlight the sensitivity of the inferred metallicity ranges to possible large offsets between instruments. Model comparisons suggest a preference for the models considering offsets, though it is inconclusive between these models. We favor the more physically plausible Gaussian prior model (i.e., Model 8) as the reference for our discussion (Section~\ref{sec:discussion}).}

\section{Discussion}\label{sec:discussion}

\subsection{Comparison Between Retrieval Methods}\label{subsec:compare}
\subsubsection{Results Comparison}
In this Section, we compare the results from the preferred PLATON and AURA models. These include the fiducial model with partial clouds and Gaussian instrumental offsets for PLATON (Section~\ref{subsec:offsets}) and the Gaussian instrumental offset model for AURA (Model 8; Section~\ref{subsec:AURA_shifts}).

The similarities reveal the most robust conclusions of our analysis, since they are retrieved despite the many different assumptions that went into each method. Notably, both retreivals robustly find a metal-rich atmosphere with metallicity (defined as  O/H) inconsistent with the solar metallicity at  $>$2-$\sigma$. Both methods find a decisive ($>$4.8-$\sigma$) water vapor detection, and at least a moderate detection ($>$2.7-$\sigma$) of a non-haze gas absorption feature in the optical. Further, both PLATON and AURA retrievals are consistent with a mostly clear atmosphere, with neither finding strong evidence of haze or uniform, high-altitude grey clouds.

Though the atmospheric properties derived from PLATON and AURA are similar, there are noteworthy differences. AURA infers a cooler limb temperature at 100~mbar ($1320^{+270}_{-200}\,$K compared to ${1710^{+100}_{-80}}\,$K for PLATON) as well as a lower metallicity of ${\log_{10}{Z/Z_{\odot}} = 1.46^{+0.53}_{-0.68}}$ compared to ${\log_{10}{Z/Z_{\odot}} = 2.33^{+0.23}_{-0.25}}$ for PLATON, a difference of 1.3-$\sigma$. This translates to ${29^{+69}_{-23}\times\,}$Z$_{\odot}$ for AURA and ${214^{+149}_{-88}\times\,}$Z$_{\odot}$. The optical absorber also differs: AURA determines the best description of the STIS feature to be absorption from sodium and AlO, whereas PLATON prefers some combination of TiO and VO absorption. Finally, AURA makes no claim on the C/O ratio { as it is a free retrieval framework and no C-bearing species are detected nor meaningfully constrained. On the other hand,} the chemical equilibrium assumption allows PLATON to find a 3-$\sigma$ upper limit on the C/O ratio of C/O$<0.83$.

\begin{figure}[t]
\centering
\includegraphics[width=0.329\textwidth, keepaspectratio]{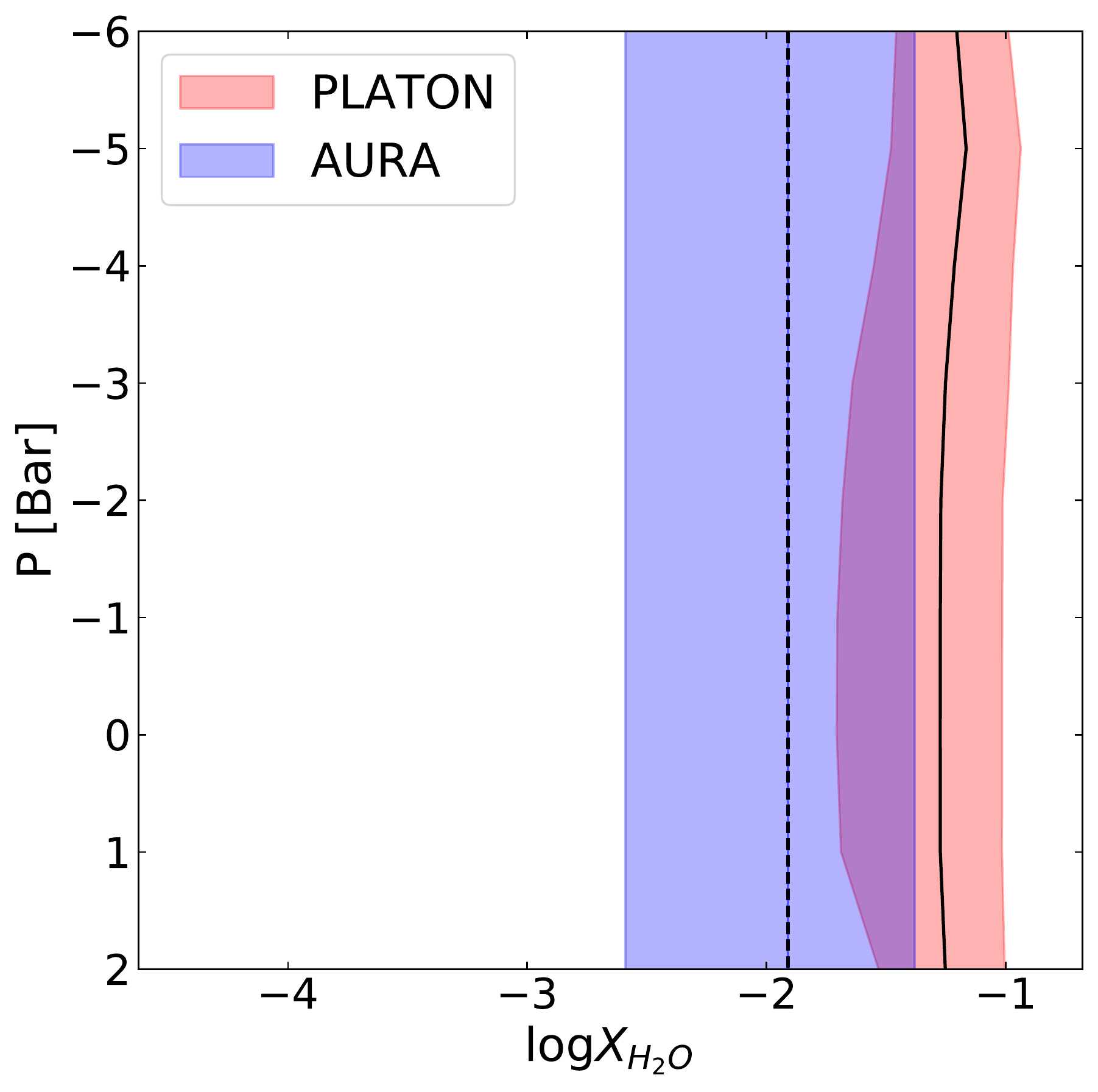}
\includegraphics[width=0.329\textwidth, keepaspectratio]{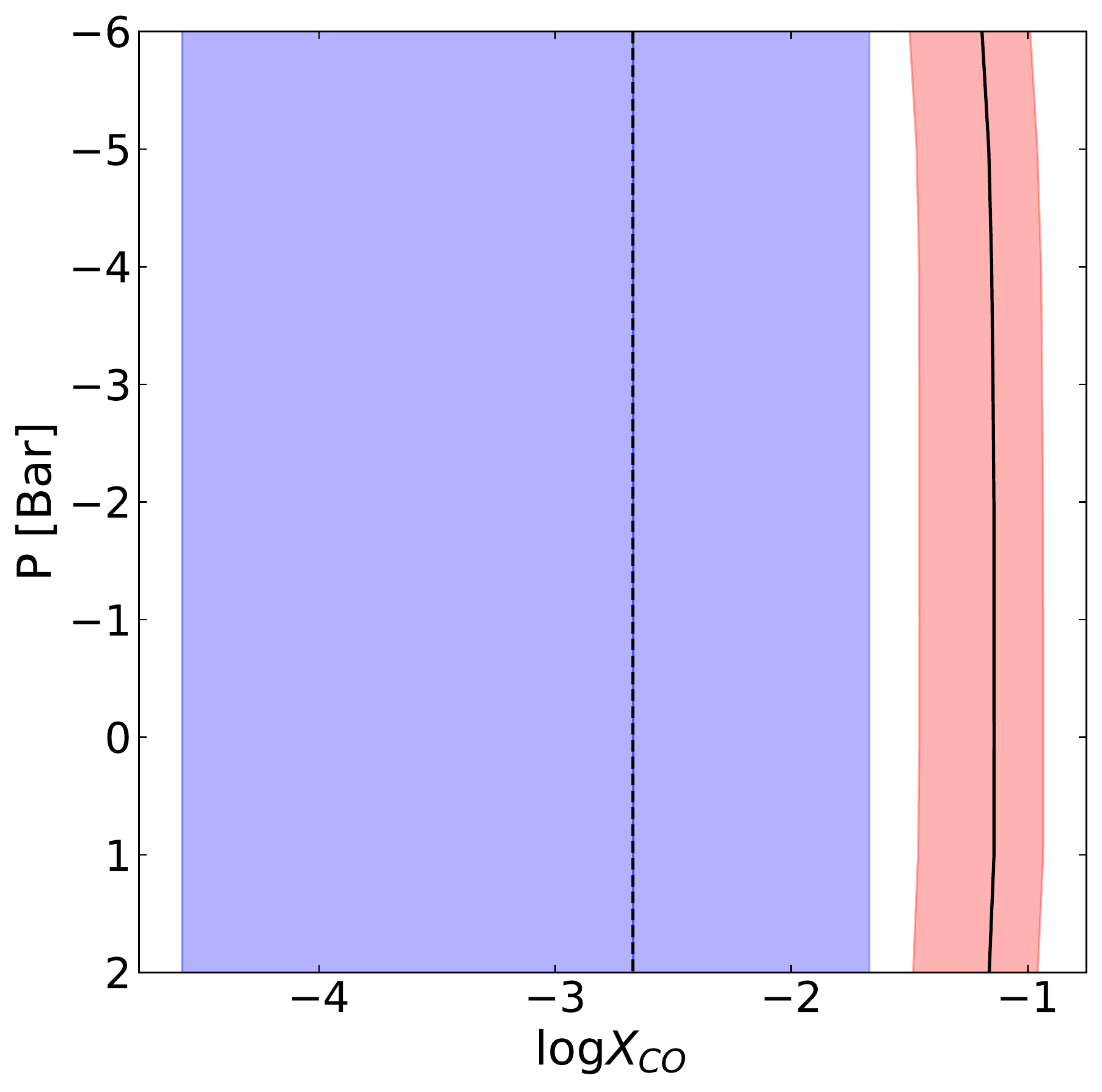}
\includegraphics[width=0.329\textwidth, keepaspectratio]{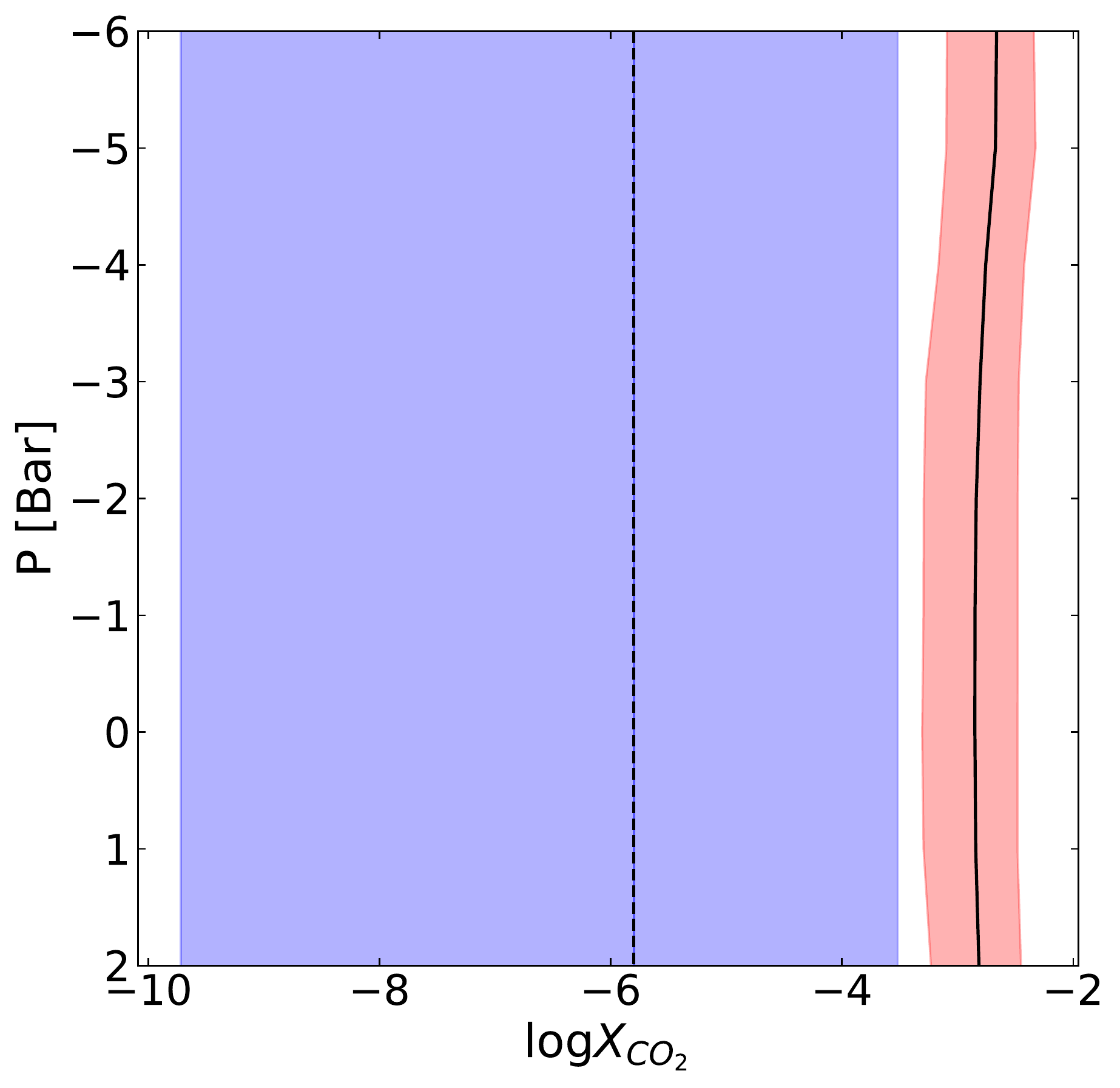}
\includegraphics[width=0.329\textwidth, keepaspectratio]{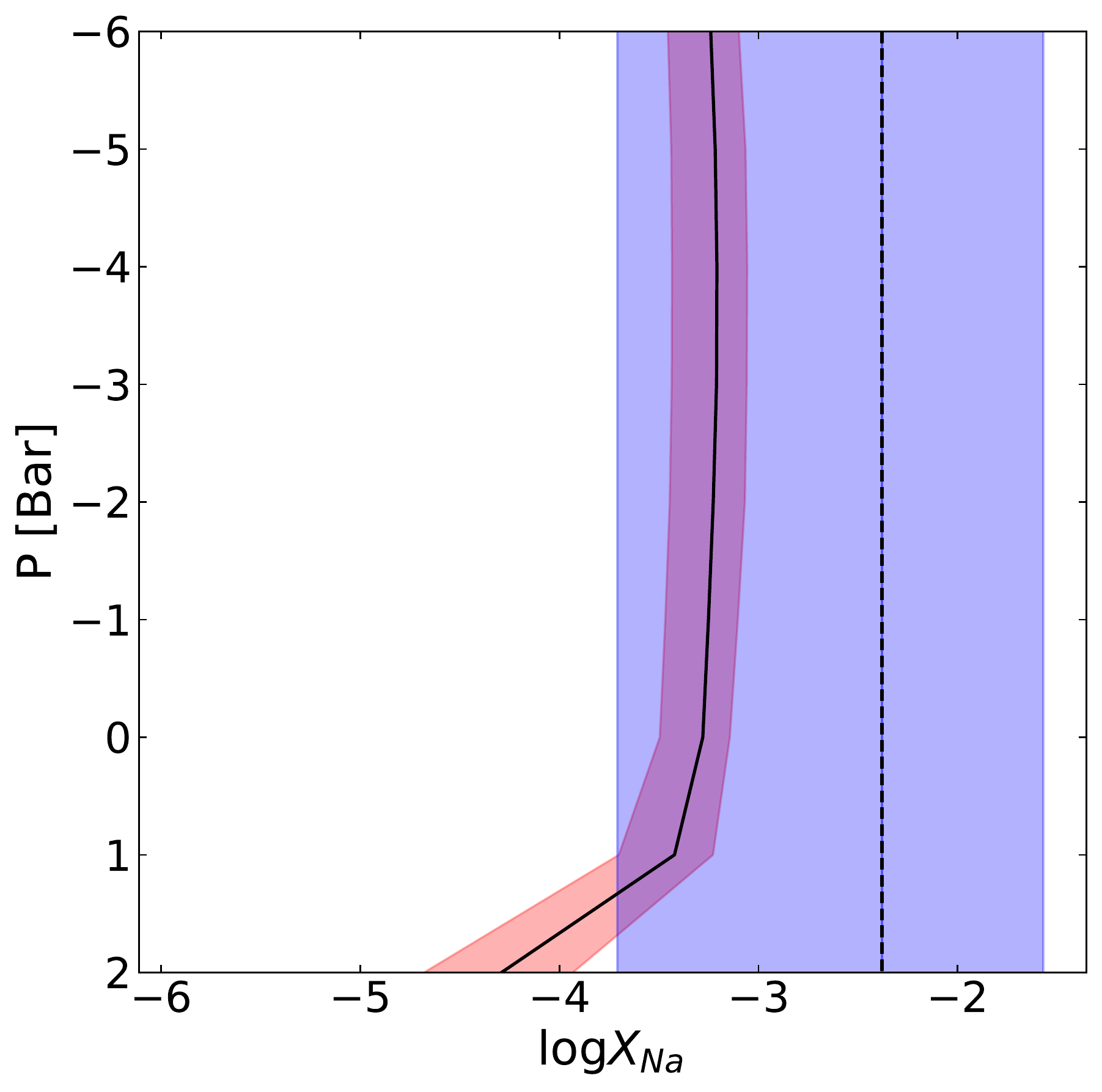}
\includegraphics[width=0.329\textwidth, keepaspectratio]{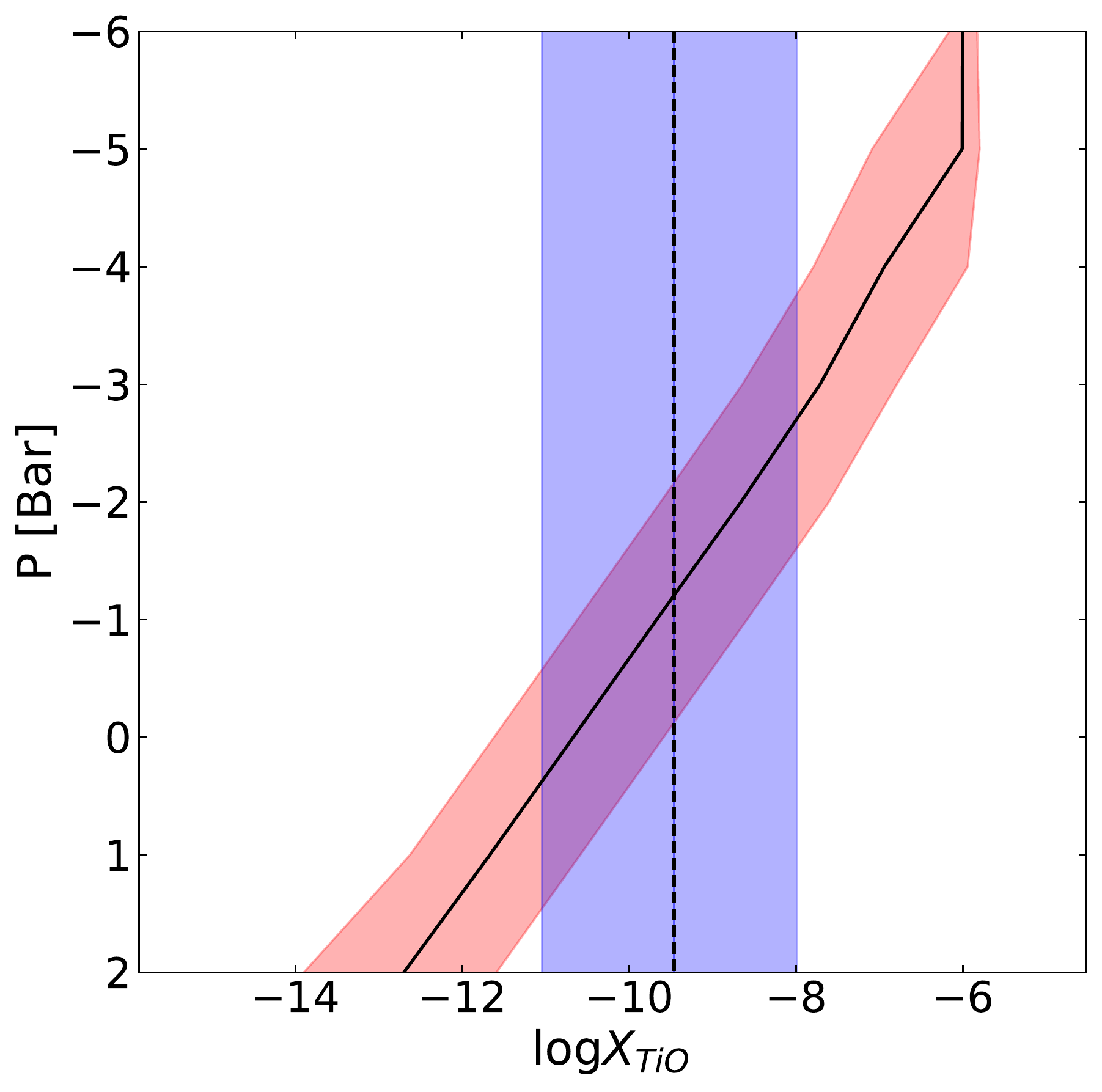}
\includegraphics[width=0.329\textwidth, keepaspectratio]{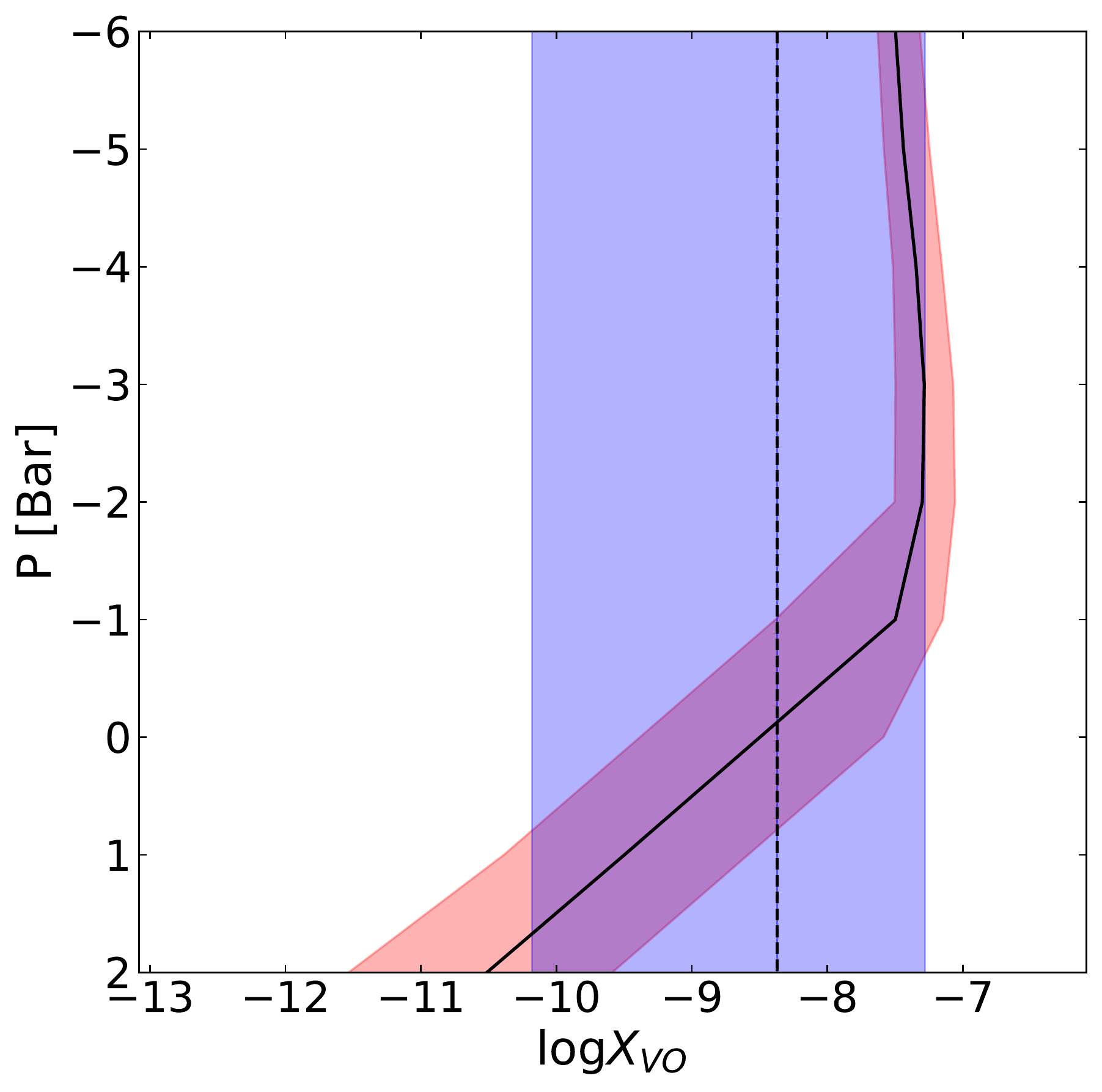}
\caption{Abundance profiles for PLATON (red) and AURA (blue) for four relevant gaseous species. PLATON abundance profile distributions are derived by sampling the posterior 200 times, calculating the abundance profiles for each species for each sample, and finding the median value (solid black line) with 1-$\sigma$ uncertainties at each pressure layer. AURA assumes abundances to be constant with pressure. The median retrieved value (dashed black line) and 1-$\sigma$ uncertainty range are shown.}
\label{fig:abundances}

\end{figure}

To further contextualize the results, we added the functionality to retrieve the abundance profiles of relevant molecules in PLATON. We show abundance profiles for {six} spectroscopically relevant species from AURA which are also included in PLATON --- H$_2$O, {CO, CO$_2$}, Na, TiO, and VO --- in Figure~\ref{fig:abundances}. {We emphasize the enforcing chemical equilibrium narrows the abundance constraints, and we are not reporting these abundances. Instead, they should be interepreted as the expected abundance profiles under the conditions of stable chemical equilibirum for the reported temperature, metallicity, and C/O ratio. As an example, we find no observational constraint on CO, but its abundance is well defined under chemical equilibrium for the temperatures and metallicities that we do observationally constrain via the water feature. Still,} these profiles provide a useful baseline for comparison to free-chemistry retrieval abundances.  

The abundance profiles for the optical-wavelength absorbers reflect the disagreement on the primary gas absorber: AURA prefers Na, and so it retrieves $\sim10\times$ more Na than PLATON and significantly less TiO and VO. {Note that the decreasing TiO and VO abundance with increasing pressure for PLATON is due to those molecules condensing out of the atmosphere.} PLATON's {inferred} water abundance is typically a few times greater than AURA's, reflecting the difference in inferred metallicities. {CO$_2$ and CO are unconstrained by AURA, while PLATON finds a high abundance of CO$_2$ is consistent with the \textit{Spitzer} observations. This difference is expected, given that PLATON finds weak evidence of CO$_2$ while AURA found none. Interestingly, this may relate to AURA's only chemical constraint, which is that CO$2$ must be less abundant than CO and H$_2$O due to the inferred temperatures (Section~\ref{subsec:AURA}).}

In total, AURA finds a cooler atmosphere with less oxygen but a large sodium enrichment to explain the optical absorption, while PLATON finds a hotter, higher oxygen{-abundance} atmosphere with TiO/VO absorption in the optical. 

\subsubsection{Impact of Retrieval Model Assumptions}\label{subsec:assume}


The differences between a free{-chemistry} retrieval (AURA) and one constrained by chemical equilibrium (PLATON) are the natural result of the different assumptions made by each method. We therefore consider the PLATON and AURA retrievals to be two orthogonal analyses.  We examine the impact of the differences by first explicitly listing the notable assumptions in each method, and then by providing the rationale for which assumptions are driving the differences. 

The relevant methodological differences for PLATON as compared to AURA are 1) the assumption of chemical equilibrium, 2) fixing the elemental ratios between all metals other than carbon to their solar values, 3) assuming an isothermal profile for the atmosphere, 4) not including opacity from AlO, and 5) not including the \citet{Allard2019} H$_2$-broadened Na line profile. 

We find that the \citet{Allard2019} H$_2$-broadened Na line profile is the key driver in the differences between the retrievals, and flexible element abundances and chemical equilibrium also play roles. AURA is the more flexible retrieval, so we first describe its solution before addressing why PLATON differs.

AURA's lower temperature solution is preferred for being able to explain the H$_2$O feature in the WFC3 spectrum, while also explaining the STIS data with H$_2$-broadened Na absorption and capturing the \textit{Spitzer} data. AURA is able to provide a fit to the STIS data by {independently} increasing the Na abundance and by also invoking AlO at relatively low temperatures. At this lower temperature ($T\sim1300\,$K), the amount of oxygen necessary for the water abundance and scale height to explain the observed water feature is about $29\times\,$Z$_{\odot}$, with a mean molecular weight of about 2.7~AMU and a scale height of about 440~km.

Since PLATON has not yet incorporated the H$_2$-broadened Na line profile, the low temperature solution is a relatively poor fit to the STIS data. Instead, TiO/VO are needed to explain the STIS absorption feature, and these are only abundant enough in chemical equilibirum (with fixed metal ratios) at around 1650~K. At this higher temperature, a higher mean molecular weight is required for the same scale height, which must be small enough to explain the molecular feature sizes as well as the dominance of TiO/VO absorption over Rayleigh scattering. The atmospheric metallicity necessary to achieve the higher mean molecular weight is the much higher $\sim200\times$~Z$_{\odot}$. Therefore, the differences make sense in light of the stricter assumptions.

To provide more support to this idea, we compare results with a those of a third retrieval method, ATMO \citep{amundsen2014,tremblin2015,tremblin2016,tremblin2017,sing2016}, which acts as a middle ground between PLATON and AURA. ATMO's spectral retrievals can further help to gain insight into the effect of retrieval assumptions as it includes the \citet{allard2007} pressure-broadened sodium line but also has the added flexibilty of performing a free-element equilibrium-chemistry retrieval. With this assumption for the chemistry, the elemental abundances for each model are freely fit and calculated in equilbirum on the fly. Four elements were selected to vary independently, as they are major species which are also likely to be sensitive to spectral features in the data, while the rest were varied by a trace metallicity parameter ([Z$_{\rm trace}$/Z$_{\odot}$]). By separately varying the carbon, oxygen, sodium and vanadium elemental abundances ([C/C\textsubscript{$\odot$}], [O/O\textsubscript{$\odot$}], [Na/Na\textsubscript{$\odot$}], [V/V\textsubscript{$\odot$}]) we allow for non-solar compositions but with chemical equilibrium imposed such that each model fit has a chemically-plausible mix of molecules given the retrieved temperatures, pressures and underlying elemental abundances.

The resulting retrieved atmospheric parameters describe an atmosphere most consistent with the one described by AURA. ATMO prefers a temperature of 1190$^{+170}_{-120}\,$K and a metallicity (as defined by the oxygen abundance) of $\log_{10}{O/O{\odot}} = \log_{10}{Z/Z_{\odot}} = 1.53^{+0.55}_{-0.67}$, in excellent agreement with AURA's values, and consistent with PLATON's metallicity to 1.3-$\sigma$, though the retrieved temperatures differ significantly. Like AURA, ATMO finds an enhanced sodium abundance, though uncertainties are large ($\log_{10}{Na/Na{\odot}} = 1.40^{+0.75}_{-1.80}$). This supports the idea that the inclusion of H$_2$-broadened sodium line profiles and the flexibility of non-solar metal ratios --- and not necessarily the equilibirum chemistry constraint --- allow for the low-temperature, lower oxygen abundance solution found by AURA. The metallicities on all three retrievals indicate a metal-rich atmosphere and agree at the $\sim$1.3-$\sigma$ level.

Like PLATON (Section~\ref{sec:results}), ATMO also finds a subsolar C/O ratio (C/O = $0.17^{+0.53}_{-0.16}$ consistent with stellar (C/O = 0.19), though carbon is not well constrained so the uncertainties are large. The 3-$\sigma$ upper limit of 0.94 is in good agreement with PLATON's 0.83 upper limit. However, unlike PLATON or AURA, ATMO finds no evidence of optical absorbers beyond Na, and instead prefers a haze and Na to explain the STIS optical data.

\begin{figure}[t]
\centering
{
\includegraphics[width=\textwidth,keepaspectratio]{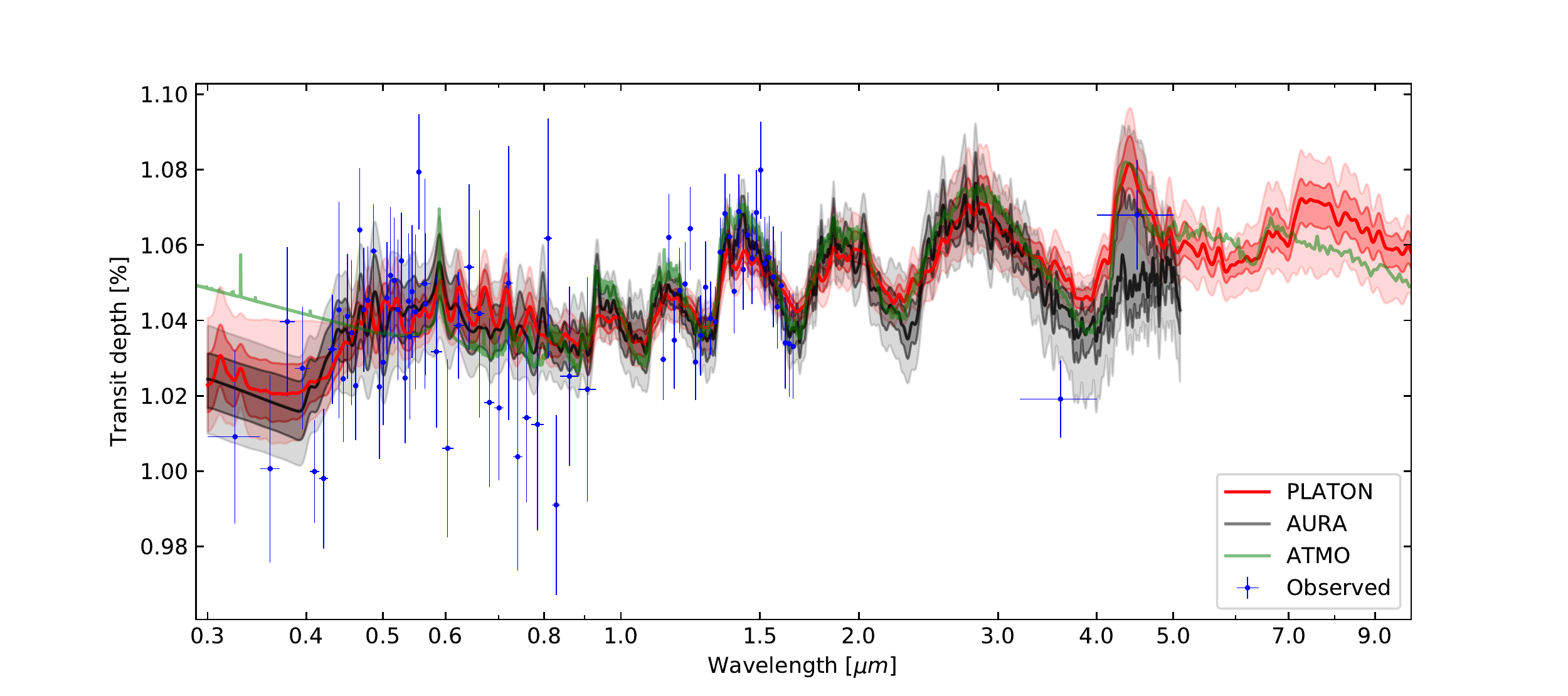}
\caption{Comparison of the median retrieved model for each retrieval method's fiducial model. 1-$\sigma$ and 2-$\sigma$ uncertainty contours are inlcluded for PLATON and AURA. PLATON and AURA are smoothed with a Gaussian filter with $\sigma=15$ for clarity. The chemical equilibrium assumption used by PLATON and ATMO allows for meaningful predictions at unobserved wavelengths, and so those models are shown out to 10\,$\mu$m.}
\label{fig:retrieval_comp}
}
\end{figure} 

Figure~\ref{fig:retrieval_comp} elucidates the differences in retrievals by showing the median retrieved fiducial model for PLATON (red), AURA (black), and ATMO (green) from 0.3--10\,$\mu$m. The 1- and 2-$\sigma$ uncertainty contours are shown for PLATON and AURA, both of which are smoothed with a Gaussian filter with $\sigma=15$ for clarity. The AURA predictions are only shown up to 5\,$\mu$m - as a free-chemistry retrieval, AURA retrieving on the 0.3--5\,$\mu$m data does not place meaningful constraints on multiple molecules with significant opacity in the 5--10\,$\mu$m range. Therefore, a prediction is not warranted.

While there are subtle differences, such as PLATON and ATMO's preference for CO$_2$ at 4.5\,$\mu$m and sodium's prominence at 0.6\,$\mu$m in the AURA and ATMO retrievals, the most obvious difference is below ~0.5\,$\mu$m, where ATMO prefers a haze instead of a metallic oxide feature. Though ATMO does not include AlO as an opacity source, this difference is likely due to different condensation schemes. PLATON uses GGchem's prescription \citep{woitke2018} such that species condense out when it is energetically favorable. AURA is a free-chemistry retrieval, so there are no restrictions on oxides being in the gaseous phase. ATMO, however, includes rainout chemistry \citep{rain2019}, such that if a species condenses at a higher pressure, that then depletes the element above that layer. It is plausible that although PLATON's condensation scheme allows TiO/VO to be in the gas phase around 1700K, ATMO's scheme does not, making the metallic oxide feature difficult to capture. 

In total, we tentatively favor AURA's derived atmospheric parameters over PLATON's, for two main reasons. First, the inclusion of the most up-to-date sodium line profiles and AlO opacity impact the retrieval. Second, constraints from interior modeling (Section~\ref{subsec:interior}), though not necessarily decisive, are consistent with AURA and in tension with PLATON. Overall, this paints a picture of an atmosphere with a supersolar --- but not necessarily superstellar --- metallicity, sodium enrichment, possible disequilibrium metallic oxides (e.g., circulated from dayside, dredged up due to vertical mixing), and a planet with a well-mixed interior and a limb temperature lower than the equilibrium temperature. 



\subsection{Comparison to Interior Modeling Metallicity Constraints}\label{subsec:interior}

Though they both describe metal-rich atmospheres, the 1-$\sigma$ retrieved atmospheric metallicities ranges from AURA and PLATON are inconsistent ($\log_{10}{Z/Z_{\odot}} = 0.78$--1.99 and $\log_{10}{Z/Z_{\odot}} = 2.08$--2.56, respectively). Further, it is questionable whether such supersolar metallicities --- especially those retrieved by PLATON -- are physically reasonable. We check the viabilitiy of these values by comparing them to atmospheric metallicity constraints from interior structure models.

\citet{thorngren2019} demonstrated how interior models can constrain atmospheric metallicity. Essentially, this is a three step process: 1) Determine what range of bulk metallicities are necessary for structure models to explain the observed radius, taking into account the planet's mass, age, heating efficiency, and parameter uncertainties, 2) set the maximum bulk metallicity to be the 3-$\sigma$ upper limit of the derived posterior distribution, and 3) set the maximum atmospheric metallicity to be equal to the maximum bulk metallicity.

The third step assumes that the atmospheric metallicity cannot be greater than the core's metallicity for significant timescales due to convection or Rayleigh-Taylor instability. They define metallicity as the ratio of all metals to hydrogen compared to the ratio in the Sun's photosphere. This is a good proxy for O/H, and so it is a valid comparison to the retrieved atmospheric metallicities. For more details on the derivation, see \citet{thorngren2019}. 

Using stellar parameters from \citet{hartman2012} (Table~\ref{tab:sysparams}), the interior structure model fit yields a bulk metal abundance ratio of Z/Z$_{\odot}=33.7\pm{9.1}$, corresponding to a maximum atmospheric metallicity of 50$\times\,$Z$_{\odot}$ (D. Thorngren, private communication). There is no significant uncertainty on this number, as it is the 3-$\sigma$ upper limit of the distribution. This is consistent with the metallicity from the AURA retrieval, but it is in tension with PLATON's retrieved metallicity --- 50$\times\,$Z$_{\odot}$ falls outside PLATON's 1-$\sigma$ range (but within 2-$\sigma$, as the metallicity distribution is asymmetric PLATON's 2-$\sigma$ lower limit is 37$\times$Z$_{\odot}$). This could indicate that the ``true'' atmospheric metallicity falls in the lower range of PLATON's retrieved metallicity, or it could be interpreted as slight evidence in support of AURA over PLATON. Either way, the atmospheric metallicity approaching the bulk metallicity indicates a well-mixed interior. Such vertical mixing could allow for micron-sized particles to stay afloat in the atmosphere, potentially facilitating gaseous metal oxide survival and Mie scattering.

\subsection{Implications for Planet Formation}\label{sec:form}

The atmospheric metallicity we retrieve for HAT-P-41b provides important constraints on the formation and migration history of the planet. At the outset, the super-solar metallicity (O/H) of $\sim$30--200$\times$~Z$_{\odot}$ requires substantial accretion of solids, beyond several Earth masses of H$_2$O ice, during the planet's evolutionary history. It is unlikely that such a large amount of volatile accretion is possible at the planet's current orbit. Therefore, the planet is unlikely to have formed in-situ \citep{Batygin2016} but instead formed far out beyond the H$_2$O snow line and migrated inward. The formation location and migration path of a giant planet can significantly affect its chemical composition. Beyond the H$_2$O snow line the gas in the protoplanetary disk is depleted in oxygen whereas the solids are enriched in oxygen \citep{Oberg2011}. Therefore, planets with high enrichment of oxygen require predominant accretion of H$_2$O-rich planetesimals while forming and migrating through the protoplanetary disk. 

The high metallicity (specifically O/H) of HAT-P-41b, therefore, supports the migration of the planet through the disk via viscous torques \citep{Madhusudhan2014b}. This is in contrast to other hot Jupiters with low O/H abundances which have been suggested to be caused by insufficient solid accretion, e.g. via disk-free migration \citep{Madhusudhan2014b} or formation via pebble accretions whereby the oxygen-rich solids are locked in the core \citep{Madhusudhan2017}. The fact that HAT-P-41b's orbit is moderately misaligned to the host star's rotation axis is also in tension with the disk migration hypothesis, since spin-orbit misalignments are considered to be evidence of disk-free migration and planet-planet interactions \citep{winn2010}. In principle, instead of disk migration, super-solar elemental abundances could be caused by accreting gas whose metallicity has been enhanced due to pebble drift \citep{Oberg2016,Booth2017}. But while pebble drift can cause metal enhancements up to $\sim10\times$~Z$_{\odot}$, much larger enhancements as constrained in the present case are unlikely to be explained by this process. More importantly, such enhancements due to pebble drift are also expected to cause high C/O ratios ($\sim$1), which may be at odds with the high H$_2$O abundace and the low C/O ratio retrieved for the planet.

Overall, the most plausible explanation for the potentially high atmospheric metallicity inferred for HAT-P-41b is formation outside the H$_2$O snowline and migration inward while accreting substantial mass in planetesimals. If confirmed, this would be a departure from other hot Jupiters observed hitherto which have generally shown low H$_2$O abundances, indicative of the low accretion efficiency of H$_2$O-rich ices that is possible for disk-free migration mechanisms \citep{Madhusudhan2014b,pinhas2019,Welbanks2019b}. Such an abundance is also a substantial departure from expectations based on Solar System giant planets. The metallicity of Jupiter in multiple elements is $\sim$1--5$\times$~Z$_{\odot}$ \cite{Atreya2016, Li2020}. With the mass of HAT-P-41~b being similar to that of Jupiter, its higher metallicity would indicate an even higher amount of solids accreted than that of Jupiter in the Solar System. 

\section{Summary}\label{sec:summary}

We have conducted a comprehensive, multi-pronged Bayesian retrieval analysis of the 0.3--5\,$\mu$m transit spectrum of HAT-P-41b derived from HST STIS (previously unpublished; Section~\ref{sec:stis}), HST WFC3 (re-analysis; Section~\ref{sec:wfc3}), and \textit{Spitzer} (independent analysis; Section~\ref{sec:spitzer}) transit observations. We determined the host star has, at most, a low level of stellar activity ($\log L_{\rm X}/L_{\rm bol}<-5.2$) using both visible and X-ray photmetric monitoring observations (Section~\ref{sec:activity}). 


We performed two complementary retrieval analyses: a relatively strict PLATON analysis (Section~\ref{subsec:platon}, Section~\ref{sec:results}) assuming chemical equilibrium and solar metal ratios (except carbon), and a more flexible AURA free-chemistry retrieval (Section~\ref{subsec:AURA}, Section~\ref{subsec:AURA_main}). Both methods' fiducial models are excellent fits to the entire transit spectrum. We further tested an array of more complicated models (Sections~\ref{sec:platon_retrieval} and \ref{subsec:AURA_shifts}), including instrumental transit depth biases (offsets), parametric rayleigh scattering, partial cloud coverage, Mie scattering (PLATON only), and stellar activity (PLATON only). We find the conclusions to be insensitive to model choice within a paradigm.

Despite PLATON and AURA's differing model assumptions, priors, and even opacity sources, we find several shared conclusions between the two methods (Section~\ref{subsec:compare}). Both PLATON and AURA retrieve a high atmospheric metallicity (O/H) that is inconsistent with Z$_{\odot}$ to greater than 2-$\sigma$ ($\log_{10}{Z/Z_{\odot}} = 1.46^{+0.53}_{-0.68}$ compared to $\log_{10}{Z/Z_{\odot}} = 2.33^{+0.23}_{-0.25}$, respectively). They also both are consistent with a haze-free and cloud-free atmosphere, and both find a decisive water vapor detection and at least suggestive evidence of an optical absorption feature. 
We futher confirm the result by performing a middle-ground retrieval, ATMO, and find results generally consistent with AURA's (Section~\ref{subsec:assume}). We determine the inclusion of H$_2$-broadened sodium opacity impacts the retrieved metallicities. While we consider AURA to be more physically plausible due to its consistency with interior modeling constraints and inclusion of H$_2$-broadened sodium opacity, we present the results from both PLATON and AURA as assumption-dependent orthogonal analyses. Overall, our study emphasizes the importance of comparative retrievals with different forward modeling, prior, and model selection assumptions in order to best contextualize presented results. 

{ Acknowledgements} The authors are grateful to Michael Zhang and Yayaati Chachan for many helpful discussions about PLATON. We thank Daniel Thorngren for providing the interior modeling metallicity constraint, and Daniel Kitzmann for providing the refractive indices of the Mie scattering condensates. We thank Eliza Kempton for productive conversations. This work is based on observations from the Hubble
Space Telescope, operated by AURA, Inc. on behalf of
NASA/ESA. This work also includes observations made
with the Spitzer Space Telescope, operated by the Jet
Propulsion Laboratory, California Institute of Technology
under a contract with NASA. KBS acknowledges support from CRESST, and funding from HST grant HST-GO-14767. AMM acknowledges support from HST grant HST-GO-14260, and the GSFC Sellers Exoplanet Environments Collaboration (SEEC), which is funded in part by the NASA Planetary Science Division’s Internal Scientist Funding Model. JSF acknowledges support from the Spanish State Research Agency project AYA2016-79425-C3-2-P. This work has made use of data from the European Space Agency (ESA) mission
{\it Gaia} (\url{https://www.cosmos.esa.int/gaia}), processed by the {\it Gaia}
Data Processing and Analysis Consortium (DPAC,
\url{https://www.cosmos.esa.int/web/gaia/dpac/consortium}). Funding for the DPAC
has been provided by national institutions, in particular the institutions
participating in the {\it Gaia} Multilateral Agreement.
\software{IRAF \citep{iraf1,iraf2}, SciPy \citep{scipy}, Matplotlib \citep{matplotlib}, nestle (\url{https://github.com/kbarbary/nestle}), BATMAN \citep{kreidberg2015a}, Kapetyn \citep{kapetyn}, Corner.py \citep{foremanmackey2016}, AstroPy \citep{astropy}, PLATON \citep{zhang2019}, Dynesty \citep{higson2019}, NumPy \citep{numpy}, Pandas \citep{pandas}}

\section{Postscript}
After our retrieval analysis was complete, we were made aware of a recently accepted paper which used a model-grid fit to claim that a combined transit spectrum using HST's WFC3 UV/Visible channel (UVIS) and independently derived \textit{Spitzer} depths is best fit by a high metallicity atmosphere \citep{wakeford2020}. While we were aware that a manuscript was being submitted comparing STIS and UVIS data, we were unaware of the atmospheric metallicity claim. Since we were unaware of that result during our analysis, our conclusions were not biased by the results claimed in that paper. 

\bibliography{bibtex}

\appendix

\section{HST Spectrophotometric Light Curve Fits}
\begin{figure}[b]
\centering
{
\includegraphics[width=0.75\textwidth,keepaspectratio]{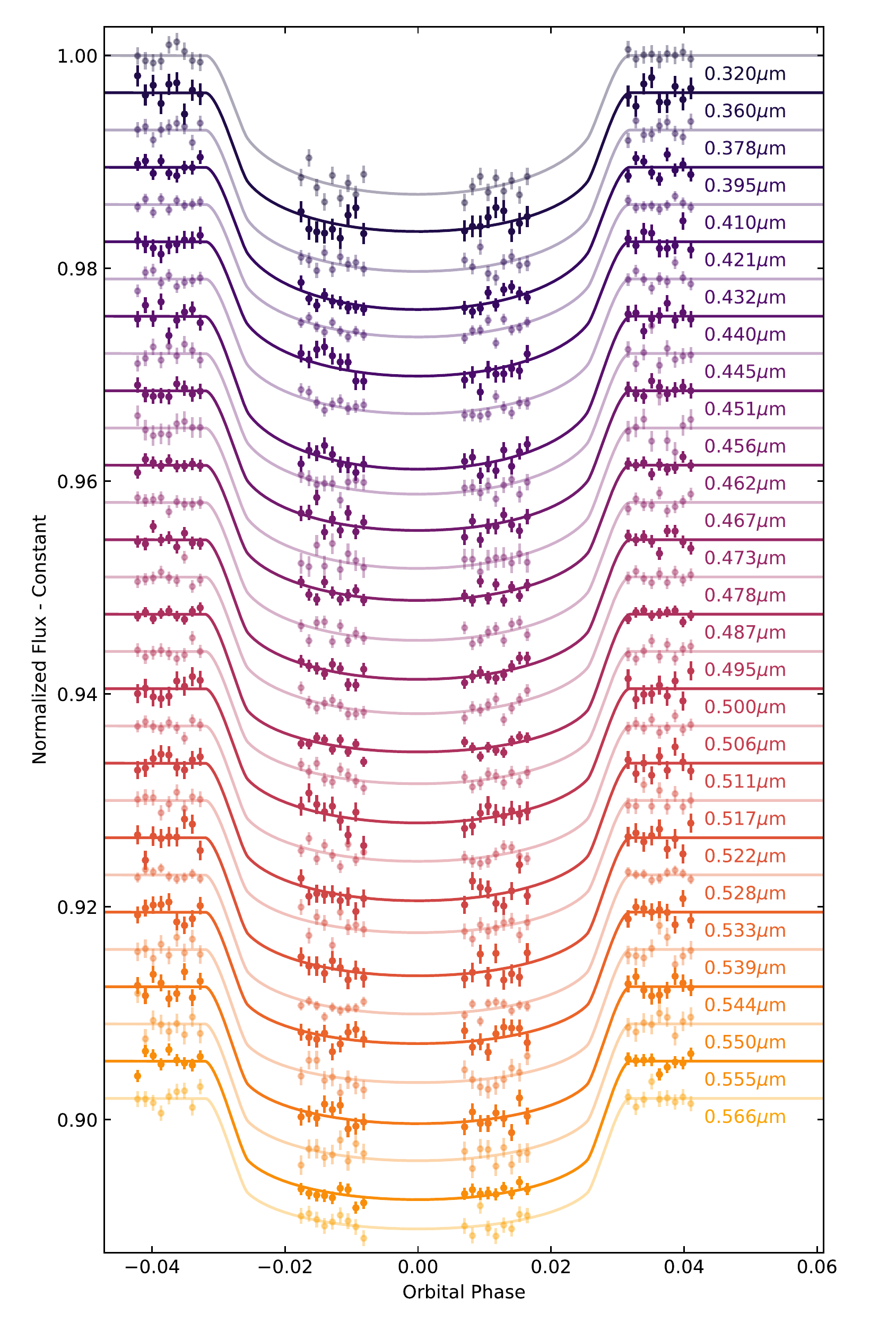}
\caption{Spectral light curves for STIS G430L, visit 83.}
\label{fig:stis83_bins}
}
\end{figure} 

\begin{figure}[b]
\centering
{
\includegraphics[width=0.75\textwidth,keepaspectratio]{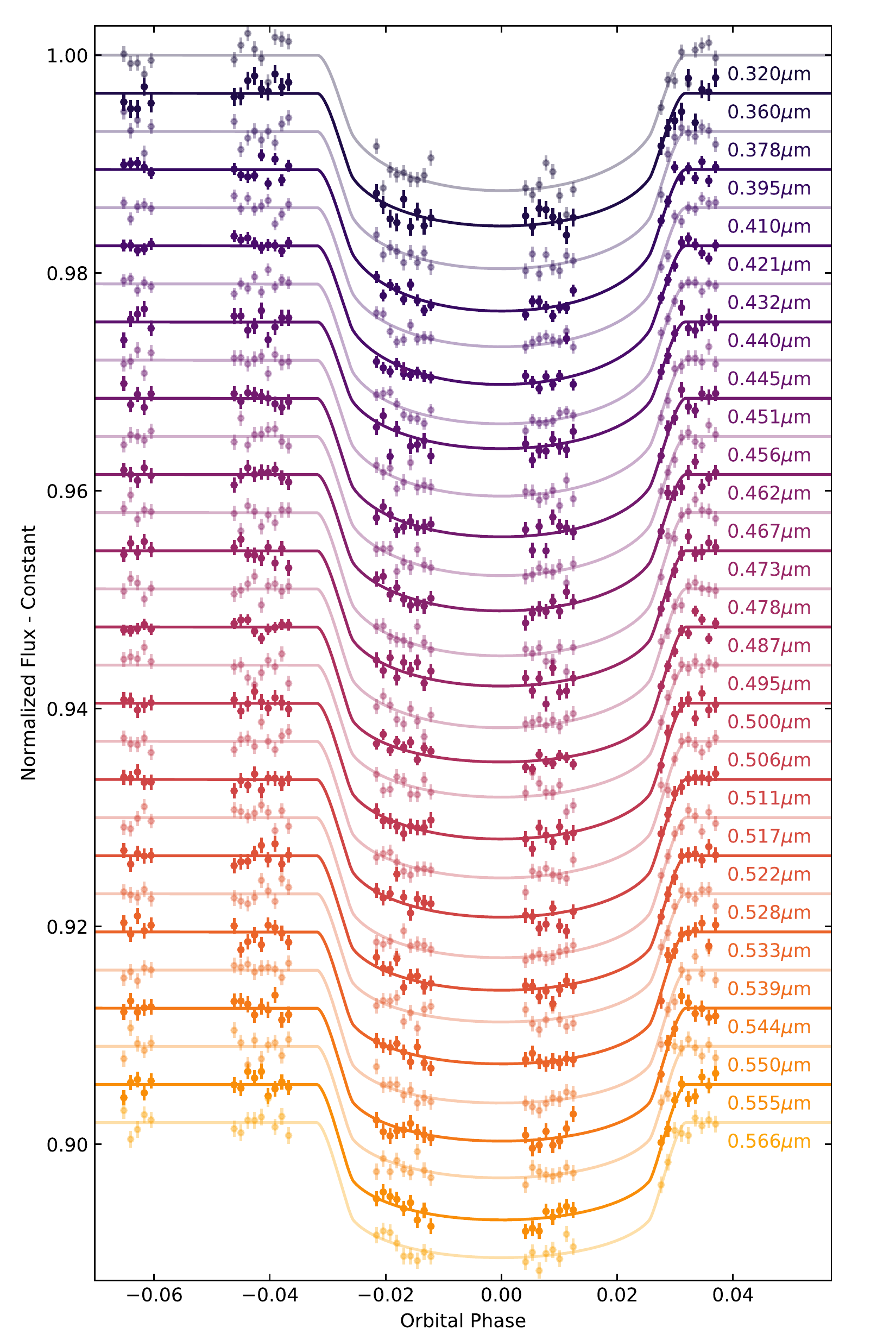}
\caption{Spectral light curves for STIS G430L, visit 84.}
\label{fig:stis84_bins}
}
\end{figure} 

\begin{figure}[b]
\centering
{
\includegraphics[width=0.75\textwidth,keepaspectratio]{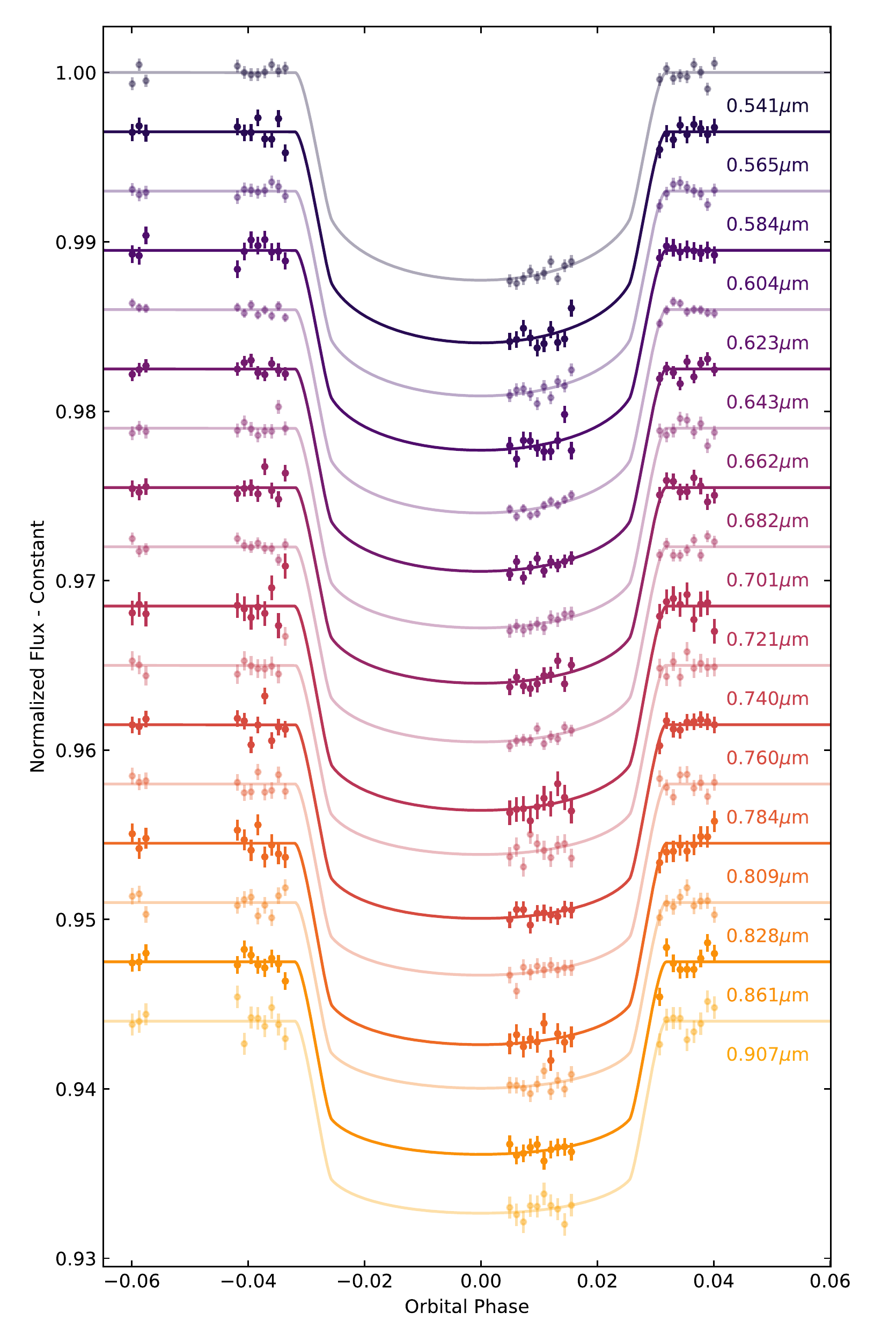}
\caption{Spectral light curves for STIS G750L, visit 85.}
\label{fig:stis85_bins}
}
\end{figure} 

\begin{figure}[b]
\centering
{
\includegraphics[width=0.75\textwidth,keepaspectratio]{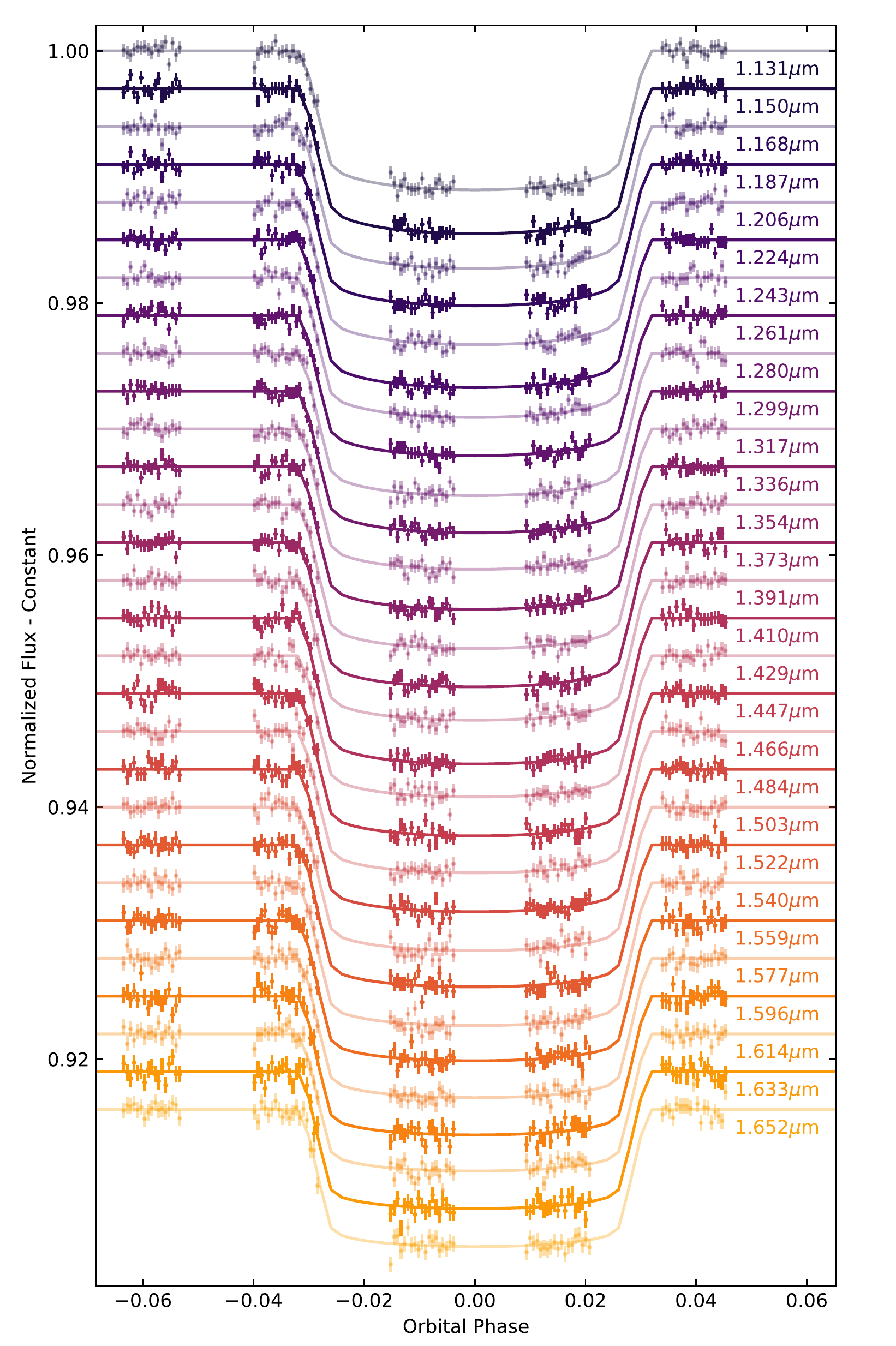}
\caption{Spectral light curves for single WFC3 visit.}
\label{fig:wfc_bins}
}
\end{figure} 

\end{document}